\documentclass[12pt]{article}
\pdfoutput=1
\usepackage{jheppub}
\usepackage[skip=2pt,font=footnotesize]{caption}
\usepackage[utf8]{inputenc}
\usepackage{verbatim}
\usepackage{amsmath}
\usepackage{amssymb}
\usepackage{amsthm}
\usepackage{amsfonts}
\usepackage{makecell}
\usepackage{tabularx}
\usepackage{comment}
\usepackage{subfigure}
\usepackage{float}

\usepackage{hyperref}
\usepackage{psfrag}
\usepackage{comment}

\def\1{{\rm 1-loop}}

\newcommand\numberthis{\addtocounter{equation}{1}\tag{\theequation}}
\usepackage{braket}

\def\<{\langle}
\def\>{\rangle}

\newcommand \s {\sigma}
\newcommand   \f  {\phi}

\newcommand{\bea}{\begin{eqnarray}}
\newcommand{\eea}{\end{eqnarray}}

\def\O{{\cal O}}

\newcommand {\be} {\begin {equation}}
\newcommand {\ee} {\end {equation}}

\newcommand {\bes} {\begin {equation*}}
\newcommand {\ees} {\end {equation*}}

\renewcommand{\Re}{{\text{Re}\hspace{.05cm}}}
\renewcommand{\Im}{{\text{Im}\hspace{.05cm}}}

\newcommand{\beq}{\begin{equation}}
\newcommand{\eeq}{\end{equation}}

\def\be{ \begin{equation} }
\def\ee{ \end{equation} }

 \newmuskip\pFqmuskip

\usepackage{enumitem}

\newlist{primenumerate}{enumerate}{1}
\setlist[primenumerate,1]{label={3$'$.}}

\newcommand*\pFq[6][8]{%
  \begingroup 
  \pFqmuskip=#1mu\relax
  \mathcode`\,=\string"8000
  \begingroup\lccode`\~=`\,
  \lowercase{\endgroup\let~}\pFqcomma
  {}_{#2}F_{#3}{\left[\genfrac..{0pt}{}{#4}{#5};#6\right]}%
  \endgroup
}
\newcommand{\pFqcomma}{\mskip\pFqmuskip}

\makeatletter
\renewcommand{\@maketitle}{
\newpage
 \begin{center}%
  {\large\bfseries \@title \par}%
 \end{center}%
 \par} \makeatother

\numberwithin{equation}{section}

\begin{document}
\hfill \hbox{{CALT-TH-2020-032}}

\title{CFT Unitarity and the AdS Cutkosky Rules}

\author[a]{David Meltzer}
\author[b]{and Allic Sivaramakrishnan}

\affiliation[a]{Walter Burke Institute for Theoretical Physics, California Institute of Technology,\\ Pasadena, California, 91125}
\affiliation[b]{Department of Physics and Astronomy, University of Kentucky, Lexington, KY, 40506}

\emailAdd{dmeltzer@caltech.edu}
\emailAdd{allicsiva@uky.edu}

\abstract{ 
We derive the Cutkosky rules for conformal field theories (CFTs) at weak and strong coupling. 
These rules give a simple, diagrammatic method to compute the double-commutator that appears in the Lorentzian inversion formula. We first revisit weakly-coupled CFTs in flat space, where the cuts are performed on Feynman diagrams. We then generalize these rules to strongly-coupled holographic CFTs, where the cuts are performed on the Witten diagrams of the dual theory. In both cases, Cutkosky rules factorize loop diagrams into on-shell sub-diagrams and generalize the standard S-matrix cutting rules. These rules are naturally formulated and derived in Lorentzian momentum space, where the double-commutator is manifestly related to the CFT optical theorem. Finally, we study the AdS cutting rules in explicit examples at tree level and one loop. In these examples, we confirm that the rules are consistent with the OPE limit and that we recover the S-matrix optical theorem in the flat space limit. The AdS cutting rules and the CFT dispersion formula together form a holographic unitarity method to reconstruct Witten diagrams from their cuts.
}

\maketitle

\section{Introduction}
\label{sec:intro}
The modern S-matrix program has transformed our understanding of quantum field theory (QFT). On-shell methods have driven progress in scattering amplitudes by simplifying computations and revealing new physics. Instead of working with individual Feynman diagrams, which can be numerous and gauge-dependent, one works in terms of on-shell building blocks. Constraints from locality, causality, and unitarity then help determine the full amplitude. In particular, unitarity methods allow us to calculate loop-level amplitudes from lower-loop on-shell quantities \cite{Bern:1994zx,Bern:1994cg}.\footnote{We refer the reader to \cite{Bern:2011qt,Elvang:2013cua,Dixon:2013uaa, Henn:2014yza,Arkani-Hamed:2016byb,Cheung:2017pzi} for references and reviews.} Unitarity is a powerful tool for exploring loop-level structure in the S-matrix that is not manifest at the level of the action, including Yangian symmetry \cite{ArkaniHamed:2010kv}, color-kinematic duality, and the double-copy property of gravity theories \cite{Bern:2008qj,Bern:2010ue}. The arsenal of modern amplitude methods brings difficult computations into reach, opening paths to novel ultraviolet physics \cite{Bern:2018jmv}.

By contrast, we understand far less about perturbation theory in the absence of an S-matrix. For example, how do structures uncovered in flat space amplitudes generalize to theories in Anti-de Sitter (AdS) space? While the S-matrix can be recovered from the flat space limit of AdS/CFT \cite{Susskind:1998vk,Polchinski:1999ry,Heemskerk:2009pn,Gary:2009ae,Penedones:2010ue, Raju:2012zr}, one cannot define an S-matrix in global AdS itself as an overlap of in- and out-states \cite{Balasubramanian:1999ri}. Instead, the asymptotic observables are the correlation functions of the boundary CFT. Like flat-space amplitudes, boundary correlators obey notions of locality, causality, and unitarity, and so it is natural to expect that S-matrix technology may be applicable to AdS/CFT. Furthermore, the program of studying AdS/CFT correlators using an on-shell approach has made remarkable progress, for example by leveraging the constraints of crossing symmetry \cite{Heemskerk:2009pn} and writing correlators in Mellin space \cite{Mack:2009mi,Penedones:2010ue,Rastelli:2016nze}. However, AdS analogues of basic S-matrix ideas remain unknown. 

In this paper, we study the following question: what are the on-shell building blocks of $1/N$ perturbation theory in AdS? Our starting point will be the Cutkosky rules, i.e. that the discontinuity of a Feynman diagram can be calculated by cutting internal lines \cite{Cutkosky:1960sp,Eden:1966dnq}. These cuts place lines on shell by replacing the time-ordered propagator with the corresponding Wightman, or on-shell, propagator. The cutting rules underpin the optical theorem for the S-matrix, ensuring that the discontinuity of an amplitude factorizes into products of on-shell sub-amplitudes. As we review\footnote{For modern work on the flat-space cutting rules see \cite{Abreu_2014,Abreu_2017,Bourjaily:2020wvq}.} and show in explicit examples, these and more general cutting rules follow from basic Lorentzian properties of QFT correlators \cite{Veltman:1963th} and persist in curved space. 

The cutting rules we explore are directly related to the Lorentzian inversion formula \cite{Caron-Huot:2017vep}, a centerpiece of modern CFT unitarity methods. The inversion formula is a CFT generalization of the Froissart-Gribov formula for the S-matrix \cite{Gribov:1961fr,Froissart:1961ux} and has generated recent progress in the study of higher-dimensional CFTs. For example, the inversion formula proves the existence of Regge trajectories and leads to a dispersion formula for CFT four-point functions \cite{Carmi:2019cub}. In the CFT dispersion formula, the four-point function is reconstructed from its much simpler \textit{double-commutator}.\footnote{As with the dispersion formula for scattering amplitudes in QFT, there are possible polynomial ambiguities affecting operators of bounded spin.} In the context of AdS/CFT, the double-commutator reduces the loop order. That is, the double-commutator of an $L$-loop one-particle irreducible Witten diagram can be computed in terms of $(L-1)$-loop data. The CFT dispersion formula then provides a way to bootstrap loop-level physics purely from tree-level data. The double-commutator then plays the same role in the CFT dispersion formula that the discontinuity of the amplitude plays in the S-matrix dispersion formula \cite{Eden:1966dnq}.

Previously, it was unclear how the double-commutator could be computed via Cutkosky rules in CFTs, at either weak or strong coupling. The goal of this work is to derive these rules for both classes of CFTs. To compute the double-commutator, we will classify the corresponding \textit{unitarity cuts} of Feynman diagrams in weakly-coupled CFTs and of Witten diagrams in the AdS dual of holographic CFTs. In the holographic case, the set of allowed cuts agrees with the previous bulk analysis \cite{Meltzer:2019nbs}. We also show that, for certain kinematics, the CFT optical theorem computes the double-commutator appearing in the inversion formula.
The relationships between the cutting rules, the CFT optical theorem, and the inversion formula are all elementary properties of CFTs and therefore extend the S-matrix unitarity method to a wider class of theories. 

To derive the cutting rules, we will follow a somewhat historical route and import Veltman's derivation of the flat-space cutting rules, via the largest-time equation \cite{Veltman:1963th}, to AdS. This approach\footnote{An alternative strategy would be to derive the Feynman rules for the CFT double-commutator directly using time-folded contours i.e. the Schwinger-Keldysh formalism \cite{Schwinger:1960qe,Keldysh:1964ud,Chou:1984es,Stanford:2015owe,Haehl:2016pec,Haehl:2017qfl,Murugan:2017eto}.} gives a simple derivation of the cutting rules and makes manifest that AdS unitarity methods are a direct generalization of the standard flat space methods. In practice, our approach amounts to replacing the double-commutator with a simpler out-of-time-ordered correlator.

While conformal field theories are typically studied in position space, one theme of this work is that Lorentzian momentum space is convenient for studying unitarity. For instance, the derivation of the allowed cuts takes a simple form in momentum space. A cut diagram has a natural interpretation in momentum space as well: it is the gluing of two sub-diagrams via a phase space integral, which is equivalent to summing over physical exchanged states. 
The study of AdS/CFT correlators in momentum space is also motivated by their relation to cosmological observables, see e.g.
\cite{Maldacena:2002vr,Maldacena:2011nz,Mata:2012bx,Kundu:2014gxa,Ghosh:2014kba,Arkani-Hamed:2015bza,Kundu:2015xta,Sleight:2019mgd,Sleight:2019hfp,Sleight:2020obc,Arkani-Hamed:2017fdk,Arkani-Hamed:2018bjr,Benincasa:2018ssx,Benincasa:2019vqr,Arkani-Hamed:2018kmz,Baumann:2019oyu,Baumann:2020dch}, and the study of on-shell AdS recursion relations \cite{Raju:2010by,Raju:2011mp,Raju:2012zr,Raju:2012zs}.\footnote{For further work on AdS/CFT in momentum space see \cite{Isono:2018rrb,Isono:2019wex,Farrow:2018yni,Lipstein:2019mpu,Albayrak:2018tam,Albayrak:2019asr,Albayrak:2019yve,Albayrak:2020isk,Albayrak:2020bso}.} 
In this work we will only study the cutting rules in momentum space, although they are also applicable to AdS/CFT correlators in position and Mellin space.

Finally, it is useful to compare our approach to other studies of unitarity in AdS/CFT. One well-established method to compute loops in AdS is via the bootstrap equations. Here one determines the operator product expansion (OPE) of the CFT at tree level by solving the crossing equations \cite{Heemskerk:2009pn}. One can then plug this data into the loop-level crossing equations \cite{Aharony:2016dwx}, or equivalently use it to compute the double-commutator \cite{Caron-Huot:2017vep, Alday:2017vkk}, to solve for loop-level OPE data. This gives a boundary method to bootstrap loops in AdS purely from tree-level data. A related approach is the Euclidean bulk method \cite{Ponomarev:2019ofr,Meltzer:2019nbs}. In this approach, one determines the boundary OPE data by studying Witten diagrams themselves and working in Euclidean signature. The split representation for bulk-to-bulk propagators \cite{Leonhardt:2003qu,Costa:2014kfa} and on-shell conditions in CFT spectral space together lead to an efficient method to study the double-commutator. The bulk and boundary methods, which are ultimately equivalent \cite{Meltzer:2019nbs}, give the double-commutator in terms of the boundary OPE data. In this work we instead study AdS in Lorentzian signature and express the double-commutator directly as a sum over cut Witten diagrams, bypassing the boundary OPE. 
The cuts we study factorize AdS diagrams using bulk normalizable modes and therefore make AdS locality and factorization manifest.
As we review in Section \ref{sec:ReviewCutkosky}, the definition of a cut diagram we use here is a direct generalization of the one used in flat-space unitarity methods.

\subsection*{Summary of results}
We will now give a summary of the main results, followed by an outline of the paper. Unless stated otherwise, we will study scalar field theories in flat space and AdS throughout this work.\footnote{Strictly speaking, the boundary correlators for a QFT in AdS define a conformal theory that does not have a stress-tensor and is therefore non-local \cite{Heemskerk:2009pn,Paulos:2016fap}. However, all of our results can be generalized to study theories of gravity in AdS with a local CFT dual.} The cutting rules will be derived for CFTs in general spacetime dimension and we will only specialize to specific dimensions when computing examples.
The derivations of the cutting rules in weakly and strongly-coupled CFTs will be essentially the same and rely on working in Lorentzian signature. Specifically, we will need the following properties of Lorentzian CFTs:
\begin{enumerate}
\item \textbf{Positive Spectrum.} The space of physical states $|\Psi(k)\>$ have momentum $k$ lying in the forward lightcone, $k^{2}\leq0$ and $k^0\geq0$. We use the mostly-plus signature for the metric $\eta_{\mu\nu}$.
\item \textbf{Causality.} Operators commute at spacelike separation:
\begin{align}
[\f(x),\f(y)]=0 \quad \text{ for } \quad (x-y)^{2}>0~. 
\end{align}
\item \textbf{CFT Optical Theorem.}\footnote{While this identity is valid in general QFTs, we will refer to this as the CFT optical theorem to avoid confusion with the S-matrix optical theorem.} We use the following combinatoric identity \cite{Gillioz:2016jnn,Gillioz:2018kwh,Gillioz:2018mto} for partially time-ordered operators:
\begin{align}
\sum\limits_{r=0}^{n}(-1)^r\sum\limits_{\sigma\in\Pi(r,n-r)}\overline{T}[\f(x_{\sigma_1})\ldots\f(x_{\sigma_r})]T[\f(x_{\sigma_{r+1}})\ldots\f(x_{\sigma_{n}})]=0~. \label{eq:opticalV0}
\end{align}
Here $\overline{T}$ and $T$ are the (anti-)time-ordering symbols and $\Pi(r,n-r)$ is the set of partitions of $\{1,\ldots,n\}$ into two sets of size $r$ and $n-r$.
\end{enumerate}
The CFT optical theorem can be verified at low points by using the definition of the (anti-)time-ordering symbol and checking that the $\theta$-functions cancel. It then follows for all $n$ by induction. 

To derive CFT cutting rules, we begin by using the CFT optical theorem to relate the real part of a time-ordered four-point function to a double-commutator \cite{Polyakov:1974gs}:
\begin{align}
-2~\Re \<T[\f(k_1)\f(k_2)\f(k_3)\f(k_4)]\>=\<[\f(k_3),\f(k_4)]_{A}[\f(k_1),\f(k_2)]_{R}\>~, \label{eq:FromReToDC}
\end{align}
where the subscripts $A,R$ indicate the advanced and retarded commutators,
\begin{align}
[\f(x_1),\f(x_2)]_{A}=\theta(x_2^0-x_1^0)[\f(x_1),\f(x_2)]~,
\\
[\f(x_1),\f(x_2)]_{R}=\theta(x_1^0-x_2^0)[\f(x_1),\f(x_2)]~.
\end{align}
We will refer to the right-hand side of \eqref{eq:FromReToDC} as the causal double-commutator. This is the same double-commutator that appears in the CFT inversion formula and can be computed by taking a double-discontinuity of the correlator in cross-ratio space \cite{Caron-Huot:2017vep}.\footnote{The fact that the inversion formula involves a causal double-commutator is manifest in \cite{ssw,Kravchuk:2018htv},  where the causal restrictions are put into the definition of the integration region.} The identity \eqref{eq:FromReToDC} only holds for the kinematics
\begin{align}
&k_{i}^{2}>0, \quad (k_i+k_j)^{2}>0 \quad \text{except for} \quad k_1+k_2\in V_+, \quad k_3+k_4 \in V_{-}~, \label{eq:kinematics}
\end{align}
where $V_{\pm}$ are the closed forward and backward lightcones,
\begin{align}
V_{\pm}=\{k \ | \ k^{2}\leq0, \ \pm k^{0}\geq0\}~.
\end{align}

\begin{figure}
\begin{center}
\includegraphics[scale=.25]{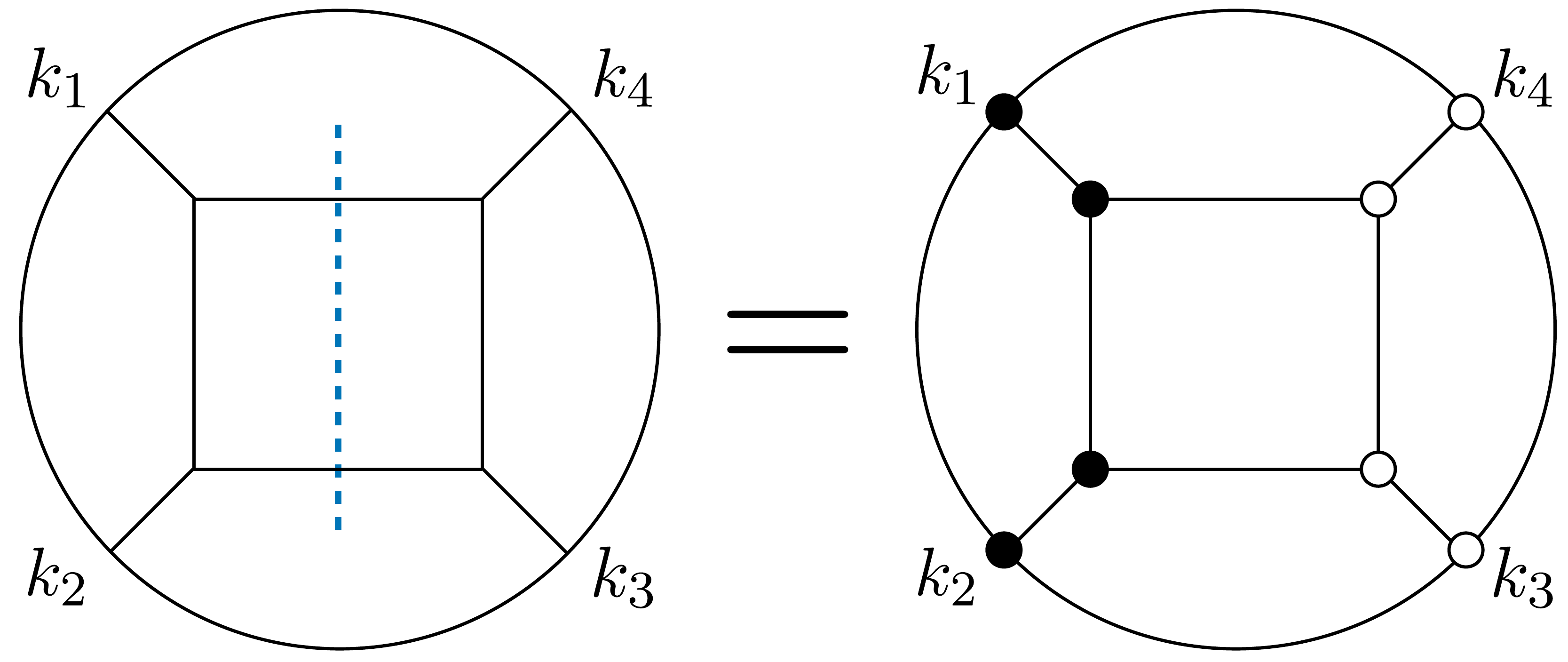}
\end{center}
\caption{The map from the cutting to coloring rules for the AdS box diagram.}
\label{AdS_Box_Example}
\end{figure}

Next, we derive the diagrammatic rules for the left-hand side of \eqref{eq:FromReToDC} by generalizing Veltman's derivation of the Cutkosky rules for the S-matrix \cite{Veltman:1963th}. Veltman's derivation is based on the largest-time equation, which is a general relationship between Feynman diagrams for Lorentzian QFTs. Crucially, this relationship holds in both flat space and AdS. The largest-time equation involves a set of ``cut" graphs, which come from introducing black and white vertices for both internal and external points in the original diagram.\footnote{These are the circling rules of \cite{Veltman:1963th} and should not be confused with on-shell diagrams.} The largest-time equation states that the sum over all possible colorings of the vertices vanishes. This is simply the graphical version of the CFT optical theorem \eqref{eq:opticalV0}. Finally, there is a one-to-one correspondence between the set of non-vanishing graphs with these two types of vertices and the cut graphs of Cutkosky \cite{Cutkosky:1960sp,Eden:1966dnq,Veltman:1963th} (e.g. see figure \ref{AdS_Box_Example}).

The relation \eqref{eq:FromReToDC} turns the cutting rules for $\Re \<T[\f\f\f\f]\>$ into the cutting rules for the causal double-commutator $\<[\f,\f]_{A}[\f,\f]_{R}\>$ in the restricted kinematics \eqref{eq:kinematics}. To analytically continue to general momenta, we use that the retarded commutator $[\f(x_1),\f(x_2)]_{R}$ is only non-zero if $x_{1}$ is in the causal future of $x_2$. Using standard arguments for Laplace transforms of Wightman functions, this causality condition in position space translates into an analyticity property in momentum space \cite{Streater:1989vi,Haag:1992hx}. Using this property for both commutators, we analytically continue away from the restricted kinematics \eqref{eq:kinematics} and prove the cutting rules for the causal double-commutator with general momenta. Here we are analytically continuing only the double commutator and not $\Re \<T[\f\f\f\f]\>$, which in general differs from $\<[\f,\f]_{A}[\f,\f]_{R}\>$ for generic momenta.

For the reader interested in the final result, we will now summarize the AdS cutting rules for $\<[\f(k_3),\f(k_4)]_{A}[\f(k_1),\f(k_2)]_{R}\>$ in the Poincar\'e patch. This double-commutator is only non-zero for $k_1+k_2\in V_{+}$, so the momentum necessarily runs from the left to the right. Aside from satisfying momentum conservation, the other momenta are left generic. The cutting rules for an individual Witten diagram are as follows:
\begin{enumerate}
\item Draw a cut that crosses only bulk-to-bulk propagators. For each of these cut propagators, replace the time-ordered, or Feynman, propagator $G_{\Delta}(k,z_i,z_j)$ by the corresponding on-shell propagator $G^{+}_{\Delta}(k,z_i,z_j)$.
\item For each propagator to the (right) left of the cut, use the (anti-)time-ordered propagator. 
\item For each internal vertex to the left of the cut, multiply by $ig$. For internal vertices to the right of the cut, multiply by $-ig$. Here $g$ is the bulk coupling.
\item Sum over all cuts consistent with momentum conservation.
\end{enumerate}
The AdS on-shell propagator, $G^{+}_{\Delta}(k,z_i,z_j)$, is a two-point Wightman function for a free scalar in AdS. This is precisely the same structure as in flat space: the on-shell propagator in flat space is by definition a free-field two-point Wightman function, which is a $\delta$-function in momentum space. The standard cutting rules for the S-matrix correspond to replacing cut lines by two-point Wightman functions. In the examples we study, we will find that cut AdS diagrams reduce to the corresponding cut scattering amplitudes in the flat space limit, confirming that we have generalized the flat-space methods to AdS.

\begin{figure}
\begin{center}
\includegraphics[scale=.25]{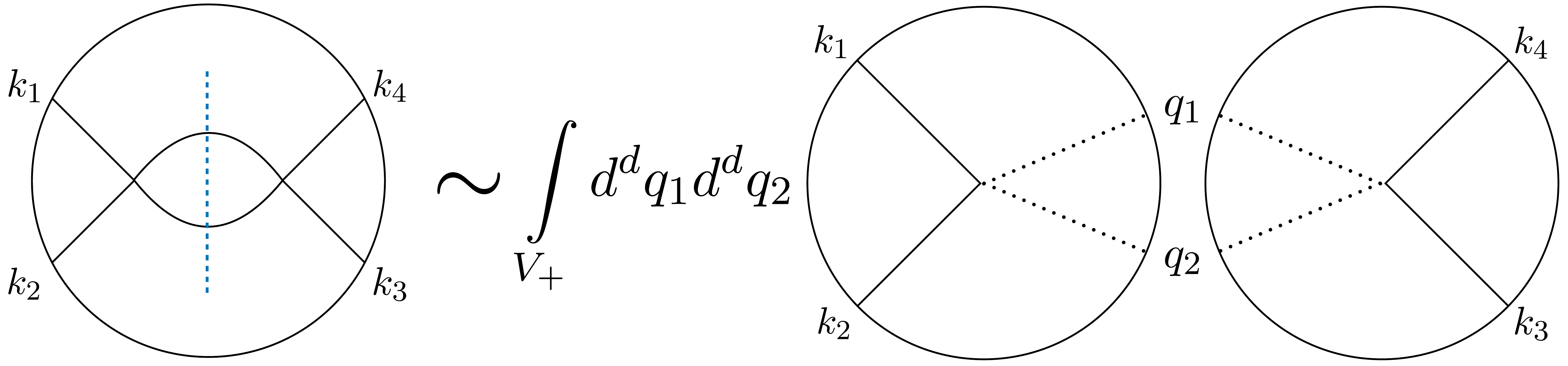}
\caption{A cut bubble is the on-shell gluing of contact diagrams: the undotted lines are Feynman propagators and the dotted lines are Wightman, or on-shell, propagators.}
\label{fig:Cutbubble_Transition}
\end{center}
\end{figure}

In AdS we also find that the on-shell propagator $G^{+}_{\Delta}(k,z_1,z_2)$ has a simple split representation in terms of the on-shell, or Wightman, bulk-to-boundary propagator $K^{+}_{\Delta}(k,z)$:
\begin{align}
G^{+}_{\Delta}(k,z_1,z_2)&\propto (\sqrt{-k^{2}})^{d-2\Delta}K^{+}_{\Delta}(k,z_1)K^{+}_{\Delta}(k,z_2)~.
\end{align}
Diagrams with on-shell bulk-to-boundary propagators are known as \textit{transition amplitudes} and have been studied in the context of recursion relations \cite{Raju:2010by,Raju:2011mp,Raju:2012zr,Raju:2012zs}. The on-shell bulk-to-boundary propagators correspond to normalizable solutions of the bulk equations of motion \cite{Balasubramanian:1998sn,Balasubramanian:1998de,Balasubramanian:1999ri}. This aligns with our interpretation of a cut diagram as a sum over states. We have drawn this schematically in figure \ref{fig:Cutbubble_Transition} and provide more details in Section \ref{sec:UnitarityCutsAdS}.

\subsection*{Outline}

In Section \ref{sec:Polyakov} we review the CFT optical theorem and how to relate the real part of a four-point function to the causal double-commutator. In Section \ref{sec:ReviewCutkosky} we use this identity and the largest-time equation to derive the double-commutator cutting rules for weakly coupled theories in flat space. In Section \ref{sec:UnitarityCutsAdS} we generalize this argument to correlation functions in AdS/CFT and introduce the transition amplitudes. In Section \ref{sec:Examples} we discuss the OPE and flat space limit of cut AdS diagrams, and give examples at tree and loop level. We conclude with a summary and discussion of future directions in Section \ref{sec:Discussion}. In Appendix \ref{app:largest_time} we review the largest-time equation. In Appendix \ref{app:Analyticity_k_Space} we summarize the analyticity properties of the double-commutator in momentum space. In Appendix \ref{sec:FeynmanTree} we give a short derivation of the Feynman tree theorem in AdS. Finally, in Appendix \ref{sec:SKDerivation} we give an alternative derivation of the cutting rules using the Schwinger-Keldysh formalism.

\section{CFT Unitarity Conditions}
\label{sec:Polyakov}

In this section we review how unitarity conditions apply to the full correlator and derive \eqref{eq:FromReToDC}. This identity is useful because it relates the real part of a time-ordered correlator, which can be computed via a set of cutting rules for theories with a weakly-coupled description, to the causal double-commutator, which is the central element of AdS/CFT unitarity methods \cite{Aharony:2016dwx,Caron-Huot:2017vep,Meltzer:2019nbs}. 
This section will be a review of \cite{Polyakov:1974gs}, although we will follow the presentation of \cite{Gillioz:2016jnn,Gillioz:2018kwh,Gillioz:2018mto}. The identities reviewed here will hold for all causal unitary QFTs and do not rely on assuming either conformal invariance or weak coupling.\footnote{The arguments in this section will rely on using $\theta$-function identities, which can be subtle in general non-perturbative QFTs. However, all the identities derived in this section can also be derived by using the axiomatic definition of the time-ordered product \cite{Bogolyubov:1990kw}, see \cite{Meltzer:2021bmb} for a review.}

\begin{figure}
\begin{center}
\includegraphics[scale=.2]{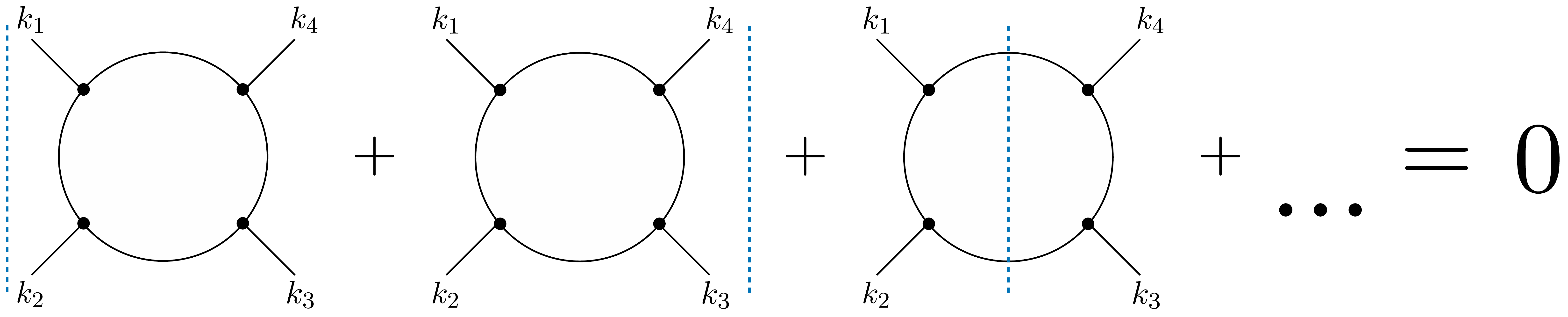}
\end{center}
\caption{Optical theorem for a QFT four-point function. The grey disk represents a general correlation function and not a Feynman diagram. The external lines label the momentum of the external operator. Operators to the (right) left of the blue line are (anti-)time-ordered in the four-point functions given in \eqref{eq:CombCor}.}
\label{fig:graphicaloptical}
\end{figure}

The proof of \eqref{eq:FromReToDC} follows from the CFT optical theorem and the positive spectrum condition. The four-point version of the CFT optical theorem states \cite{Gillioz:2016jnn,Gillioz:2018kwh,Gillioz:2018mto}:
\begin{align}
0=\<\overline{T}[\f(k_1)\f(k_2)\f(k_3)\f(k_4)]\>\hspace{.1cm}+\hspace{.1cm}&\<T[\f(k_1)\f(k_2)\f(k_3)\f(k_4)]\> 
\nonumber \\ +\hspace{.1cm}&\<\overline{T}[\f(k_3)\f(k_4)]T[\f(k_1)\f(k_2)]\> \nonumber 
\nonumber  \\ 
-\hspace{.1cm}&\<\f(k_1)T[\f(k_2)\f(k_3)\f(k_4)]\>
\nonumber \\   \vspace{.1cm}
-\hspace{.1cm}&\<\overline{T}[\f(k_2)\f(k_3)\f(k_4)]\f(k_1)\>
+\left( \text{partitions} \right)
~,\label{eq:CombCor}
\end{align}
where for convenience we have gone to momentum space and suppressed the other partitions of the external operators. This equation is shown graphically in figure \ref{fig:graphicaloptical} and says that summing over all ``cuts" of a four-point function vanish. What we call a cut here refers to how the four external operators are grouped under the (anti-)time-ordering symbols. For example, the first three correlators of \eqref{eq:CombCor} correspond to the first three diagrams shown in figure \ref{fig:graphicaloptical}. In Sections \ref{sec:ReviewCutkosky} and \ref{sec:UnitarityCutsAdS} we explain how to compute cuts of Feynman and Witten diagrams.

To simplify this equation, it is useful to choose the momentum to lie in the configuration \eqref{eq:kinematics}. The salient feature of these kinematics is that only the sum $k_1+k_2\in V_+$ and is therefore on shell. By restricting to this configuration, we ensure that only $s$-channel cuts contribute to $\Re\<T[\f\f\f\f]\>$. This means only the three diagrams shown in figure \ref{fig:graphicaloptical} are non-zero. The other cut diagrams (not shown) all vanish. To show this, we will use the identities:
\begin{align}
\f(k_i)|0\>&=0 \quad \text{ if } \ k_{i}\notin V_+~,
\\
T[\f(k_i)\f(k_j)]|0\>&=0 \quad \text{ if } \ k_i+k_j\notin V_+~.
\end{align}
Both equalities follow from the fact that all the states in the physical Hilbert space, $\mathcal{H}$, have momentum in the forward lightcone $V_{+}$. For the chosen kinematics, the CFT optical theorem then becomes
\begin{align}
-2\hspace{.1cm}\Re \<T[\f(k_1)\f(k_2)\f(k_3)\f(k_4)]\>=\<\overline{T}[\f(k_3)\f(k_4)]T[\f(k_1)\f(k_2)]\>~, \label{eq:CFTOptical}
\end{align}
where we used that $\<T[\f\f\f\f]\>+\<\overline{T}[\f\f\f\f]\>$ gives twice the real part of the time-ordered correlator.
This unitarity condition is analogous to its S-matrix counterpart. Writing the S-matrix as $\mathcal{S}=1+i\mathcal{T}$, the unitarity condition $\mathcal{S}^{\dagger}\mathcal{S}=1$ implies that\footnote{Writing the non-trivial piece as $i\mathcal{T}$ explains why we take the imaginary piece for the S-matrix but the real part for the four-point function.}
\begin{align}
2\hspace{.1cm}\Im(\mathcal{T})=\mathcal{T}^{\dagger}\mathcal{T}~. \label{eq:SMatrixOptical}
\end{align}
In both \eqref{eq:CFTOptical} and \eqref{eq:SMatrixOptical}, the left-hand side is found by taking a discontinuity while the right-hand side has a naturally factorized form in terms of two lower-loop objects.\footnote{The relation (\ref{eq:CombCor}) is also used in axiomatic studies of QFT to prove unitarity of the S-matrix \cite{Schweber:1961zz}.} More precisely, \eqref{eq:CFTOptical} is the correlation function analogue of studying the $s$-channel discontinuity of a scattering amplitude. 

To make a connection with the inversion formula, we need to replace the time-ordering symbol with the retarded commutator. To do this, we use
\begin{align}
T[\f(k_1)\f(k_2)]|0\>-&[\f(k_1),\f(k_2)]_{R}|0\>
\nonumber \\ &=\int d^{d}x_1d^{d}x_2 e^{i(k_1\cdot x_1 +k_2\cdot x_2)}\bigg(\theta(t_1-t_2)\f(x_1)\f(x_2)+\theta(t_2-t_1)\f(x_2)\f(x_1)
\nonumber \\ &\hspace{1.85in}-\theta(t_1-t_2)(\f(x_1)\f(x_2)-\f(x_2)\f(x_1))\bigg)|0\>
\nonumber \\ &=\int d^{d}x_1d^{d}x_2 e^{i(k_1\cdot x_1 +k_2\cdot x_2)}\f(x_2)\f(x_1)|0\>=\f(k_2)\f(k_1)|0\>=0~,
\end{align}
where the final equality follows from having $\f(k)|0\>=0$ when $k^2>0$. In other words,
both $T[\f(k_1)\f(k_2)]$ and $[\f(k_1),\f(k_2)]_{R}$ have the same action on the vacuum for spacelike momenta. With the same kinematics, a similar argument gives
\begin{align}
\<0| \overline{T}[\f(k_3)\f(k_4)]-&\<0| [\f(k_3),\f(k_4)]_{A}=0~.
\end{align}
Finally, we arrive at
\begin{align}
-2~\Re \<T[\f(k_1)\f(k_2)\f(k_3)\f(k_4)]\>&=\<\overline{T}[\f(k_3)\f(k_4)]T[\f(k_1)\f(k_2)]\>
\nonumber \\ &=\<[\f(k_3),\f(k_4)]_{A}[\f(k_1),\f(k_2)]_{R}\>~, \label{eq:FromReToDCV2}
\end{align}
in the configuration \eqref{eq:kinematics}. While \eqref{eq:FromReToDCV2} is a property of general QFTs, our primary focus will be computing the left-hand side of \eqref{eq:FromReToDCV2} in theories with a weakly-coupled description.

Using \eqref{eq:FromReToDCV2}, we can now compute $\<[\f,\f]_{A}[\f,\f]_{R}\>$ in the restricted configuration \eqref{eq:kinematics} from our knowledge of $\Re \<T[\f\f\f\f]\>$. Moreover, once the causal double-commutator is known for these momenta, it can be analytically continued to general configurations.
As we review in Appendix \ref{app:Analyticity_k_Space}, the position-space causality conditions for this double-commutator imply analyticity properties in momentum space \cite{Polyakov:1974gs,Streater:1989vi,Haag:1992hx}. We can choose the three independent momenta to be $k_1$, $k_4$, and $k_1+k_2$ and then analytically continue in $k_1$ and $k_4$ according to
\begin{align}
k_{i}\rightarrow k_{i}-i\eta_{i}, \quad \text{for} \quad i=1,4 \quad  \text{and} \quad \eta_i\in V_+ ~.
\end{align} 
The causal double-commutator is analytic in this region, and we can continue to $k_i^{2}<0$ to recover the full causal double-commutator. The double-commutator is only non-zero for $k_1+k_2\in V_+$, so we do not need to relax any conditions on this variable.

\section{Cutting Rules at Weak Coupling}
\label{sec:ReviewCutkosky}
In this section we study \eqref{eq:FromReToDCV2} for weakly-coupled QFTs. As an example, we consider a real scalar field $\phi$ with a non-derivative interaction $g\phi^n$. We will review the derivation of the cutting rules for the connected part of the time-ordered correlator, following the method of \cite{Veltman:1963th}. The new result is the combination of these cutting rules with the identity \eqref{eq:FromReToDCV2} to derive the corresponding cutting rules for the causal double-commutator. We will also explain why this double-commutator is a simpler object to study than the real part of a time-ordered correlator. By working this case out in detail, the generalization to AdS will be manifest. We will not need to assume conformal invariance in this section, but due to the role of the double-commutator in the inversion formula, this is an interesting case to study.

We will need the Feynman propagator $\Delta_{F}(x)$ and the two-point Wightman functions $\Delta^{\pm}(x)$ for the free field $\phi$,
\begin{align}
\Delta_{F}(x)&=\<T[ \f(x)\f(0)]\>_{\text{free}}=\int\frac{d^{d}k}{(2\pi)^{d}}\frac{-i}{k^{2}+m^{2}-i\epsilon}e^{ik\cdot x}  \nonumber
\\ &=\theta(x^0)\Delta^{+}(x)+\theta(-x^0)\Delta^{-}(x)~, \label{eq:time_ordered}
\\ \Delta^{\pm}(x)&=\<\f(\pm x)\f(0)\>_{\text{free}}=\int \frac{d^{d}k}{(2\pi)^{d}}2\pi \theta(\pm k^0)\delta(k^2+m^2)e^{ik\cdot x}~. 
\end{align}
Similarly, for the anti-time-ordered propagator we have:
\begin{align}
\Delta_{F}^*(x)=\<\overline{T}[ \f(x)\f(0)]\>_{\text{free}}&=\int\frac{d^{d}k}{(2\pi)^{d}}\frac{i}{k^{2}+m^{2}+i\epsilon}e^{ik\cdot x} \nonumber
\\ &=\theta(x^0)\Delta^{-}(x)+\theta(-x^0)\Delta^{+}(x)~.
\label{eq:time_ordered_star}
\end{align}
An important difference between \eqref{eq:time_ordered} and \eqref{eq:time_ordered_star} is the opposite signs of the  $i\epsilon$. We will use $\Delta^{-}(-x)=(\Delta^{-}(x))^*=\Delta^{+}(x)$ to express all quantities in terms of $\Delta^{+}$. The free-field two-point Wightman functions correspond to on-shell propagators, which in momentum space are given by
\begin{align}
\Delta^{+}(k)=2\pi \delta(k^2+m^2)\theta(k^0)~.
\end{align}
We will also refer to $\Delta^{+}(k)$ as the Wightman propagator. In the standard S-matrix unitarity method, cut propagators are placed on shell by replacing a propagator with a $\delta$-function, which we see corresponds to $\Delta_{F}\rightarrow \Delta^{+}$.

We will now review Veltman's derivation of the cutting rules, which are formulated in terms of the Wightman propagators. For simplicity, we start with the following one-loop correction to the two-point function $\<\f(x_1)\f(x_2)\>$ in $\phi^3$ theory,
\vspace{-.3in}
\begin{center}
\begin{eqnarray}
\includegraphics[scale=.33]{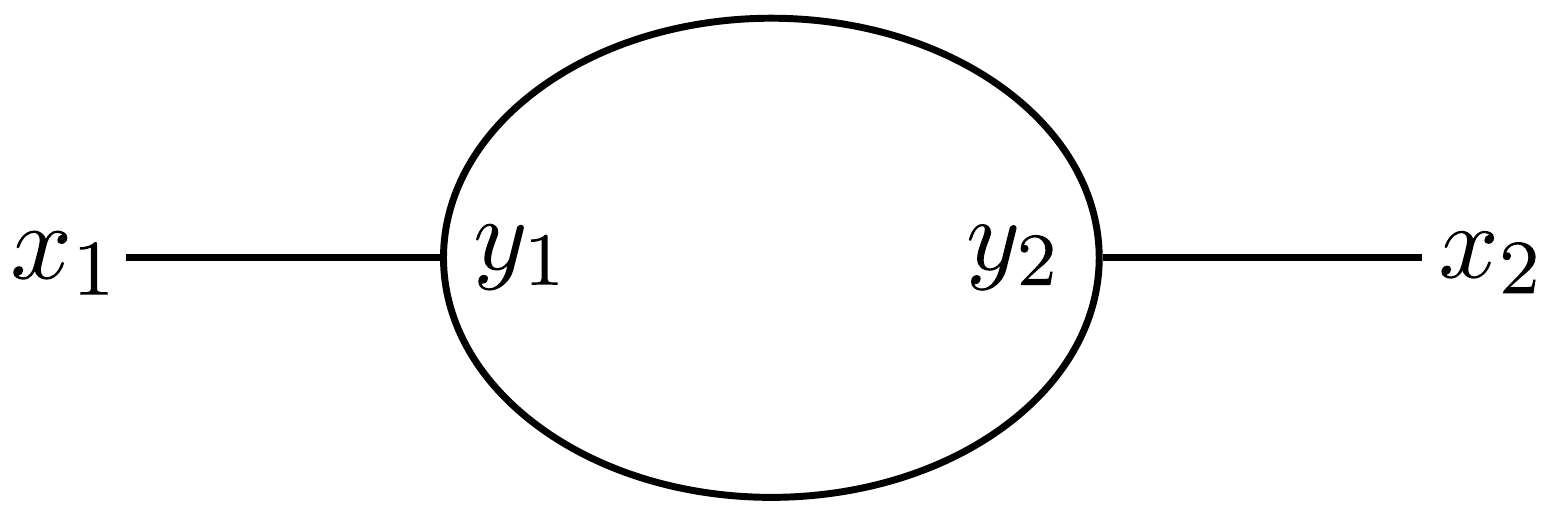}~.
\end{eqnarray}
\end{center}
As usual, the Feynman rules tell us to assign a factor of $ig$ to each interaction vertex, a Feynman propagator $\Delta_{F}(x_{ij})$ to each line, and then to integrate over the internal points $y_i$. When relevant, we also need to include symmetry factors.

From any Feynman diagram, $F(x_1,\ldots,x_n)$, with $n$ external points and $m$ internal points, we generate $2^{n+m}$ new graphs by introducing two distinct vertices, which we label with white and black dots. We will refer to both internal and external points as the vertices of the diagram.  The new collection of ``decorated'' graphs, $\widehat{F}_{q}(x_i)$ with $q=1,\ldots,2^{n+m}$,  are defined via the following rules:
\begin{enumerate}
\item For each internal vertex of either color, multiply by $ig$.
\item For each white vertex, either internal or external, multiply by $-1$.
\item For lines connecting two black vertices, $x_{i}$ and $x_{j}$, use $\Delta_{F}(x_{ij})$.
\\ For lines connecting two white vertices, $x_i$ and $x_j$, use $\Delta_{F}^{*}(x_{ij})$.
\\ For lines connecting a white vertex, $x_i$, and a black vertex, $x_j$, use $\Delta^{+}(x_{ij})$.
\end{enumerate}
For simplicity, we assume no line begins and ends at the same point, i.e. no propagator has a vanishing argument.\footnote{To take into account loop corrections, one should instead work with the renormalized propagator. Then the same cutting rules carry over, but now with the corresponding renormalized on-shell propagator.} These rules were first given in \cite{Veltman:1963th} for amputated Feynman diagrams to give an alternative derivation of the Cutkosky rules. We will not need the specific map between the label $q$ and the assignment of black and white vertices as most decorated diagrams vanish by momentum conservation. Instead, we will find that the non-zero decorated diagrams are in one-to-one correspondence with the allowed unitarity cuts of the original diagram.

To find the unitarity cuts of a diagram, we need to introduce the largest-time equation \cite{Veltman:1963th,tHooft:1973wag,Veltman:1994wz}:
\begin{align}
\sum\limits_{q=1}^{2^{m+n}}\widehat{F}_{q}(k_1,\ldots,k_n)=0~.  \label{eq:Largesttime}
\end{align}
We give a short derivation of this identity in Appendix \ref{app:largest_time}. To explain its connection to the cutting rules, we need to isolate two terms in the sum (\ref{eq:Largesttime}): the graph with all black vertices, $\widehat{F}_{q=1}(k_i)$, and the graph with all white vertices, $\widehat{F}_{q=2^{m+n}}(k_i)$. Their relation to the original Feynman diagram is
\begin{align}
\widehat{F}_{q=1}(k_1,\ldots,k_n)&=F(k_1,\ldots,k_n),
\\
\widehat{F}_{q=2^{n+m}}(k_1,\ldots,k_n)&=(-1)^nF^*(k_1,\ldots,k_n)~. \label{eq:ComplexConjF}
\end{align}
To show \eqref{eq:ComplexConjF}, recall that using white vertices amounts to letting $ig\rightarrow -ig$ and \\ $\Delta_{F}(x)\rightarrow \Delta_{F}^{*}(x)$. This replacement generates the complex conjugated graph and the overall factor of $(-1)^n$ comes from using white vertices for the external points. If we pull out these two graphs, the largest-time equation says
\begin{align}
F(k_1,\ldots,k_n)+(-1)^n F^*(k_1,\ldots,k_n)=-\sum\limits_{q=2}^{2^{n+m}-1}\widehat{F}_{q}(k_1,\ldots,k_n)~.  \label{eq:unintLargesttimeV2}
\end{align}
Note that for $n$ even or odd, the left-hand side is the real or imaginary part of the correlation function respectively.\footnote{For the S-matrix we always amputate external lines, in which case the $(-1)^n$ is not present and we always take the real part of the amputated diagram.}

To use \eqref{eq:unintLargesttimeV2} we must na\"ively study a large set of diagrams that grows exponentially in the number of vertices. Moreover, it is also not yet manifest how these diagrams are related to the usual cutting rules. Fortunately, most of the $\widehat{F}_{q}(k_i)$ will vanish by momentum conservation. For example, consider the following two diagrams:
\begin{equation}
\includegraphics[scale=.3]{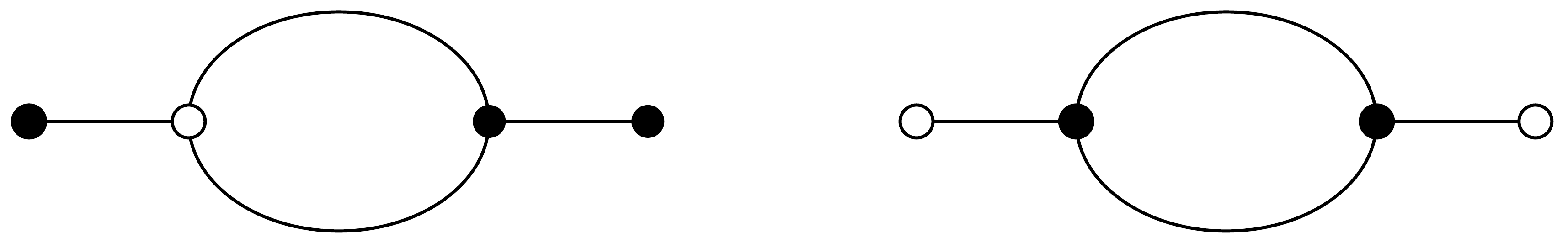}~.
\label{eq:graphvanishes}
\end{equation}
The first graph vanishes because positive timelike momentum is flowing into the white vertex from each propagator. The second graph vanishes because all the momenta is flowing out of the bubble. In order to have a non-zero decorated graph, we need both black and white external vertices, which serve as the ``source" and ``sink" for the momenta. 

The classic result of Veltman \cite{Veltman:1963th} is that the set of non-zero colorings of vertices is in one-to-one correspondence with the allowed unitarity cuts of the diagram, where a unitarity cut splits a diagram into two on-shell sub-diagrams. The map between cut diagrams and the coloring rules given above is as follows:
\begin{equation}
\includegraphics[scale=.3]{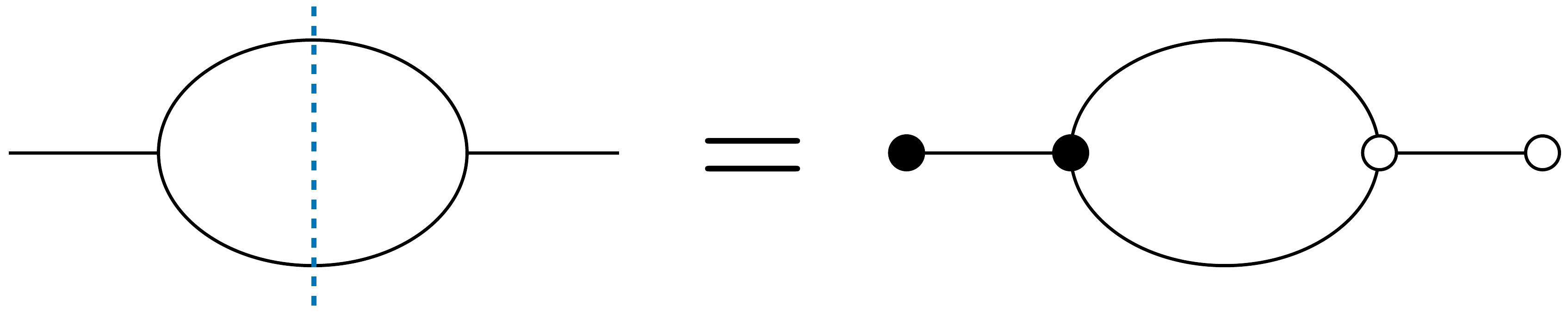} \label{eq:CutToVert}.
\end{equation}
That is, we cut a diagram in two and use black vertices to the left of the cut and white vertices to the right. Then the cut lines correspond to the on-shell, Wightman propagators. When we draw general cuts there can be an ambiguity about the assignment of vertices, i.e. on which side to place which vertices. However, there is only one choice that is non-zero for a given choice of external momenta.  For that reason, we will leave the assignment of black and white vertices implicit when drawing cut diagrams. In (\ref{eq:CutToVert}), if we take the momentum to run from left to right, then the other possible assignment vanishes.

According to \eqref{eq:unintLargesttimeV2}, we can compute the real or imaginary part of a correlation function by summing over all cuts of the diagrams. For example, the internal cut of the bubble is shown in \eqref{eq:CutToVert}. Here we are studying correlation functions with off-shell external legs, so we need to consider the external line cuts as well. These do not reduce the loop order of the diagram, as can be seen for the bubble,
\begin{equation}
\includegraphics[scale=.3]{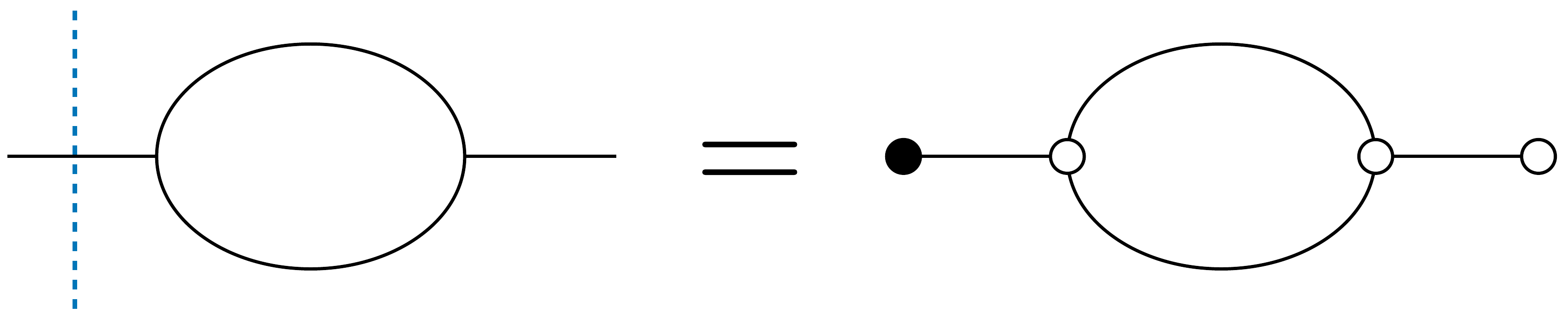}~.
\label{eq:ExtCutBubble}
\end{equation}
For the four-point function, which will be our main object of study, we can similarly have cuts passing through only the external lines,
\begin{equation}
\includegraphics[scale=.3]{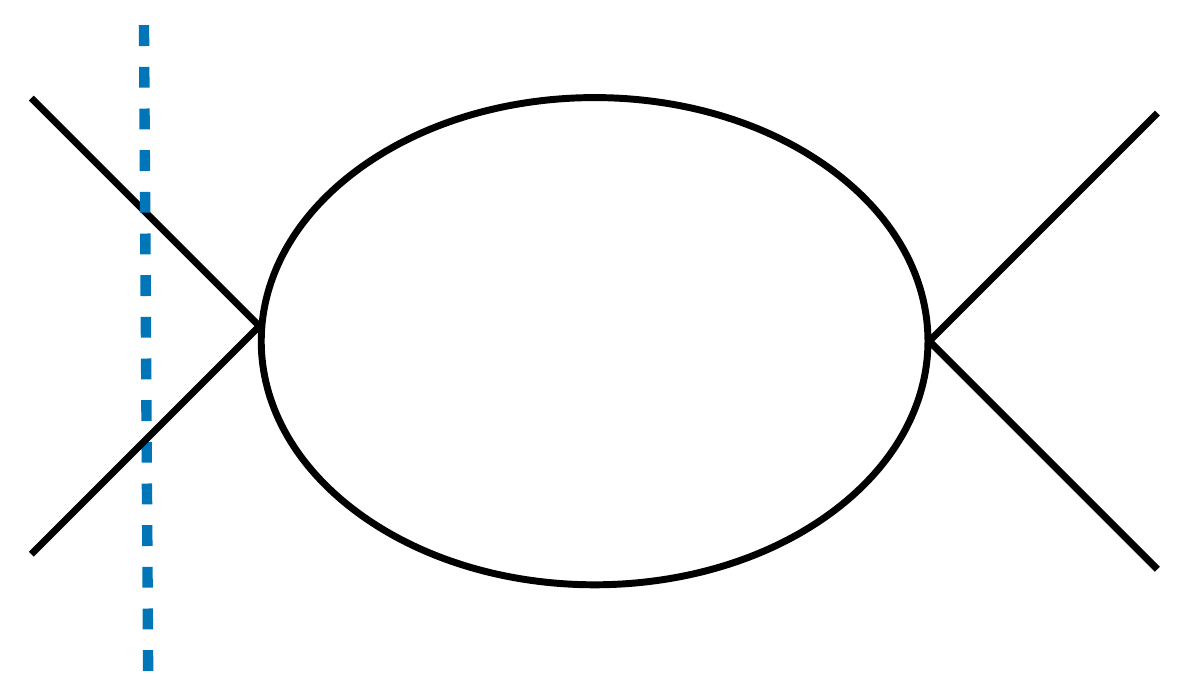}~.
\label{eq:extcut4ptBubble}
\end{equation}
For the remainder of this section, we will focus on four-point functions.

The external line cuts for $\<T[\f\f\f\f]\>$ are trivial in flat space as they simply place the external momenta on shell, i.e. $k^{2}=-m^{2}$ such that $k^0\geq 0$.\footnote{The terminology ``external cut" for $\<T[\f\f\f\f]\>$ refers to a cut of a propagator connected to an external point. We study diagrams for this correlator in particular because the same topologies will appear for single-trace correlators in holographic CFTs. Nevertheless, the cutting rules can also be studied for more general correlators of composite operators, such as $\<T[\f^2\f^2\f^2\f^2]\>$.} As long as the external momenta do not lie exactly on the mass-shell, these cuts vanish. In preparation for AdS/CFT, where the analogous external cuts are non-trivial, it useful to have a dispersion formula which depends solely on the internal cuts. As we will demonstrate shortly, the CFT dispersion formula \cite{Caron-Huot:2017vep,Carmi:2019cub} meets this criteria for four-point correlators by using the causal double-commutator as input. 

To see why the double-commutator only depends on internal cuts, we study the cutting rules in the kinematics \eqref{eq:kinematics}, where all external momenta are spacelike. In this configuration, external cuts such as \eqref{eq:extcut4ptBubble} manifestly vanish. As reviewed in Section \ref{sec:Polyakov}, the real part of a time-ordered four-point function, $\Re\<T[\f\f\f\f]\>$, is equivalent to the causal double-commutator, $\<[\f,\f]_{A}[\f,\f]_{R}\>$, in this configuration. The causal double-commutator with restricted kinematics therefore only has internal cuts. This is the first hint that the double-commutator is a simpler object to study than the real part of the correlator.

As the external momenta are all spacelike, we can go further: any internal line cut that leaves one operator to the left or right of the cut must vanish. In other words, the following class of cuts for a general Feynman diagram $F$ vanish:
\begin{center}
\begin{equation}
\includegraphics[scale=.25]{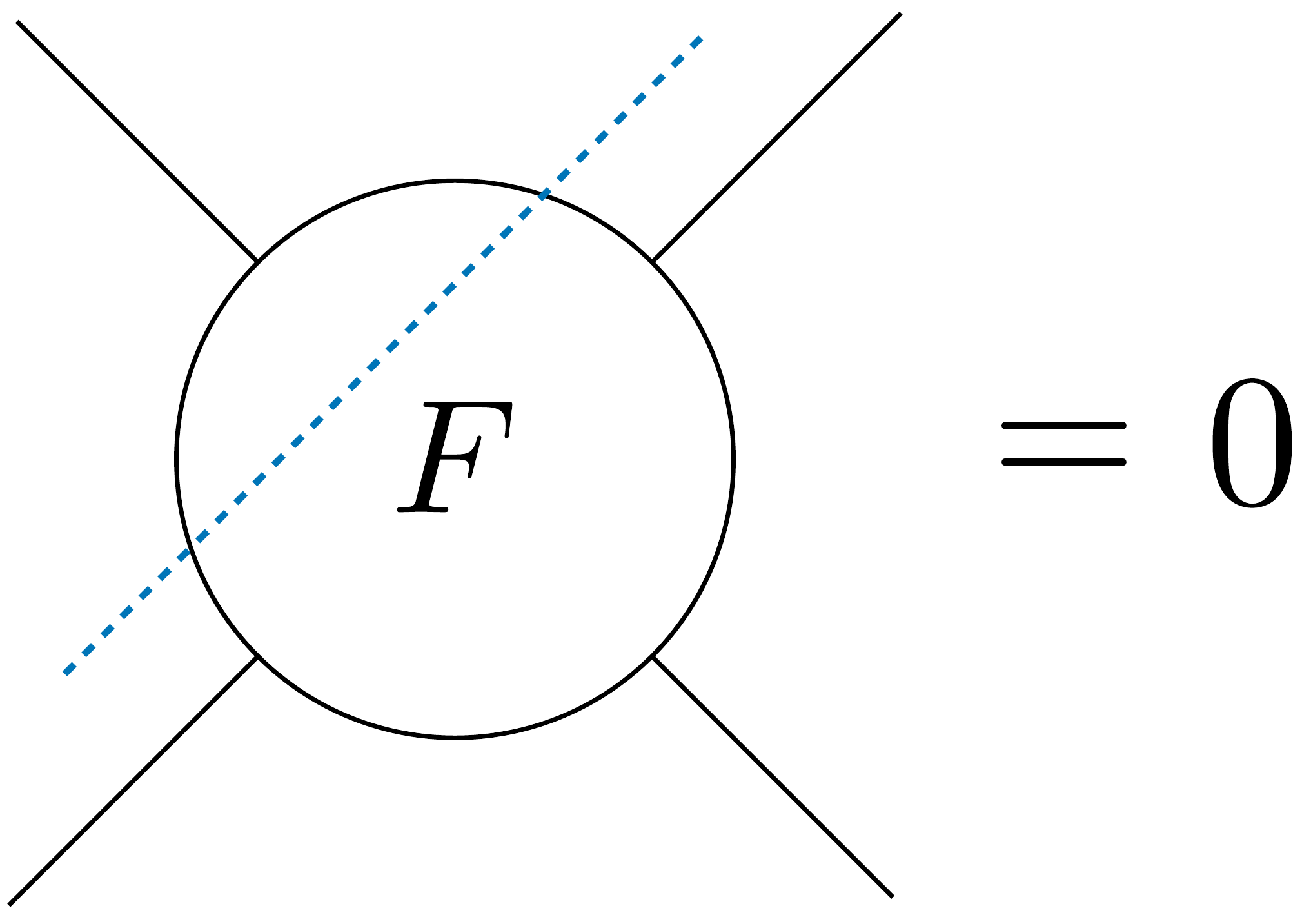}~.
\end{equation}
\end{center}
The reason is simple - we need to have timelike momenta flowing through a unitarity cut in order for it to be non-zero. With spacelike external momenta, this cannot happen if only one external point is to the left or right of the cut. The vanishing of these cuts is equivalent to terms in \eqref{eq:CombCor} such as $\<\f(k_1)T[\f(k_2)\f(k_3)\f(k_4)]\>$ being zero for spacelike $k_i$. 
To give examples, the following cuts both give zero in the kinematics \eqref{eq:kinematics}:
\begin{center}
\begin{equation}
\includegraphics[scale=.24]{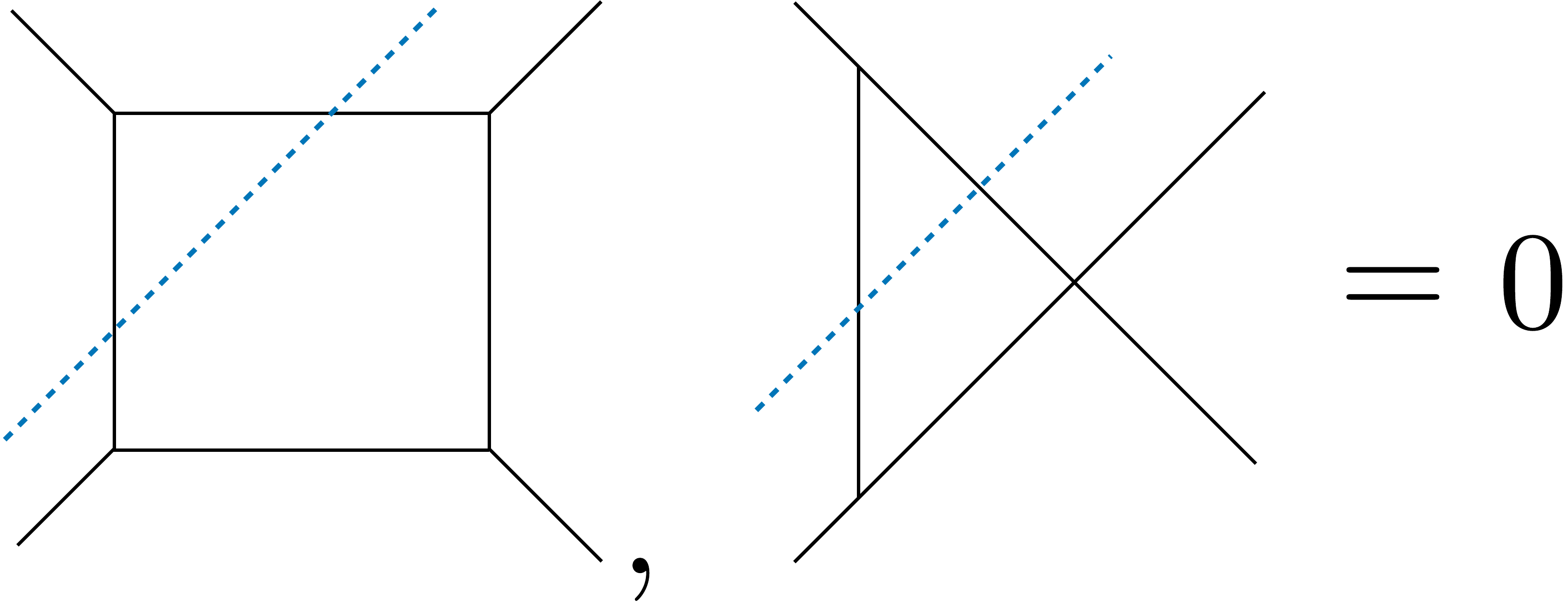}~.
\end{equation}
\end{center}
These types of cuts typically vanish when studying amplitudes because on-shell three-point amplitudes for gluons and gravitons vanish. Here all of our external lines are off-shell and these cuts vanish due to the choice of external momenta.

Finally, we will consider cuts that split the external legs into pairs, i.e. the $s$, $t$, and $u$-channel cuts familiar from S-matrix unitarity. In the kinematics \eqref{eq:kinematics}, we have $k_1+k_2\in V_{+}$ but $k_1+k_3$ and $k_1+k_4$ are spacelike. With this choice, only the $s$-channel cuts are non-zero:
\begin{equation}
\includegraphics[scale=.25]{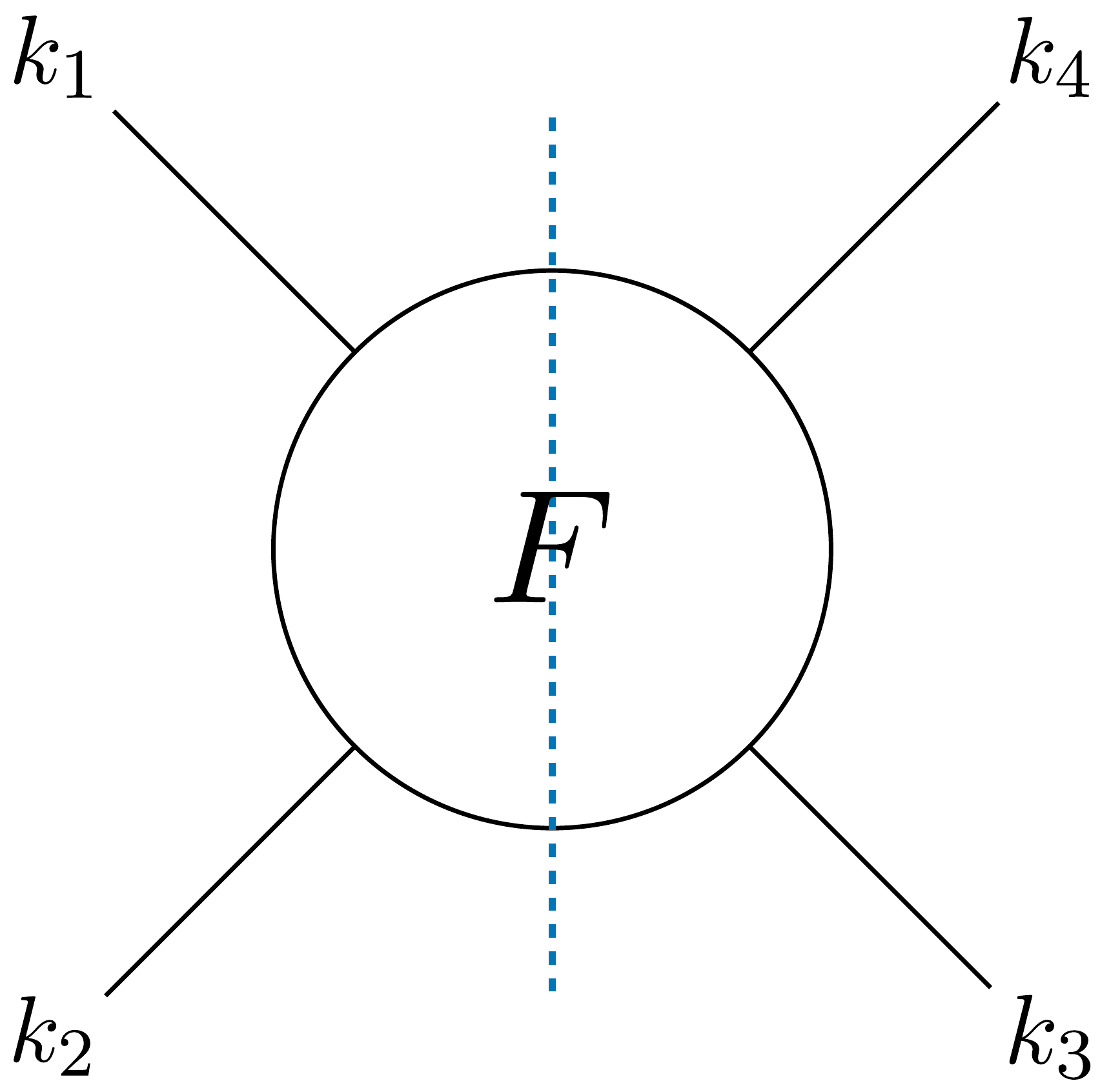}~.
\end{equation}
For example, the following cuts are all non-zero:
\begin{equation}
\includegraphics[scale=.235]{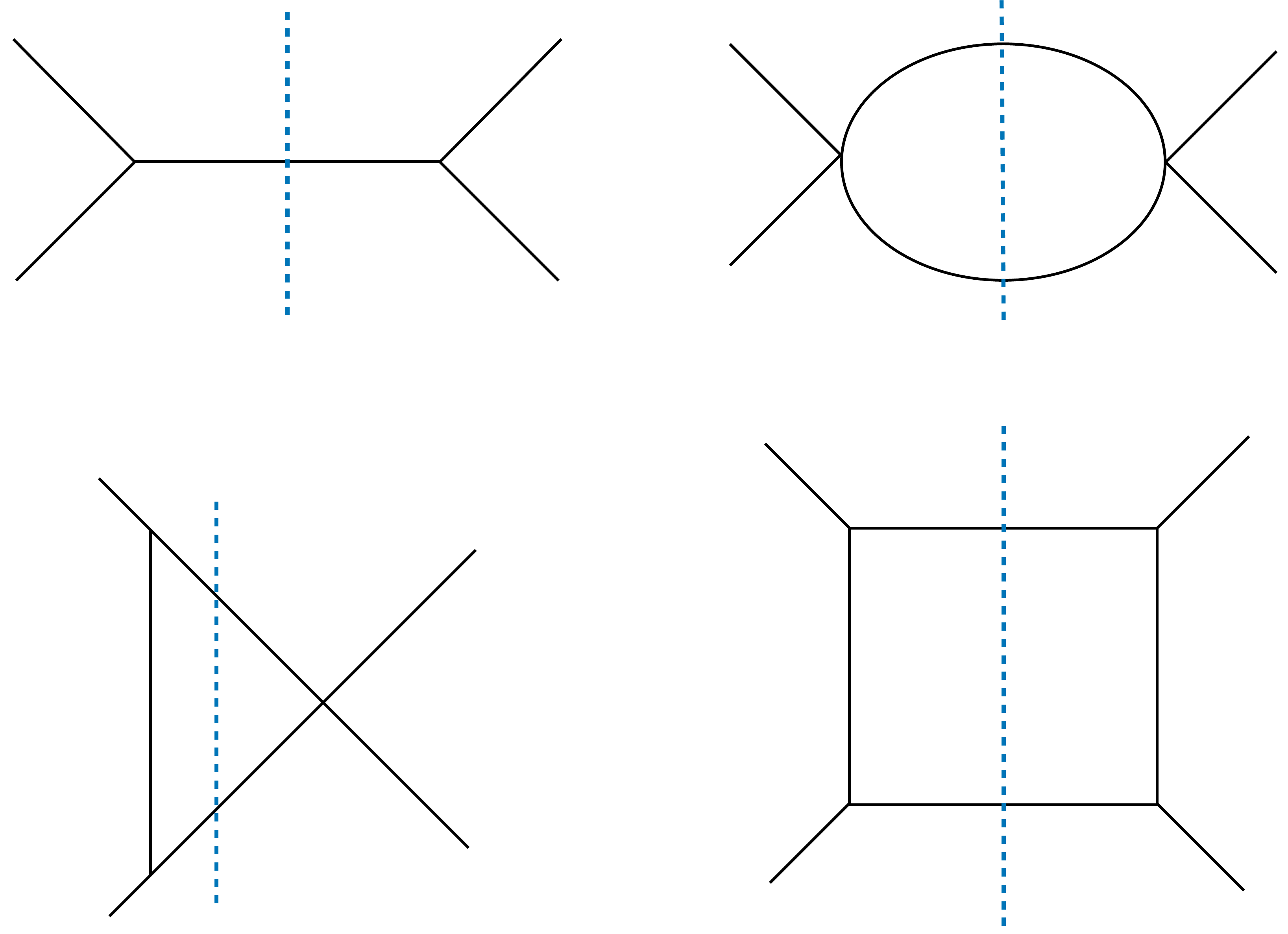}~.
\end{equation}
We can now make an explicit connection to the double-commutator. Using the largest-time equation
\begin{align}
-2~\Re F(k_1,k_2,k_3,k_4)=\sum\limits_{q=2}^{2^{n+m}-1}\widehat{F}_{q}(k_1,k_2,k_3,k_4)~,  \label{eq:unintLargesttimeV3}
\end{align}
and the identity \eqref{eq:FromReToDCV2}, we find
\begin{align}
\<[\f(k_3),\f(k_4)]_{A}[\f(k_1),\f(k_2)]_{R}\>=\sum\limits_{q=2}^{2^{n+m}-1}\widehat{F}_{q}(k_1,k_2,k_3,k_4)~.
\label{eq:DC_Cutting_Rules}
\end{align}
In other words, the sum over decorated graphs in the kinematics \eqref{eq:kinematics} computes the causal double-commutator. On the right-hand side we have written the full sum over $q$, but as emphasized earlier only a few graphs are consistent with momentum conservation. 

While we derived \eqref{eq:DC_Cutting_Rules} using the vertex coloring rules, we can summarize the result more simply.
The cut graphs that contribute to the double-commutator $\<[\f,\f]_A[\f,\f]_R\>$ are determined by working in the kinematics (\ref{eq:kinematics}) and using the following cutting rules:
\begin{enumerate}
\item For each Feynman diagram, draw a cut that passes only through internal lines. 
\\ For each line that is cut, use the on-shell propagator $\Delta^{+}(k)$.
\item For all propagators to the left of the cut, use $\Delta_{F}(k)$. 
\\
For all propagators to the right of the cut, use $\Delta_{F}^{*}(k)$.
\item For each internal vertex multiply by $ig$.
\\  For each vertex to the right of the cut, internal or external, multiply by an additional $-1$.
\item Sum over cuts consistent with momentum conservation.
\end{enumerate}
As a reminder, we choose the on-shell propagators $\Delta^{+}(k)$ such that the momenta is flowing across the cut from $\f(k_1)$ and $\f(k_2)$ to $\f(k_3)$ and $\f(k_4)$. For cut four-point functions, all external points come in pairs and we can replace the $3^{\text{rd}}$ rule by:
\\
\\
\hphantom{>>} 3$'$. For each internal vertex to the left of the cut multiply by $ig$.
\\ \hphantom{>>>.....}For each internal vertex to the right of the cut multiply by $-ig$.
\\
\\
However, when studying higher-point functions it will be important to keep track of how many external points lie to the right of the cut. One can restore the previous vertex coloring rules by assigning black and white vertices to the left and right of the cut respectively, see for example \eqref{eq:CutToVert}. 

We were careful to work with kinematics (\ref{eq:kinematics}) in order to classify the cut diagrams that contribute to the double-commutator. Once we have classified and computed these cuts, the argument reviewed in Section \ref{sec:Polyakov} allows us to analytically continue the final result to general kinematics. While using spacelike momenta and studying the cutting rules for $\Re\<T[\f\f\f\f]\>$ are not strictly necessary to derive the cutting rules for the double-commutator, we find this to be a particularly simple approach. 

As an aside, we can also derive the cutting rules by assuming $k_i^{2}>0$ and applying the identity
\begin{align}
\<\overline{T}[\f(k_3)\f(k_4)]T[\f(k_1)\f(k_2)]\>=\<[\f(k_3),\f(k_4)]_{A}[\f(k_1),\f(k_2)]_{R}\>~.\label{eq:TTbtodDisc}
\end{align}
As we reviewed in Section \ref{sec:Polyakov}, this follows from the positive spectrum condition. Then
the partially time-ordered correlation function on the left can be computed using the Schwinger-Keldysh rules \cite{Schwinger:1960qe,Keldysh:1964ud,Chou:1984es,Haehl:2016pec,Haehl:2017qfl} and one arrives at the same set of cutting rules for the double-commutator.\footnote{The relation between the Schwinger-Keldysh formalism and unitarity cuts is also given in section 11 of
\cite{Haehl:2016pec}.} In this approach working with spacelike momenta is useful as well: we only need a single time-fold to compute the left-hand side of \eqref{eq:TTbtodDisc} while for the right-hand side, for generic momenta, we need two time-folds \cite{Haehl:2017qfl}. We explain how to derive the cutting rules from the Schwinger-Keldysh formalism in Appendix \ref{sec:SKDerivation}.

\section{Unitarity Cuts in AdS/CFT}
\label{sec:UnitarityCutsAdS}
\subsection{Cutting Rules}
In this section we will generalize the analysis of Section \ref{sec:ReviewCutkosky} to AdS$_{d+1}$/CFT$_{d}$, the main application of interest in this work. The generalization is straightforward as the derivation of the cutting rules did not rely on the explicit form of the propagators. Instead, it followed from general features of Lorentzian two-point functions. The cutting rules will therefore also hold for weakly coupled theories in AdS. Our aim here is to discuss how the cutting rules generalize, connect to previous work, and give the explicit expressions necessary for later computations.

We will study a bulk scalar field $\Phi$ that has a non-derivative interaction $g\Phi^n$ and is dual to the boundary scalar operator $\phi$. We work in the Poincar\'e patch of AdS with the standard metric 
\begin{align}
ds^{2}=\frac{dz^{2}+\eta_{\mu\nu}dx^{\mu}dx^{\nu}}{z^{2}}~,
\end{align}
where we again take $\eta_{\mu\nu}$ to be mostly plus, and $z=0$ is the boundary of AdS. Finally, we will only study the connected Witten diagrams for correlation functions of the single-trace operator, $\<T[\f(x_1)...\f(x_n)]\>$.

We begin by expanding the Feynman bulk-to-bulk propagator $G_{\Delta}(x_1,z_1;x_2,z_2)$ in terms of the Wightman bulk-to-bulk propagators,
\begin{align}
G_{\Delta}(x_1,z_1;x_2,z_2)=&\<T[\Phi(x_1,z_1)\Phi(x_2,z_2)]\>_{\text{free}}
\nonumber \\=&\theta(x^0_1-x^0_2)G_{\Delta}^{+}(x_1,z_1;x_2,z_2)+\theta(x^0_2-x^0_1)G_{\Delta}^{+}(x_2,z_2;x_1,z_1)~,
\\
G_{\Delta}^{+}(x_1,z_1;x_2,z_2)=&\<\Phi(x_1,z_1)\Phi(x_2,z_2)\>_{\text{free}}~,
\end{align}
where $\Delta$ is the scaling dimension of the boundary scalar $\f$.
As in flat space, the Wightman propagators are defined to be the free-field two-point Wightman functions. These will again correspond to the on-shell propagators.

We now Fourier transform in the flat $x^{\mu}$ directions to use the AdS/CFT momentum-space propagators. The bulk-to-bulk propagators take the form \cite{Liu:1998ty}:
\begin{align}
G_{\Delta}(k,z_1,z_2)&= -i(z_1 z_2)^{\frac{d}{2}}\int\limits_{0}^{\infty} dp \hspace{.07cm} p\frac{\mathcal{J}_{\nu}(pz_1)\mathcal{J}_{\nu}(pz_2)}{k^{2}+p^{2}-i\epsilon}~,
\\
G^{\pm}_{\Delta}(k,z_1,z_2)&=\pi(z_1z_2)^{\frac{d}{2}}\mathcal{J}_{\nu}(\sqrt{-k^{2}}z_1)\mathcal{J}_{\nu}(\sqrt{-k^{2}}z_2)\theta(-k^{2})\theta(\pm k^0)~, \label{eq:PosBB}
\end{align}
where $\mathcal{J}_{\nu}$ is the Bessel function of the first kind and $\nu=\Delta-d/2$.\footnote{When studying the Euclidean principal series we usually write $\Delta=\frac{d}{2}+i\nu$, but to be consistent with previous work on AdS/CFT momentum space we use $\Delta=\frac{d}{2}+\nu$.} The bulk-to-boundary propagator, $K_{\Delta}(k,z)$, is then defined by taking one point to the boundary:
\begin{align}
K_{\Delta}(k,z)&=\lim\limits_{z'\rightarrow 0}z'^{-\Delta}G_{\Delta}(k,z,z')
\nonumber \\ &=-i\frac{1}{2^{\nu}\Gamma(1+\nu)}z^{\frac{d}{2}}(\sqrt{k^{2}})^{\nu}\mathcal{K}_{\nu}(\sqrt{k^{2}}z)~,
\\
K^{\pm}_{\Delta}(k,z)&=\frac{\pi}{2^{\nu}\Gamma(1+\nu)}(\sqrt{-k^{2}})^{\nu}z^{\frac{d}{2}}\mathcal{J}_{\nu}(\sqrt{-k^{2}}z)\theta(-k^{2})\theta(\pm k^0)~, \label{eq:PosBb}
\end{align}
where $\mathcal{K}_{\nu}$ is the modified Bessel function of the second kind.\footnote{In general, we need to work with the regulated bulk-to-boundary propagators \cite{Freedman:1998tz},
\begin{align}
K^{\delta}_{\Delta}(k,z)=\left(\frac{z}{\delta}\right)^{d/2}\frac{\mathcal{K}_{\nu}(\sqrt{k^{2}} z)}{\mathcal{K}_{\nu}(\sqrt{k^{2}}\delta)}~,
\end{align}
where $\delta\ll1$ is the cut-off on the $z$-coordinate. However, for the discussion and examples considered in this work we can use $K_{\Delta}(k,z)$ throughout.} Here we have given the Feynman bulk-to-boundary propagator $K_{\Delta}(k,z)$ for spacelike $k$. When analytically continuing to timelike momenta, we give $k^{2}$ a small imaginary part as dictated by the $i\epsilon$ prescription. Finally, $K^{\pm}_{\Delta}(k,z)$ are the on-shell, or Wightman, bulk-to-boundary propagators.

We can now repeat the arguments of Section  \ref{sec:ReviewCutkosky} with minor changes. For completeness, we spell them out here. For a Witten diagram $W(x_1,\ldots,x_n)$ with $n$ external (boundary) points and $m$ internal (bulk) points, we can define $2^{n+m}$  new graphs $\widehat{W}_{q}(x_1,\ldots,x_n)$ by using two types of vertices. We again distinguish them by using white or black dots. The new decorated graphs are defined as follows:
\begin{enumerate}
\item For each internal vertex multiply by $ig$.
\item For each white vertex, internal or external, multiply by an additional $-1$.
\item For lines between black vertices in the bulk use $G_{\Delta}(x_i,z_i;x_j,z_j)$.
\\ For lines between white vertices in the bulk use $G_{\Delta}^{*}(x_i,z_i;x_j,z_j)$.
\\ For lines between a white vertex at $(x_i,z_i)$, and a black vertex at $(x_j,z_j)$, use $G_{\Delta}^{+}(x_i,z_i;x_j,z_j)$.
\item If a line ends on the boundary, use the appropriate bulk-to-boundary propagator.
\end{enumerate}

The only difference from the previous section is that we now have two kinds of propagators, depending on whether a point lies on the boundary or in the bulk. Here we have taken all external points to the boundary in order to study the CFT correlator $\<\f(x_1)\ldots\f(x_n)\>$. For a QFT in AdS, we can also study purely bulk correlation functions $\<\Phi(x_1,z_1)\ldots\Phi(x_n,z_n)\>$. When all external points lie in the bulk, the derivation of the cutting rules in AdS is exactly the same as in flat space. In this work however we will focus on CFT correlators.

The largest-time equation in AdS says:
\begin{align}
W(x_1,\ldots,x_n)+(-1)^nW^{*}(x_1,\ldots,x_n)=-\sum\limits_{q=2}^{2^{n+m}-1}\widehat{W}_{q}(x_1,\ldots,x_n)~, \label{eq:largest_time_witten}
\end{align}
where we pulled out the original graph and its complex conjugate. Restricting to four-point functions ($n=4$) and using the kinematics \eqref{eq:kinematics}, we can again use \eqref{eq:FromReToDC} to go from the real part of a time-ordered correlator to the causal double-commutator. The diagrammatic expansion for the double-commutator in the configuration \eqref{eq:kinematics} is:
\begin{align}
\<[\f(k_3),\f(k_4)]_{A}[\f(k_1),\f(k_2)]_{R}\>=\sum\limits_{q=2}^{2^{n+m}-1}\widehat{W}_{q}(k_1,\ldots,k_4)~.\label{eq:DC_Cutting_Rules_Witten}
\end{align}
Although the right-hand side runs over a large number of terms, only a few Witten diagrams are non-zero for our choice of momenta, just as in flat space. To simplify the presentation, we use the same cutting notation as before:
\begin{equation}
\includegraphics[scale=.25]{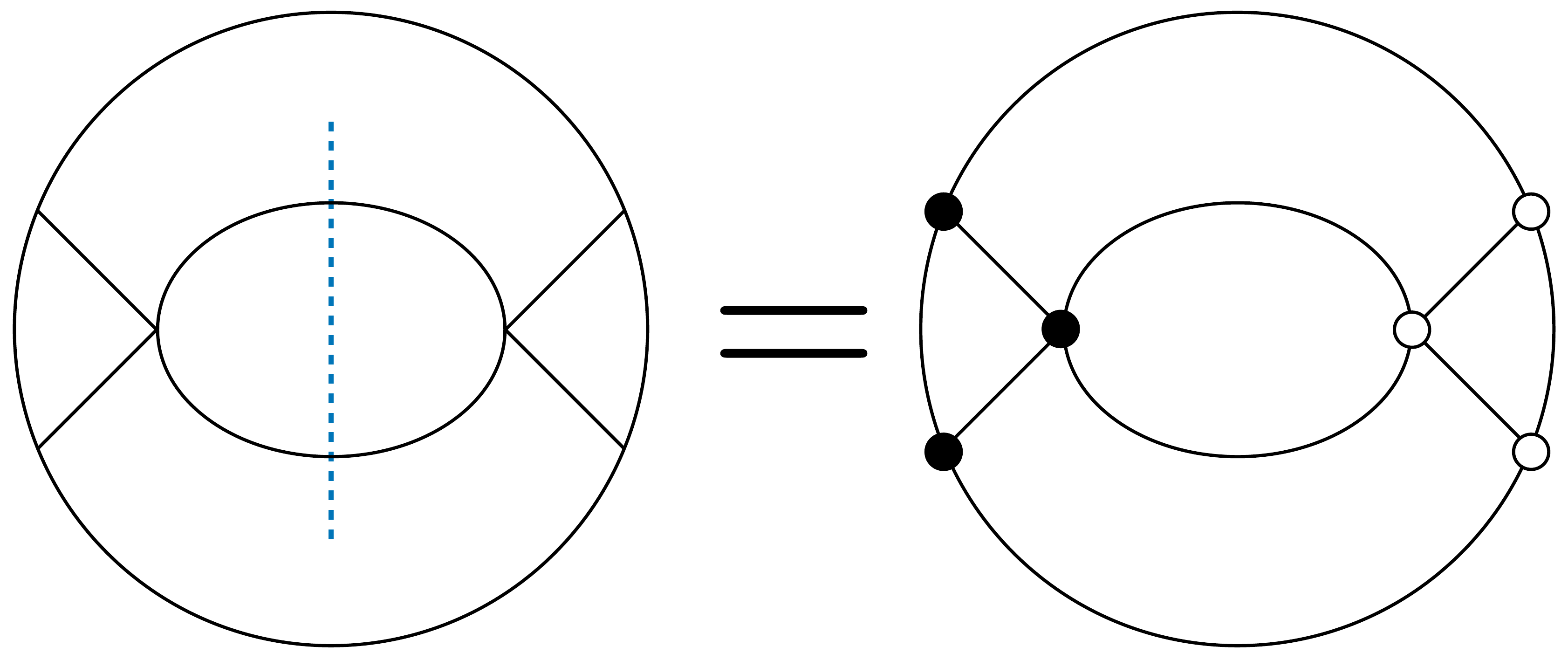}~. \label{eq:Witten_CutToVertices}
\end{equation}
Our assumption that the external momenta are spacelike implies that cuts of bulk-to-boundary propagators vanish identically. This is consistent with earlier work on unitarity cuts in AdS/CFT \cite{Fitzpatrick:2011dm,Meltzer:2019nbs}, in which it was found that internal cuts compute the ``absorptive" part of the diagram. Using external spacelike momenta also implies the cuts split the external points into two pairs. For our choice of momenta \eqref{eq:kinematics}, we must have $\{k_1,k_2\}$ to the left of the cut and $\{k_3,k_4\}$ to the right. Cuts through internal lines that leave a single external point on one side of the cut will vanish. In short, the cut structure for Witten diagrams is exactly the same as for the corresponding Feynman diagrams in flat space.

One difference in comparison to flat space is that external line cuts in AdS are less restrictive. Cutting through an external line in flat space means an external momentum must lie on the mass-shell, i.e. $k^2=-m^2$ and $k^0\geq 0$. In AdS a cut external line is non-zero as long as $k\in V_+$. Therefore, the external line cuts will contribute to $\Re\<T[\f\f\f\f]\>$ for general external timelike momenta and furthermore these cuts do not reduce the loop order of a diagram. By working with spacelike momenta, or by studying the causal double-commutator, we ensure that only internal cuts are allowed, and these do simplify the diagram.

The diagrams that contribute to the right-hand side of \eqref{eq:DC_Cutting_Rules_Witten} are found from the rules summarized in Section \ref{sec:intro}, which we repeat here for convenience:
\begin{enumerate}
\item Given a Witten diagram, draw a cut such that only bulk-to-bulk propagators are cut. For each cut propagator use the on-shell propagator $G^{+}_{\Delta}(k,z_i,z_j)$.
\item For all propagators to the left of the cut, use $G_{\Delta}(k,z_i,z_j)$. \\ For all propagators to the right of the cut, use $G^*_{\Delta}(k,z_i,z_j)$.
\item For each internal vertex multiply by $ig$.
\\ For each vertex to the right of the cut, multiply by an additional $-1$.
\item Sum over all cuts consistent with momentum conservation.
\end{enumerate}
To revert to the vertex assignment rules, we again use black and white dots for all vertices to the left and right of the cut respectively, see e.g. \eqref{eq:Witten_CutToVertices}. 
The rules given in Section \ref{sec:intro} were specialized to four points, in which case we can ignore factors of $-1$ from external points as they always come in pairs. Here we have given the rules for a general $n$-point Witten diagram as we will later study cut five-point diagrams.

\subsection{AdS Transition Amplitudes}
\label{sec:transitionamps}
Our methods rely on using Lorentzian signature, and studying Lorentzian AdS allows us to interpret the cut diagrams in terms of a sum over states, or equivalently a phase-space integral. Specifically, we will relate the cut propagators to normalizable solutions to the bulk equations of motion. This provides another sense in which a cut diagram is on shell and allows us to make a connection with the CFT optical theorem.

To set the stage, recall the unitarity condition for the flat-space S-matrix:
\begin{align}
\<\textrm{out}|\Im( \mathcal{T})|\textrm{in}\>=\<\textrm{out}|\mathcal{T}^{\dagger}\mathcal{T}|\textrm{in}\>~.
\end{align}
Inserting a complete set of states, the right-hand side factorizes as
\begin{align}
\<\textrm{out}|\Im(\mathcal{T})|\textrm{in}\>=\sum\limits_{\Psi}\<\textrm{out}|\mathcal{T}^{\dagger}|\Psi\>\<\Psi|\mathcal{T}|\textrm{in}\>~.
\end{align}
It is well-known that there is a one-to-one map between the allowed unitarity cuts of a diagram and the physical states $|\Psi\>$ that can be exchanged. 
As reviewed earlier, see \eqref{eq:FromReToDCV2}, the CFT statement of unitarity is
\begin{align}
-2~\Re \<T[\f(k_1)\f(k_2)\f(k_3)\f(k_4)]\>=\<\overline{T}[\f(k_3)\f(k_4)]T[\f(k_1)\f(k_2)]\>~,  \label{eq:FromReToTTb}
\end{align}
in the momentum configuration \eqref{eq:kinematics}. Once again, the right-hand side factorizes when we insert a complete set of states. However, it may not be clear what the map is between the states exchanged and the bulk cutting procedure. In other words, which basis of the AdS/CFT Hilbert space are we picking out with our cuts? As we will demonstrate, the natural set of states are the normalizable modes of the Poincar\'e patch, which are also used to define Poincar\'e transition amplitudes \cite{Balasubramanian:1999ri}. 

In Lorentzian AdS, the bulk equations of motion have both normalizable and non-normalizable solutions \cite{Avis:1977yn,Breitenlohner:1982jf,Breitenlohner:1982bm}. The normalizable modes are quantized to obtain the bulk Hilbert space and the non-normalizable modes are classical, non-fluctuating backgrounds. The bulk normalizable and non-normalizable solutions are dual to boundary states and sources in the CFT respectively \cite{Balasubramanian:1998sn,Balasubramanian:1998de}.
To find these solutions, we solve the equations of motion for a scalar $\Phi$,
\begin{align}
(\Box -m^2)\Phi=0~,
\end{align}
by working in momentum space. For 
spacelike momenta, $k^{2}>0$, there is a single solution that is regular in the interior of AdS:
\begin{align}
\Phi(k,z)=\phi_0 z^{d/2}\mathcal{K}_{\nu}(\sqrt{k^{2}}z)~.
\end{align}
For timelike momenta $k^{2}<0$, there are two solutions:
\begin{align}
\Phi(k,z)=\phi_0 z^{d/2}\mathcal{J}_{\nu}(\sqrt{-k^{2}}z)~,
\\
\Phi(k,z)=\phi_0 z^{d/2}\mathcal{Y}_{\nu}(\sqrt{-k^{2}}z)~.
\end{align}
Here $\mathcal{J}$ and $\mathcal{K}$ are the Bessel functions defined earlier, and $\mathcal{Y}$ is a Bessel function of the second kind. The $\mathcal{J}$ solution gives a normalizable mode while the $\mathcal{Y}$ and $\mathcal{K}$ solutions give non-normalizable modes. In the limit $z\rightarrow 0$, the normalizable solutions scale like $z^{\Delta}$ while the non-normalizable solutions scale like $z^{d-\Delta}$. Correlation functions are computed by choosing non-normalizable modes for all the external legs of the Witten diagram.

The connection between the AdS cutting rules and the Hilbert space can be seen from a ``split representation''. It is well-known that time-ordered bulk-to-bulk propagators in AdS can be expressed as \cite{Leonhardt:2003qu,Penedones:2010ue}:
\begin{align}
G_{\Delta}(k,z_1,z_2)=\int\limits_{-\infty}^{\infty} d\omega P(\omega,\Delta)K_{\frac{d}{2}+i\omega}(k,z_1)K_{\frac{d}{2}-i\omega}(k,z_2)~, \label{eq:EucSplitRep}
\\
P(\omega,\Delta)=\frac{1}{\omega^{2}+\left(\Delta-\frac{d}{2}\right)^{2}}\frac{\omega^{2}}{\pi}~.  \label{eq:Pdef}
\end{align}
That is, the Feynman bulk propagator is a spectral integral over the corresponding bulk-to-boundary propagators.
By comparing \eqref{eq:PosBB} and \eqref{eq:PosBb}, we observe that the on-shell bulk-to-bulk propagator also has a simple split representation:
\begin{align}
G^{+}_{\Delta}(k,z_1,z_2)&=\frac{2^{2\nu}\Gamma(1+\nu)^{2}}{\pi(\sqrt{-k^{2}})^{2\nu}}K^{+}_{\Delta}(k,z_1)K^{+}_{\Delta}(k,z_2)~. \label{eq:LorSplitRep}
\end{align}
Unlike the split representation for time-ordered propagators, we do not have a spectral integral.\footnote{The split representation \eqref{eq:EucSplitRep} is the basis of the Euclidean analysis in \cite{Meltzer:2019nbs,Ponomarev:2019ofr}. There, putting a line on shell corresponds to closing the $\omega$ integral on the pole in $P(\omega,\Delta)$. This produces bulk-to-boundary propagators of dimension $\Delta$ and $d-\Delta$, so the OPE of the resulting diagram has unphysical ``shadow" operators. Projecting these out yields the double-commutator. The Lorentzian split representation \eqref{eq:LorSplitRep} uses the on-shell propagators, so this projection is not required.} We can also identify the overall factor in \eqref{eq:LorSplitRep} as a boundary, Wightman, two-point function. Taking both points of the bulk-to-bulk on-shell propagator to the boundary yields:\footnote{Here we are dropping analytic terms in $k$ that contribute to contact terms in position space.}
\begin{align}
\<\!\<\f(-k)\f(k)\>\!\>=\frac{\pi}{2^{2\nu}\Gamma(1+\nu)^{2}}(\sqrt{-k^{2}})^{2\nu}\theta(-k^{2})\theta(k^0)~,
\end{align}
where we use the notation
\begin{align}
\<\f(k_1)\ldots\f(k_n)\> \equiv (2\pi)^{d}\delta^{d}(k_1+\ldots+k_n)\<\!\<\f(k_1)\ldots\f(k_n)\>\!\>~.
\end{align}
As before, $\Delta=d/2+\nu$ is the dimension of the boundary scalar $\phi$. We can then write the on-shell propagator as:\footnote{Another way to derive this is to consider the two-point Wightman function in free-field theory, $\<\Phi(k_1,z_1)\Phi(k_2,z_2)\>_{\text{free}}$, and expand the fields in terms of creation and annihilation operators for the normalizable Poincar\'e modes \cite{Balasubramanian:1999ri}.}
\begin{align}
G^{+}_{\Delta}(k,z_1,z_2)&=K^{+}_{\Delta}(k,z_1)\frac{1}{\<\!\<\f(-k)\f(k)\>\!\>}K^{+}_{\Delta}(k,z_2)~. \label{eq:SplitLorentzian}
\end{align}
As shown in \eqref{eq:PosBb}, $K^{+}_{\Delta}(k,z) \sim \mathcal{J}(\sqrt{-k^2} z)$, and so the bulk-to-bulk on-shell propagator factorizes into a product of normalizable modes. 
We can now see explicitly that cutting bulk-to-bulk propagators inside a Witten diagram produces two sub-diagrams glued together via on-shell bulk-to-boundary propagators with the correct normalization. The on-shell condition restricts the momentum $k$ to lie in the forward lightcone, $V_+$, and this turns the momentum integral into a phase space integral. Finally, dividing by the two-point CFT Wightman function gives the correct normalization for the exchanged states.

The relation between the bulk and boundary descriptions of unitarity becomes clear when we work in terms of the ``transition amplitudes'', $\<\Psi_{q'}|T[\f(k_1)\ldots\f(k_n)]|\Psi_{q}\>$ \cite{Balasubramanian:1999ri}.\footnote{While transition amplitudes are the standard name in AdS/CFT, these are more precisely the analogues of flat space form factors.} In the Poincar\'e patch, the states are defined via boundary conditions on the past and future Poincar\'e horizons. For the transition amplitudes studied here, the states $|\Psi_{q}\>$ and $\<\Psi_{q'}|$ are defined in terms of a collection of normalizable modes with momenta $q_1,\ldots,q_r$ and $q'_1,\ldots,q'_s$ such that $q_{i}$, $q'_{j}\in V_{+}$. The $q_i$ and $q'_j$ are incoming and outgoing momenta respectively. In practice, these transition amplitudes are computed by replacing some of the time-ordered bulk-to-boundary propagators in a standard Witten diagram with the corresponding on-shell propagators, i.e. the normalizable modes \cite{Balasubramanian:1998de,Balasubramanian:1998sn}. From \eqref{eq:SplitLorentzian}, we see that cutting a bulk-to-bulk propagator produces sub-diagrams with normalizable external lines. To be concrete, we can consider the cut of a tree-level exchange diagram in $\Phi^3$ theory,
\begin{equation}
\includegraphics[scale=.22]{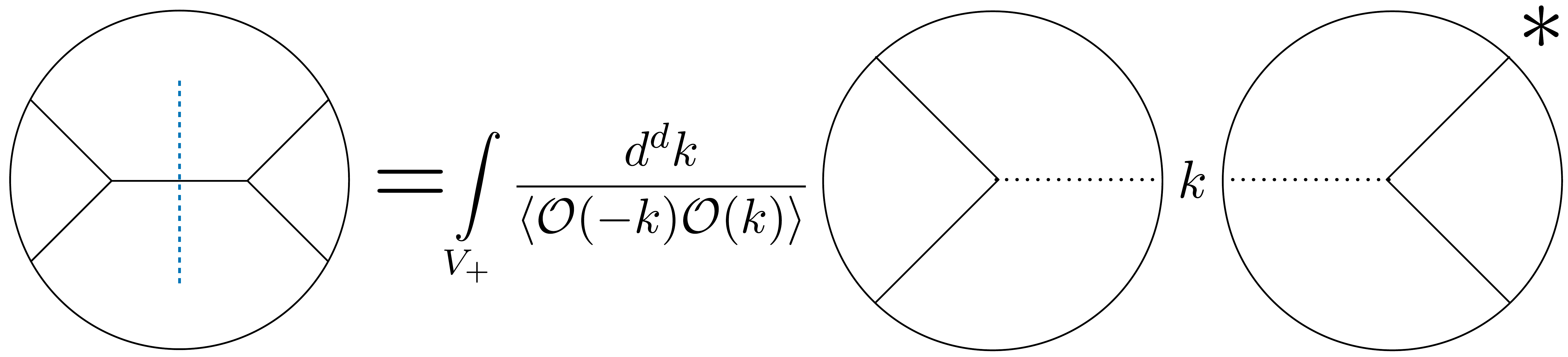}~.
\label{eq:Cut_Tree_Lorentzian}
\end{equation}
The dotted lines on the right-hand side of \eqref{eq:Cut_Tree_Lorentzian} are the on-shell bulk-to-boundary propagators, $K^{+}_{\Delta}(k,z)$, while the undotted lines are the corresponding Feynman propagators, $K_{\Delta}(k,z)$. Following the cutting rules, we also complex conjugate the three-point Witten diagram to the right of the cut. Specifically, we find:
\begin{align*}
-2~\Re  W'_{\phi,\text{exch}}(k_1,\ldots,k_4)=g^{2}\int\limits_{0}^{\infty}\frac{dz_1dz_2}{z_1^{d+1}z_2^{d+1}}&K_{\Delta}(k_1,z_1)K_{\Delta}(k_2,z_1)G^{+}_{\Delta}(k_{12},z_1,z_2)
\\
&K^*_{\Delta}(k_3,z_2)K^*_{\Delta}(k_4,z_2), \label{eq:Cut_Tree_Lorentzian_eqn}
\numberthis
\end{align*}
where $k_{ij}=k_i+k_j$ and the prime means we drop the overall momentum conserving $\delta$-function,
\begin{align}
W(k_1,\ldots,k_4)=(2\pi)^{d}\delta^d(k_1+\ldots+k_4)W'(k_1,\ldots,k_4)~.
\end{align}
Using \eqref{eq:SplitLorentzian} we can rewrite this as 
\begin{align*}
-2~\Re  W'_{\phi,\text{exch}}(k_1,\ldots,k_4)=& g^{2} \int\limits_{0}^{\infty}\frac{dz_1dz_2}{z_1^{d+1}z_2^{d+1}}K_{\Delta}(k_1,z_1)K_{\Delta}(k_2,z_1)K^{+}_{\Delta}(k_{12},z_1)
\\
&\frac{1}{\<\!\<\f(-k_{12})\f(k_{12})\>\!\>}K^{+}_{\Delta}(k_{12},z_2)K^{*}_{\Delta}(k_3,z_2)K^{*}_{\Delta}(k_4,z_2)~. 
\numberthis
\label{eq:ReTreeExch}
\end{align*}
Denoting the three-point transition amplitude as
\begin{align}
\mathcal{R}_{3-\textrm{pt}}(k_1,k_2 | k)= ig\int\limits_{0}^{\infty}\frac{dz}{z^{d+1}}&K_{\Delta}(k_1,z)K_{\Delta}(k_2,z)K^{+}_{\Delta}(k,z)~,
\end{align}
we see that the cut Witten diagram is a product of transition amplitudes:
\begin{align}
-2~\Re  W'_{\phi,\text{exch}}(k_1,\ldots,k_4)=\mathcal{R}_{3-\textrm{pt}}(k_1,k_2| k_{12}) \frac{1}{\<\!\<\f(-k_{12})\f(k_{12})\>\!\>} \mathcal{R}^*_{3-\textrm{pt}}(k_3,k_4|k_{12} )~. \label{eq:ReWtoTransition}
\end{align}
The final result and ordering agree with inserting a resolution of the identity in the right-hand side of \eqref{eq:FromReToTTb}. Specifically, we can insert a complete set of single-particle states, which we label as $|\Psi_{k}\>$, into \eqref{eq:FromReToTTb} to find: 
\begin{align}
\<\overline{T}[\f(k_3)\f(k_4)]T[\f(k_1)\f(k_2)]\>=& \int\limits_{V_+}\frac{d^{d}k}{(2\pi)^{d}}\frac{\<0|\overline{T}[\f(k_3)\f(k_4)]|\Psi_{k}\>\<\Psi_{k}|T[\f(k_1)\f(k_2)]|0\>}{\<\Psi_{k}|\Psi_{k}\>}
\nonumber \\
=&  \int\limits_{V_+}\frac{d^{d}k}{(2\pi)^{d}}\frac{\<\Psi_{k}|T[\f(k_3)\f(k_4)]|0\>^*\<\Psi_{k}|T[\f(k_1)\f(k_2)]|0\>}{\<\Psi_{k}|\Psi_{k}\>}~. \label{eq:sumpoincstates}
\end{align}
We can restrict to single-particle states because we are working at tree-level in the AdS theory.
The result from the cutting rules \eqref{eq:ReWtoTransition} and from inserting a complete set of states \eqref{eq:sumpoincstates} then agree due to the relations,
\begin{align}
\<\Psi_{k}|T[\f(k_1)\f(k_2)]|0\>&= (2\pi)^{d}\delta^{d}(k+k_{12})\mathcal{R}_{3-\textrm{pt}}( k_1,k_2|k)~,
\\
\<\Psi_{k}|\Psi_{k}\>&=(2\pi)^{d}\<\!\<\f(-k)\f(k)\>\!\>~.
\end{align}
This example shows that the cutting rules for Witten diagrams, which were derived using purely diagrammatic identities, have a simple correspondence with transition amplitudes defined between states on the Poincar\'e horizons.

The definition we have used for the transition amplitudes is perturbative in nature, as they are defined directly via Witten diagrams.  
In principle, one can also give a non-perturbative definition for Poincar\'e transition amplitudes via correlation functions in global AdS. We will not need this definition and will instead point the reader to \cite{Balasubramanian:1999ri,Raju:2011mp} for more details.

\subsection{Higher-Point Functions}
\label{sec:Higher_Point}
In this section we will briefly discuss the cutting rules for higher-point functions. We start by using the CFT optical theorem \eqref{eq:CombCor} for general points \cite{Gillioz:2016jnn}:
\begin{align}
&\<T[\f(x_1)\ldots\f(x_n)]\>+(-1)^n\<\overline{T}[\f(x_1)\ldots\f(x_n)]\>
\nonumber \\ &\hspace{1in}=-\sum\limits_{r=1}^{n-1}(-1)^r\sum\limits_{\substack{\sigma\in\Pi(r,n-r)}}\<\overline{T}[\f(x_{\s_1})\ldots\f(x_{\s_r})]T[\f(x_{\s_{r+1}})\ldots\f(x_{\s_n})]\>~, \label{sec:CutnPoint}
\end{align}
where we recall $\Pi(r,n-r)$ is the set of partitions of $\{1,\ldots,n\}$ into two sets of size $r$ and $n-r$. This relation tells us that the (real) imaginary parts of (even) odd-point correlators can be expressed in terms of lower-point correlators. We can then factorize the right-hand side by using a resolution of the identity.

Next, we use that the cutting rules given in Sections \ref{sec:ReviewCutkosky} and \ref{sec:UnitarityCutsAdS} also compute the real and imaginary piece for even and odd-point functions, respectively. In the cutting rules, this happens because there is a factor of $(-1)$ for each external point to the right of the cut. For general $n$-point Witten diagrams we find
\begin{align}
W(x_1,\ldots,x_n)+(-1)^nW^{*}(x_1,\ldots,x_n)=-\sum\limits_{q=2}^{2^{n+m}-1}\widehat{W}_{q}(x_1,\ldots,x_n)~.\label{eq:WittenGenNpoints}
\end{align}
This result is expected, as one can also derive \eqref{eq:WittenGenNpoints} directly from \eqref{sec:CutnPoint} using the Schwinger-Keldysh rules.

In Section \ref{sec:Polyakov} we used a special choice of kinematics to relate $\Re \<T[\f\f\f\f]\>$ to a double-commutator. The motivation there was to make a connection with the Lorentzian inversion formula \cite{Caron-Huot:2017vep,ssw,Kravchuk:2018htv}, where the same double-commutator appears. For higher-point functions the corresponding inversion formula is not known, but certain kinematics still simplify \eqref{sec:CutnPoint}. One natural choice is to set all the external momenta to be spacelike, $k_i^{2}>0$. Then the terms with $r=1 \ \text{and} \ n-1$ vanish in \eqref{sec:CutnPoint} since $\f(k_i)|0\>=0$ with this choice. At the level of Witten diagrams, this choice of momenta sets the external cuts to zero.

As an example, we can consider a five-point function with the following kinematics:
\begin{align}
k_{i}^{2}>0, \quad (k_i+k_j)^{2}>0, \quad \text{except} \quad k_1+k_2\in V_+~.
\label{eq:5ptKinematics}
\end{align}
In this case,
\begin{align}
-2i\Im \<T[\f(k_1)\ldots\f(k_5)]\>= \<\overline{T}\left[\f(k_3)\f(k_4)\f(k_5)\right]T[\f(k_1)\f(k_2)]\>~.
\end{align}
Next, we will look at the allowed cuts for an individual Witten diagram. Given the restrictive kinematics we have chosen, the momentum flowing through a cut has to be equal to $k_{12}$. For example, for the following five-point tree-level diagram,
\begin{equation}
\includegraphics[scale=.28]{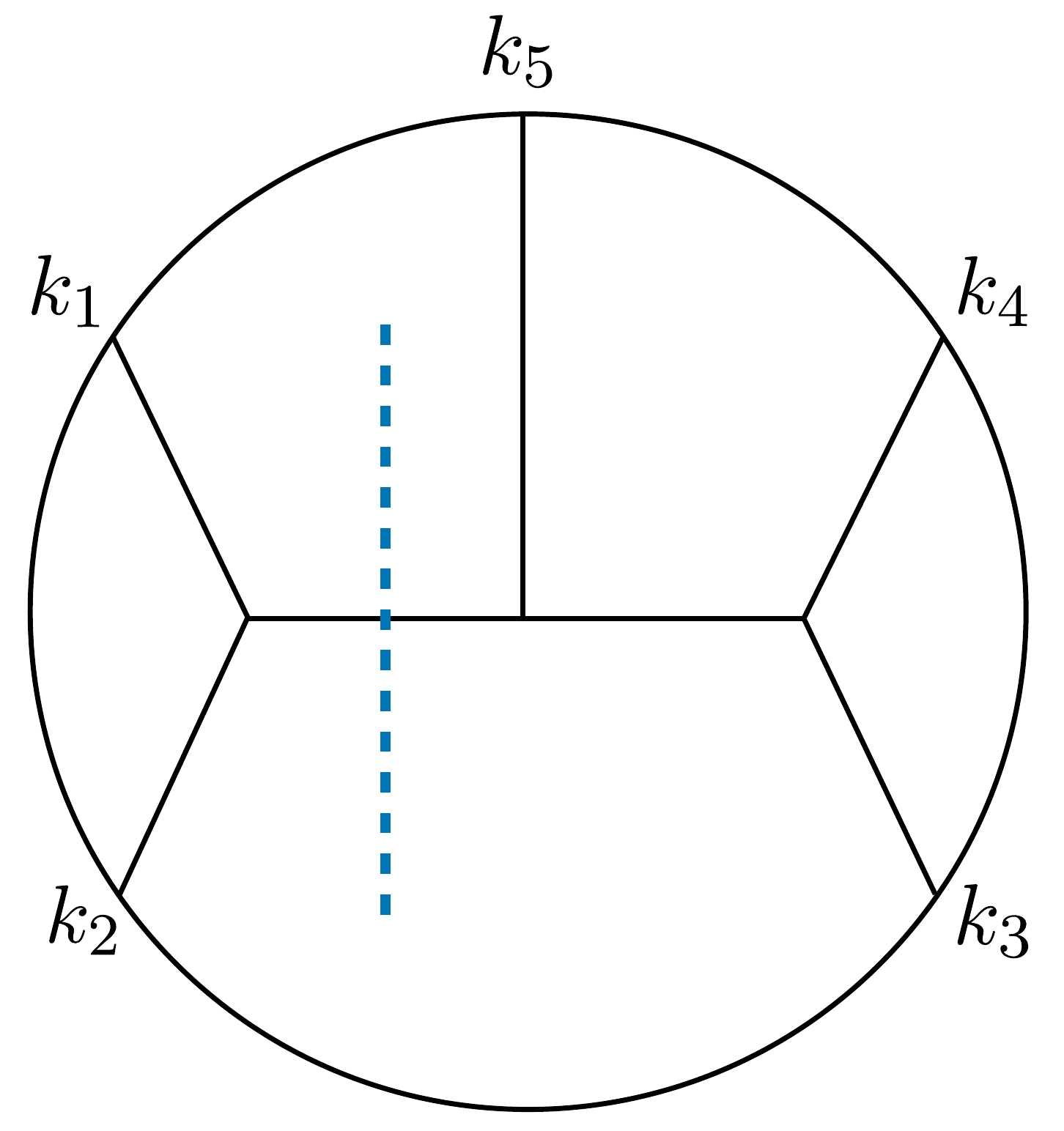}
\label{eq:fivepointdiagram}
\end{equation}
only the above cut is non-zero. Defining the three and four-point transition amplitudes as
\begin{align*}
\mathcal{R}_{3-\textrm{pt}}(k_1,k_2 | k) &= ig\int\limits_{0}^{\infty}\frac{dz}{z^{d+1}}K_{\Delta}(k_1,z)K_{\Delta}(k_2,z)K^{+}_{\Delta}(k,z)~,
\numberthis
\\
\mathcal{R}_{4-\textrm{pt}}(k_3,k_4,k_5 | k) &=-g^2\int\limits_{0}^{\infty}\frac{dz_1dz_2}{z_1^{d+1}z_2^{d+2}}K^{+}_{\Delta}(k,z_1)K_{\Delta}(k_5,z_1)G_{\Delta}(k+k_5,z_1,z_2)
\\
&~~~~~~~~~~~~~~~~~~~~~~~~~~
K_{\Delta}(k_3,z_2)K_{\Delta}(k_4,z_2)~,
\numberthis
\end{align*}
we find, in the kinematics \eqref{eq:5ptKinematics}, that
\begin{align}
-2i\hspace{.1cm}\Im \<T[\f(k_1)\ldots\f(k_5)]\>=\frac{1}{\<\!\<\f(-k_{12})\f(k_{12})\>\!\>}\mathcal{R}_{3-\textrm{pt}}( k_1,k_2 |k_{12}) \mathcal{R}^*_{4-\textrm{pt}}( k_3,k_4,k_5 | k_{12})~. 
\end{align}
We see that the cut five-point diagram can be written as the product of two transition amplitudes, in agreement with our previous analysis. 

An important open question is: what is the minimal set of reduced correlators that we need to know in order to reconstruct the full five-point function? At four points we can choose spacelike momenta to reduce $\Re \<T[\f\f\f\f]\>$ to double-commutators. There are three double-commutators we can consider, but the Lorentzian inversion formula \cite{Caron-Huot:2017vep} shows that two of them are sufficient to reconstruct the full correlator. It would be interesting to answer this question at higher points and understand the connection to the cutting rules presented here.

\section{Applications to Witten Diagrams}
\label{sec:Examples}

In this section we check our cutting rules in a variety of ways. At tree level, we confirm that our cutting rules agree with the discontinuity of the full Witten diagram. By using the momentum-space OPE \cite{Gillioz:2019lgs,Gillioz:2019iye}, we will relate the bulk cut structure to the spectrum of the dual CFT and find agreement with previous work on the OPE limit of Witten diagrams. Finally, by studying tree and loop examples, we show that cut AdS diagrams become the corresponding cut flat space diagrams in the flat space limit. This gives evidence that the flat space limit of the AdS Cutkosky are the corresponding S-matrix rules.

 \subsection{OPE and Flat Space Limits}
 \label{sec:Limits}
\subsubsection*{OPE in momentum space}
In this section we study how the bulk cutting procedure in the Poincar\'e patch is related to the standard boundary OPE. We begin with the relation
\begin{align}
-2~\Re \<T[\f(k_1)\f(k_2)\f(k_3)\f(k_4)]\>=\<\overline{T}[\f(k_3)\f(k_4)]T[\f(k_1)\f(k_2)]\>~,
\end{align}
which holds in the configuration \eqref{eq:kinematics}. To find the Lorentzian OPE \cite{Gillioz:2016jnn,Gillioz:2018kwh,Gillioz:2018mto,Gillioz:2019lgs}, we will insert a complete set of states between the pairs of ordered operators. Specifically, we will use
\begin{align}
\mathbb{I}=|0\>\<0| + \sum\limits_{\O}\int\limits_{V_{+}}\frac{d^dk}{(2\pi)^{d}}P^{\Delta_{\O}}_{\mu_1\ldots\mu_\ell,\nu_1\ldots\nu_\ell}(k)|\O^{\mu_1\ldots\mu_\ell}(k)\>\<\O^{\nu_1\ldots\nu_\ell}(-k)|~,
\label{eq:IdRes}
\end{align}
where the sum runs over all the local primary operators $\O$ of the boundary CFT.
The explicit form of the projector is
\begin{align}
P^{\Delta}_{\mu_1\ldots\mu_\ell,\nu_1\ldots\nu_\ell}(k)=&\frac{(-k^{2})^{d/2-\Delta}}{C_{\Delta}}\sum\limits_{n=0}^{\ell}\frac{2^n\ell!(\Delta-\frac{d}{2})_{n}}{n!(\ell-n)!(\Delta-\ell-d+2)_{n}}
\nonumber 
\\
& \left(\frac{1}{\ell!}\frac{k_{\mu_1}k_{\nu_1}\ldots k_{\mu_n}k_{\nu_n}}{(-k^{2})^{n}}\eta_{\mu_{n+1}\nu_{n+1}}\ldots\eta_{\mu_{\ell}\nu_{\ell}}+\text{perms} - \text{traces}\right)~.
\end{align}
The tensor $P^{\Delta}_{\mu_1\ldots\mu_\ell,\nu_1\ldots\nu_\ell}(k)$ is what appears in the two-point function for the shadow operator $\widetilde{\O}_{d-\Delta,\ell}$, i.e. for a fictitious operator of dimension $d-\Delta$ and spin $\ell$. The factor $C_{\O}$ is related to the normalization of the two-point function and is given by\footnote{To compare with eqn 2.15 of \cite{Gillioz:2018mto} we note that the operators there are unit normalized, $\<\O_{\Delta,\ell}|\O_{\Delta,\ell}\>=1$, while here we have $\<\O_{\Delta,\ell}|\O_{\Delta,\ell}\>=\frac{(\ell+\Delta-1)\Gamma(\Delta)}{2\pi^{d/2}(\Delta-1)\Gamma(\Delta+1-d/2)}$.}
\begin{align}
C_{\Delta}=\frac{ 2^{d-2 \Delta }\pi}{\Gamma \left(-\frac{d}{2}+\Delta +1\right)^2}~.
\end{align}
Finally, using \eqref{eq:IdRes} gives the momentum-space OPE \cite{Gillioz:2019lgs}
\begin{align}
\hspace{-.7cm}
\<\!\<\overline{T}[\f(k_3)\f(k_4)]T[\f(k_1)\f(k_2)]&\>\!\>=\sum\limits_{\O}\<\!\<\overline{T}[\f(k_3)\f(k_4)]\O^{\mu_1\ldots\mu_\ell}(k_{12})\>\!\> 
\nonumber \\
\times& P^{\Delta_{\O}}_{\mu_1\ldots\mu_\ell,\nu_1\ldots\nu_\ell}(k)\<\!\<\O^{\nu_1\ldots\nu_\ell}(-k_{12})T[\f(k_1)\f(k_2)]\>\!\>~.
\end{align}
As a reminder, the $\<\!\<\ldots\>\!\>$ notation means that we drop the overall momentum-conserving $\delta$-function.
To relate the bulk cutting rules to the momentum-space OPE, we will study Witten diagrams in the limit that the exchanged momentum goes to zero, $k_{12}\rightarrow 0$. In this limit, we have \cite{Gillioz:2019lgs}:
\begin{align}
\<\!\<\O(-k_{12})T[\f(k_1)\f(k_2)]\>\!\>\sim (-(k_{12})^2)^{\Delta_{\O}-d/2}(k_1^2-i\epsilon)^{\Delta-\Delta_{\O}/2-d/2}~.
\end{align}
The exchange of the operator $\O$ therefore gives the scaling
\begin{align}
\<\!\<\overline{T}[\f(k_3)\f(k_4)]T[\f(k_1)\f(k_2)]\>\!\>\bigg|_{\O}\sim  (-k_{12}^2)^{\Delta_{\O}-d/2}~, \ \label{eq:OPETTb}
\end{align}
where we used that the projector scales as $P^{\Delta}_{\mu_1\ldots\mu_\ell,\nu_1\ldots\nu_\ell}(k)\sim (-k^{2})^{d/2-\Delta}$. Using this zero-momentum limit, we will show that there is a correspondence between the cut lines of a Witten diagram and the operators that appear in the boundary OPE. That is, if we can perform a cut in which only a single propagator for the bulk scalar $\Phi$ is cut, then its dual operator $\f$ must appear in the boundary OPE. Similarly, if we can cut multiple $\Phi$ lines then the corresponding multi-trace operator built from $\f$ must appear in the OPE. 

The correspondence between bulk cuts and boundary operators is expected, both from previous work on AdS/CFT unitarity \cite{Fitzpatrick:2011dm,Aharony:2016dwx,Yuan:2017vgp,Yuan:2018qva,Meltzer:2019nbs} and from the previous discussion on cut graphs and Poincar\'e transition amplitudes. However, one subtlety is that our derivation of the cutting rules is based on quantizing the AdS theory on slices of constant Poincar\'e time. This is why there is a simple map between the cuts of a diagram and the Poincar\'e transition amplitudes. On the other hand, in order to study the OPE we quantize the CFT using radial quantization, which is dual to quantizing the AdS theory on slices of constant global time. We therefore do not expect that our Poincar\'e cuts necessarily isolate the dual single- or multi-trace operator in the boundary OPE. Instead, we will give evidence for a weaker but still useful statement: the existence of a bulk cut implies the existence of the corresponding single- or multi-trace operators in the boundary OPE.

We begin by studying the simplest non-trivial case, the exchange Witten diagram, which is given in \eqref{eq:Cut_Tree_Lorentzian}-\eqref{eq:Cut_Tree_Lorentzian_eqn} and for convenience is reproduced below,
\begin{align*}
-2~\Re  W'_{\f,\text{exch}}(k_1,\ldots,k_4)=\int\limits_{0}^{\infty}\frac{dz_1dz_2}{z_1^{d+1}z_2^{d+1}}&K_{\Delta}(k_1,z_1)K_{\Delta}(k_2,z_1)G^{+}_{\Delta}(k_{12},z_1,z_2)
\\
&K^*_{\Delta}(k_3,z_2)K^*_{\Delta}(k_4,z_2)~. 
\numberthis
\label{eq:treeRePart}
\end{align*}
The tree diagram will be a useful example for seeing the cutting rules in action and understanding the structure in more general diagrams. As expected, we will show that the boundary OPE of \eqref{eq:treeRePart} involves only the exchange of the single-trace operator $\f$ and its descendants \cite{Caron-Huot:2017vep}. 

To understand how the scaling \eqref{eq:OPETTb} emerges, we will study the limit $k_{12}\rightarrow 0$ under the $z$ integrals.\footnote{In Section \ref{subsec:4ptexchange} we will show the expected scaling emerges when we perform the $z$ integrals first.} From  \eqref{eq:treeRePart}, we see that the dependence on $k_{12}$ comes from the on-shell bulk-to-bulk propagator. In the zero-momentum limit, the propagator takes the form
\begin{align}
G^{+}_{\Delta}(k,z_1,z_2)\approx  \frac{\pi}{2^{2\nu}\Gamma(1+\nu)^{2}} (z_1z_2)^{\Delta}(-k^2)^{\Delta-d/2}~,
\end{align}
where we used the explicit expression \eqref{eq:PosBB}. Substituting this into \eqref{eq:treeRePart} and comparing to \eqref{eq:OPETTb} confirms the expected scaling due to $\f$ exchange.

We next study the bubble diagram, drawn in \eqref{eq:Witten_CutToVertices}, in the same limit. As we are performing a two-particle cut, we expect that by taking the limit $k_{12} \rightarrow0$ we will see the exchange of the double-trace operators $[\f\f]_{n,J}$ in the boundary OPE. The double-traces have the form
\begin{align}
[\f\f]_{n,J}=\f\partial^{\mu_1}\ldots\partial^{\mu_J}\f -\text{traces}~,
\\
\Delta_{n,J}=2\Delta+2n+J~.
\end{align}
We will study the leading OPE contribution, which is governed by the exchange of the scalar $[\f\f]_{0,0}$ with dimension $2\Delta$. The full expression for the cut bubble is
\begin{align}
-2~\Re  W'_{\text{bubble}}(k_1,\ldots,k_4)=\int\limits_{V_+} \frac{d^{d}\ell}{(2\pi)^{d}}\int\limits_{0}^{\infty}&\frac{dz_1dz_2}{z_1^{d+1}z_2^{d+1}}K_{\Delta}(k_1,z_1)K_{\Delta}(k_2,z_1)G^{+}_{\Delta}(\ell,z_1,z_2)\nonumber
\\ &G^{+}_{\Delta}(k_{12}-\ell,z_1,z_2) K^*_{\Delta}(k_3,z_2)K^*_{\Delta}(k_4,z_2)~.
\end{align}
We will take the limit $k_{12}\rightarrow 0$ with $k_{12}\in V_{+}$. The above expression involves on-shell propagators that are non-zero only for $k_{12}-\ell\in V_{+}$ and $\ell \in V_+$. These conditions imply that when $k_{12}\rightarrow 0$, we must also have $\ell\rightarrow0$. That is, the integration region for the phase space integral is bounded by the size of the incoming momenta. To see this explicitly, we write
\begin{align}
k_{12}=v+\ell, \qquad v\in V_{+}~.
\end{align}
Squaring both sides yields
\begin{align}
k_{12}^2=v^2+2v\cdot \ell +\ell^2~.
\end{align}
As $v$, $\ell\in V_+$, each term on the right-hand side is negative. Taking $k_{12}\rightarrow 0$ then requires that $v$, $\ell\rightarrow 0$ as well. When we take this limit, we therefore find that each on-shell propagator scales like $(-k_{12}^2)^{\Delta-d/2}$ while the shrinking phase-space integral gives an extra factor of $(-k_{12}^2)^{d/2}$. Putting this together, we find the expected scaling:
\begin{align}
-2~\Re  W'_{\text{bubble}}(k_1,\ldots,k_4)\sim (-k_{12}^{2})^{\Delta-d/2}~.
\end{align}
This pattern continues at higher loops: when we cut $n$ lines, we find the expected scaling for an $n$-trace operator. For each propagator we cut, we have a factor of $(-k_{12}^{2})^{\Delta-d/2}$ and for each loop momentum in a cut line we find a factor of $(-k_{12}^{2})^{d/2}$ from the loop measure.
\subsubsection*{Flat Space Limit}
Studying the flat space limit together with our cutting rules will provide another non-trivial consistency check. As we are working in the Poincar\'e patch, we will use the flat space limit given in \cite{Raju:2012zr}, which we review here. To define this limit, we write the Witten diagram as an independent function of the momenta $k_i$ and their norms $|k_i| \equiv \sqrt{k^{2}}$,
\begin{align}
W(k_1,|k_1|,\ldots,k_4,|k_4|)~.
\end{align}
We will assume the $d$-dimensional momentum $k$ is spacelike and then define a $(d+1)$-dimensional null vector,
\begin{align}
\tilde{k}=(k,i|k|)~.\label{eq:flat_momenta}
\end{align}
If we define the total energy as
\begin{align}
E_{T}=\sum\limits_{i}|k_i|~,
\end{align}
then the flat-space amplitude comes from a total energy pole of the Witten diagram,
\begin{align}
M(\tilde{k}_1,\ldots,\tilde{k}_{4})\propto\lim\limits_{E_{T}\rightarrow0} (E_T)^{\alpha}W(k_1,|k_1|,\ldots,k_4,|k_4|)~.
\end{align}
In general, the exact strength of the pole and proportionality factor depend on the loop order and the theory.\footnote{For related work on dS correlators see e.g. \cite{Arkani-Hamed:2017fdk,Arkani-Hamed:2018kmz,Baumann:2020dch,Benincasa:2018ssx,Arkani-Hamed:2018bjr,Benincasa:2019vqr}.}  

In the physical region, all $|k_i|$ are positive for $k$ spacelike and we do not have access to the total energy pole. To reach this pole, we instead treat the $|k_i|$ as independent complex variables and analytically continue in them.\footnote{This analytic continuation is distinct from the one used to go from spacelike to timelike momenta, where the $i\epsilon$ prescription determines how to approach the branch cut at $k^2<0$ and $|k|$ is not an independent variable. Instead one keeps $|k|=\sqrt{k^2}$, which is imaginary for timelike momenta.} However, to obtain null momenta in the flat space limit, we still need to impose that $|k_i|^2=k_i\cdot k_i$ before taking the flat space limit. The procedure is then: we first analytically continue in some of the $|k_i|$ to flip their signs and then take the limit $E_{T}\rightarrow 0$.  By using \eqref{eq:flat_momenta}, we recover the flat-space amplitude with complexified $(d+1)$-dimensional momenta. This flat space limit originates from the fact that the total energy pole comes from the $z\rightarrow \infty$ limit of the AdS integration, where the AdS integrand takes the same form as the flat-space integrand. Comparing the AdS and flat-space expressions fixes the coefficient of $|k|$ in \eqref{eq:flat_momenta} \cite{Raju:2012zr}.

To be concrete, we can consider a conformally-coupled scalar in AdS, which is dual to a boundary scalar of dimension $\Delta_{c}=\frac{1}{2}(d+1)$. The flat-space amplitude is then the residue of the total energy pole,
\begin{align}
W(k_1,|k_1|,\ldots,k_4,|k_4|)=\frac{M(\tilde{k}_{1},\ldots,\tilde{k}_{4})}{E_T}+\ldots,
\end{align} 
where the omitted terms are regular at $E_T=0$.

Before performing any analytic continuation in $|k_i|$, we find
\begin{align}
\Re W(k_1,|k_1|,\ldots,k_4,|k_4|)=\frac{\Re M(\tilde{k}_{1},\ldots,\tilde{k}_{4})}{E_T}+\ldots
\end{align}
and so we identify the discontinuity of the flat space tree-level amplitude as the coefficient of a total energy pole in $\Re W$.
One simple way to understand this limit is to write the real part of the CFT correlator as
\begin{align}
2~\Re\<T[\f\f\f\f]\> =\<T[\f\f\f\f]\>+\<\overline{T}[\f\f\f\f]\>~. \label{eq:RePartFlatSpace}
\end{align}
We can then take the flat space limit of each correlator on the right-hand side individually. The flat space limit of the time-ordered correlator gives matrix elements for $i\mathcal{T}$ \cite{Gary:2009ae,Okuda:2010ym} while the flat space limit of the anti-time-ordered correlator gives matrix elements for $-i\mathcal{T}^{\dagger}$. Their sum is then the natural object to study whose flat space limit yields $\Im(\mathcal{T})$. We then see from \eqref{eq:FromReToDC} that for certain kinematics the total energy pole in the causal double-commutator computes the discontinuity of a flat space amplitude \cite{Alday:2017vkk,Alday:2018kkw,Bissi:2020wtv}. We will verify this explicitly in the following sections.

\subsection{Four-Point Scalar Exchange}
\label{subsec:4ptexchange}
To make the previous discussions more concrete, we will now consider explicit examples of cut Witten diagrams. For simplicity, we consider diagrams with external conformally-coupled scalars $\phi_{c}$ with dimension $\Delta_{c}=\frac{1}{2}(d+1)$. When calculating the real part of a four-point Witten diagram, we will always work in the kinematics \eqref{eq:kinematics}. Computing the real part of a diagram is then equivalent to taking a discontinuity with respect to $k_{12}^{2}$ across the branch cut at $k_{12}^{2}<0$. In contrast to the flat space limit, when computing this discontinuity we impose that $|k_i|=\sqrt{k_{i}^{2}}$ and similarly for $k_{ij}$. 

One benefit of using conformally-coupled scalars is that the bulk-to-boundary propagator takes a simple form,
\begin{align}
K_{\Delta_c}(k,z)= -i z^{\frac{d-1}{2}}e^{-|k|z}~.
\end{align}
First, we will consider an exchange diagram for $\<\f_c\f_c\f_c\f_c\>$ where the exchanged scalar $\mathcal{O}$ has arbitrary dimension $\Delta_{\mathcal{O}}$:
\begin{align}
-2~\Re W'_{\O,\text{exch}}(k_1,\ldots,k_4)=g^{2}\int\frac{dz_1dz_2}{z_1^{d+1}z_{2}^{d+1}}&K_{\Delta_c}(k_1,z_1)K_{\Delta_c}(k_2,z_1)G^+_{\Delta_{\O}}(k_{12},z_1,z_2)
\nonumber \\ &K^*_{\Delta_c}(k_3,z_2)K^*_{\Delta_c}(k_4,z_2)~.\label{eq:cutExchV2}
\end{align}
The $z$ integrals can be evaluated and we find
\begin{align}
-2~\Re W'_{\O,\text{exch}}(k_1,\ldots,k_4)=&
g^2\frac{\pi   2^{d-2 \Delta_{\O}} \Gamma (\Delta_{\O}-1)^2 }{\Gamma \left(-\frac{d}{2}+\Delta_{\O}+1\right)^2}
\frac{
(-k_{12}^{2})^{\Delta_{\O}-\frac{d}{2}} \theta(-k^{2})\theta(k^0)
}
{
 ((|k_1|+|k_2|) (|k_3|+|k_4|))^{\Delta_{\O}-1}
}
\nonumber \\
&\, _2F_1\left(\frac{\Delta_{\O}-1}{2},\frac{\Delta_{\O}}{2};\Delta_{\O}-\frac{d}{2}+1;\frac{k_{12}^{2}}{(|k_1|+|k_2|)^2}\right) 
\nonumber \\
& _2F_1\left(\frac{\Delta_{\O}-1}{2},\frac{\Delta_{\O}}{2};\Delta_{\O}-\frac{d}{2}+1;\frac{k_{12}^{2}}{(|k_3|+|k_4|)^2}\right)~.
\end{align}
We see that when $k_{12}^{2}\rightarrow 0$, the Witten diagram scales like 
\begin{align}
-2~\Re W'_{\O,\text{exch}}(k_1,\ldots,k_4)\sim (-k_{12}^{2})^{\Delta_{\O}-\frac{d}{2}}~,
\end{align}
which corresponds to the exchange of $\O$ in the boundary CFT. Expanding the ${}_{2}F_1$ hypergeometric functions yields additional powers that correspond to descendants of $\O$. We use the notation $k_{12}^{2}$ instead of $|k_1+k_2|^2$ to make the analytic continuation in these variables clearer. This will also distinguish them from $|k_i|$, which are analytically continued to obtain the flat space limit.

To verify that our cutting rules give $-2\hspace{.1cm}\Re W$, we will consider a case where the Witten diagram can be computed in full and then take its discontinuity. One simple example is $d=5$ and $\Delta_{\O}=\Delta_{c}=3$. Assuming $k_i$ and $k_{12}$ are spacelike, we find
\begin{align}
W'_{\f_c,\text{exch}}(k_1,\ldots,k_4)=&\int\limits_{0}^{\infty} dz_1dz_2\int\limits_{0}^{\infty} dp\frac{2ig^2}{\pi}\frac{ \sin (p z_1) \sin (p z_2) e^{-(|k_1|+|k_2|)z_1 -(|k_3|+|k_4|)z_2 }}{  \left(k_{12}^{2}+p^2\right)}
\nonumber \\ =&\int\limits_{0}^{\infty}dp\frac{2ig^2}{\pi}\frac{ p^2}{ \left(k_{12}^{2}+p^2\right) \left((|k_1|+|k_2|)^2+p^2\right) \left((|k_3|+|k_4|)^2+p^2\right)}
\nonumber \\ =&\frac{i g^2}{\left(\sqrt{k_{12}^{2}}+|k_1|+|k_2|\right) \left(\sqrt{k_{12}^{2}}+|k_3|+|k_4|\right) (|k_1|+|k_2|+|k_3|+|k_4|)}~. \label{eq:exch_confScalar_d5}
\end{align}
Next, to go to the kinematics \eqref{eq:kinematics} we need to take $k_{12}^{2}<0$. From  \eqref{eq:exch_confScalar_d5} we see that there is a square root branch cut for timelike $k_{12}$. To compute the discontinuity across the cut, we take the difference between taking $k_{12}^{2}$ negative and real from below and above the real line in the complex $k_{12}^{2}$ plane. This yields  
\begin{align}
-2~\Re W'_{\f_c,\text{exch}}(k_1,\ldots,k_4)=\frac{2 g^2 \sqrt{-k_{12}^{2}}}{\left(k_{12}^{2}-(|k_1|+|k_2|)^2\right) \left(k_{12}^{2}-(|k_3|+|k_4|)^2\right)}~.  \label{eq:exch_confScalar_d5_Cut}
\end{align}
This agrees with the cutting rules \eqref{eq:cutExchV2} when we set $d=5$ and $\O=\phi_c$. 

Next, we study the flat space limit for the exchange diagram. The total energy pole in \eqref{eq:exch_confScalar_d5} appears explicitly, and its residue gives the flat space amplitude:
\begin{align}
\lim\limits_{E_T\rightarrow 0} E_T \hspace{.1cm} W'_{\f_c,\text{exch}}(k_1,\ldots,k_4)
 =-\frac{ig^{2}}{s}~, \label{eq:flat_space_Exchange}
\end{align}
where we identified the flat space Mandelstam invariant 
\begin{align}
s=(|k_1|+|k_2|)^{2}-k_{12}^{2}~. \label{eq:flatspaceinvariant}
\end{align}
By contrast, the real part of the Witten diagram as given in \eqref{eq:exch_confScalar_d5_Cut} does not have a total energy pole and then appears to vanish in the flat space limit. In order to capture the discontinuity of the flat space amplitude \eqref{eq:flat_space_Exchange}, which is given by $\delta(s)$, we need to implement the $i\epsilon$ prescription more carefully when analytically continuing the norms $|k_i|$. To see how the $\delta$-function emerges in the flat space limit, it will be convenient to use regulated $\delta$-functions in the cut propagators:
\begin{align}
G^{+,\epsilon}_\Delta(k,z_1,z_2)&=2\pi (z_1z_2)^{\frac{d}{2}}\int\limits_{0}^{\infty} dp \hspace{.1cm} p \mathcal{J}_{\nu}(pz_1)\mathcal{J}_{\nu}(pz_2)\delta^{\epsilon}(k^2+p^2)\theta(k^0)\theta(-k^{2})~,
\\
\delta^{\epsilon}(x)&=\frac{1}{\pi}\frac{\epsilon}{x^2+\epsilon^2}~.
\end{align}
Using this expression for the on-shell propagator in the cut tree-diagram gives
\begin{align*}
-2~\Re W'_{\f_c,\text{exch}}&(k_1,\ldots,k_4)
\\
=&\int\limits_{0}^{\infty} dp \frac{4 g^2}{\pi}\frac{ p^2  }{ \left((|k_1|+|k_2|)^2+p^2\right) \left((|k_3|+|k_4|)^2+p^2\right)}\frac{\epsilon}{\left(\left(k_{12}^{2}+p^2\right)^2+\epsilon ^2\right)}~.
\numberthis
\end{align*}
As the integrand is symmetric under $p\rightarrow -p$, we can extend the $p$ integration to $(-\infty,\infty)$ and evaluate the integral by closing the contour in the upper-half of the complex $p$ plane. We observe that there are four poles:
\begin{align}
p&=i(|k_1|+|k_2|)~, \label{eq:totEPt1}
\\
p&=i(|k_3|+|k_4|)~,\label{eq:totEPt2}
\\
p&=-\sqrt{-k_{12}^{2}-i\epsilon}~,\label{eq:nopolept1}
\\
p&=\sqrt{-k_{12}^{2}+i\epsilon}~.\label{eq:nopolept2}
\end{align}
Picking up the poles \eqref{eq:totEPt1} and \eqref{eq:totEPt2} will lead to a total energy pole in the final answer while the poles \eqref{eq:nopolept1} and \eqref{eq:nopolept2} will reproduce our earlier expression, which does not contain a total energy pole. Closing the $p$ contour and taking the limit $E_T\rightarrow 0$ before taking $\epsilon\rightarrow 0$ then yields the expected result:
\begin{align}
\lim\limits_{\epsilon\rightarrow 0}\lim\limits_{E_T\rightarrow 0}-2E_T~\Re W'_{\f_c,\text{exch}}(k_1,\ldots,k_4)&= \lim\limits_{\epsilon\rightarrow 0}\frac{2 \epsilon g^2}{\epsilon^2+\left((|k_1|+|k_2|)^2-k_{12}^{2}\right)^2}\theta(k_{12}^{0})
\nonumber \\ &=2\pi g^{2}\delta(s)\theta(k_{12}^{0})~.
\end{align}
We see that when sitting on a total energy pole, the real part of the AdS/CFT correlator reorganizes itself into a cut flat space amplitude. 

One noteworthy aspect of this flat space limit is that on the $E_{T}=0$ pole, the norms of the CFT momenta are identified with an emergent $(d+1)^{\text{th}}$ component of the external momenta. Heuristically, these norms can be thought of as the ``radial" momenta in the AdS dual. We see from \eqref{eq:flatspaceinvariant} that the emergent component can be identified with the energy of the respective particle in flat space. An emergent energy variable is natural in the study of dS correlators, where we expect to see bulk time emerge from the study of purely spatial correlators \cite{Arkani-Hamed:2015bza,Arkani-Hamed:2017fdk}. Here we are working in Lorentzian AdS, which already has a notion of time and energy, and the extra component is more naturally identified with a complex spatial momentum. 

\subsection{Four-Point Gauge Boson Exchange}

We will now repeat the previous analysis for Yang-Mills in AdS to illustrate the process for spinning fields. We will compute the cut of the Yang-Mills exchange diagram, both from the cutting rules and by taking a discontinuity of the full diagram. 

Following \cite{Raju:2010by,Raju:2011mp} we work in the axial gauge $A_z^a=0$, where $a$ is the color index. Throughout this section we drop the color indices, although it is straightforward to restore them. In the axial gauge, $\epsilon \cdot k=0$ for physical bulk modes, where $\epsilon$ is the polarization vector.\footnote{We hope it is clear from context where $\epsilon$ stands for a polarization vector and where it gives the $i\epsilon$ prescription for time-ordered propagators.} 
The Yang-Mills bulk-to-bulk propagators are
\begin{align}
G_{\mu\nu}^{\text{YM}}(k,z_1,z_2) &= -i (z_1 z_2)^{\frac{d-2}{2}}  \int\limits_0^\infty dp ~ p \frac{\mathcal{J}_{\frac{d-2}{2}}(p z_1)\mathcal{J}_{\frac{d-2}{2}}(p z_2)}{k^2+p^2 - i \epsilon}\mathcal{P}_{\mu\nu}(k,p)~,\label{eq:YMPropBB}
\\
G_{\mu\nu}^{\text{YM},\pm}(k,z_1,z_2) &= \pi (z_1 z_2)^{\frac{d-2}{2}}  \mathcal{J}_{\frac{d-2}{2}}(p z_1)\mathcal{J}_{\frac{d-2}{2}}(p z_2)\mathcal{P}_{\mu\nu}(k,\sqrt{-k^{2}})\theta(-k^{2})\theta(\pm k^0)~, \label{eq:YMPropBBW}
\end{align}
where we have defined the tensor
\begin{align}
\mathcal{P}_{\mu\nu}(k,p)=\eta_{\mu\nu}+\frac{k_{\mu}k_{\nu}}{p^{2}}~.
\end{align}
Taking one point to the boundary, we then find the bulk-to-boundary propagators\footnote{To obtain the bulk-to-boundary propagator, we have dropped terms analytic in $k$ that contribute to contact terms in position space.}
\begin{align}
K_{\mu\nu}^{\text{YM}}(k,z)&=-i\frac{1}{\Gamma(\frac{d}{2})2^{d/2-1}}(\sqrt{k^{2}}z)^{\frac{d-2}{2}}\mathcal{K}_{\frac{d-2}{2}}(\sqrt{k^{2}}z)\mathcal{P}_{\mu\nu}(k,\sqrt{-k^{2}})~, \label{eq:YMPropbB}
\\
K_{\mu\nu}^{\text{YM},\pm}(k,z)&=\frac{\pi}{\Gamma(\frac{d}{2})2^{d/2-1}}(\sqrt{k^{2}}z)^{\frac{d-2}{2}}\mathcal{J}_{\frac{d-2}{2}}(\sqrt{k^{2}}z)\mathcal{P}_{\mu\nu}(k,\sqrt{-k^{2}})\theta(-k^{2})\theta(\pm k^0)~.  \label{eq:YMPropbBW}
\end{align}
We see in \eqref{eq:YMPropBB}, \eqref{eq:YMPropbB}, and \eqref{eq:YMPropbBW} that the factor $\mathcal{P}_{\mu\nu}(k,\sqrt{-k^2})$ projects onto directions orthogonal to $k$.\footnote{In \cite{Raju:2011mp} the factor $\mathcal{P}_{\mu\nu}(k,-k^2)$ is not included in the bulk-to-boundary propagators as the condition $\epsilon\cdot k =0$ is imposed on the external polarizations.} The on-shell bulk-to-bulk propagator therefore factorizes into a product of on-shell bulk-to-boundary propagators. This is the same structure we saw earlier for scalar propagators in Section \ref{sec:transitionamps}. Finally, the cubic vertex is
\begin{align}
\mathcal{V}^{\mu\nu\rho}(k_1,k_2,k_3)=\frac{i}{\sqrt{2}}(\eta^{\mu\nu}(k_1-k_2)^\rho+\eta^{\nu\rho}(k_2-k_3)^{\mu}+\eta^{\rho\mu}(k_3-k_1)^\nu)~,
\end{align}
which takes the same form as the flat-space vertex factor. The full tree-level exchange diagram is then\footnote{The factors of $z_i^4$ come from using the inverse metric to contract the vertices and propagators.}:
\begin{align*}
W^{\text{YM}}_{\text{exch},\mu_1\ldots\mu_4}(k_1,k_2,k_3,k_4) &=g^{2} \int \frac{dz_1  dz_2 }{z_1^{d+1}z_2^{d+1}}  
K^{\text{YM}}_{\mu_1\nu_1}(k_1,z_1)K^{\text{YM}}_{\mu_2\nu_2}(k_2,z_1)
z_1^4 \mathcal{V}^{\nu_1\nu_2\rho} (k_1,k_2,k_{12})
\\
&
G^{\text{YM}}_{\rho\sigma}(k_{12},z_1,z_2)
z_2^4\mathcal{V}^{\nu_3\nu_4\sigma}(k_3,k_4,-k_{12})
K^{\text{YM}}_{\mu_3\nu_3}(k_3,z_2)K^{\text{YM}}_{\mu_4\nu_4}(k_4,z_2)~.
\numberthis
\end{align*}
To make the notation more compact, we will contract the external indices with polarization vectors
\begin{align}
W^{\text{YM}}_{\text{exch}}(k_1,k_2,k_3,k_4)=\epsilon^{\mu_1}_1\ldots\epsilon^{\mu_4}_4W^{\text{YM}}_{\text{exch},\mu_1\ldots\mu_4}(k_1,k_2,k_3,k_4)~.
\end{align}
The condition $\epsilon_i\cdot k_i=0$ trivializes the projector in the bulk-to-boundary propagators (\ref{eq:YMPropbB}) and (\ref{eq:YMPropbBW}). Finally, by specializing to $d=3$ one can perform the $p$ and $z$ integrals in closed form. This computation was carried out in \cite{Albayrak:2018tam} so, accounting for differences in normalization, we will quote the final result:
\begin{align}
W^{\text{YM}}_{\text{exch}}(k_1,k_2,k_3,k_4)=-ig^{2}&
\frac{\mathcal{V}^{12\rho}(k_1,k_2,-k_{12})\mathcal{V}^{34\sigma}(k_3,k_4,k_{12})}
{(\sqrt{k_{12}^{2}}+|k_1|+|k_2|)(\sqrt{k_{12}^{2}}+|k_3|+|k_4|)E_T}
\nonumber \\
&\left(\eta_{\rho\sigma}+\frac{(\sqrt{k_{12}^{2}}+E_T)(k_{12})_\rho (k_{12})_\sigma}{\sqrt{k_{12}^{2}}(|k_1|+|k_2|)(|k_3|+|k_4|)}\right)~,\label{eq:YMAdS4Full}
\end{align}
where $\mathcal{V}^{12\rho}=(\epsilon_{1})_\mu (\epsilon_{2})_\nu \mathcal{V}^{\mu\nu\rho}$. As a consistency check, we can take the flat space limit:
\begin{align}
\lim\limits_{E_T\rightarrow 0}E_{T}\hspace{.1cm}W^{\text{YM}}_{\text{exch}}(k_1,k_2,k_3,k_4)=\frac{ig^{2}}{s}&\mathcal{V}^{12\rho}(k_1,k_2,-k_{12})\mathcal{V}^{34\sigma}(k_3,k_4,k_{12})
\nonumber
\\
&\left(\eta_{\rho\sigma}-\frac{(k_{12})_\rho (k_{12})_\sigma}{(k_{12}\cdot n)^{2}}\right)~,
\end{align}
where $n=(0,0,0,1)$. This matches the flat-space amplitude, where we recall the vertices $\mathcal{V}^{\mu\nu\rho}$ only have indices in the first three directions.  

We can now compute the real part of \eqref{eq:YMAdS4Full} by taking the discontinuity and using the cutting rules applied to spinning particles. Using the cutting rules yields
\begin{align}
-2~\Re W^{\text{YM}}_{\text{exch},\mu_1...\mu_4}(k_1,k_2,k_3,k_4)=-g^{2} \int & \frac{dz_1  dz_2 }{z_1^{d+1}z_2^{d+1}}  
K^{\text{YM}}_{\mu_1\nu_1}(k_1,z_1)K^{\text{YM}}_{\mu_2\nu_2}(k_2,z_1)
\mathcal{V}^{\nu_1\nu_2\rho} (k_1,k_2,k_{12})
\nonumber \\
&
G^{+\text{YM}}_{\rho\sigma}(k_{12},z_1,z_2)
\mathcal{V}^{\nu_3\nu_4\sigma}(k_3,k_4,-k_{12})
\nonumber
\\
&
K^{*\text{YM}}_{\mu_3\nu_3} (k_3,z_2) K^{*\text{YM}}_{\mu_4\nu_4} (k_4,z_2) ~,
\end{align}
where the overall minus sign on the right-hand side comes because the vertices include a factor of $i$. Evaluating the $z$ integrals and contracting with the polarization vectors gives
\begin{align}
-2~\Re W^{\text{YM}}_{\text{exch}}(k_1,k_2,k_3,k_4)=&-2g^{2}\frac{\sqrt{-k_{12}^{2}}}{(k_{12}^{2}-(|k_1|+|k_2|)^{2})(k_{12}^{2}-(|k_3|+|k_4|)^{2})}
\nonumber \\ &\mathcal{V}^{12\rho}(k_1,k_2,k_{12})\mathcal{V}^{34\sigma}(k_3,k_4,-k_{12})\left(\eta_{\rho\sigma}-\frac{k_{12,\rho}k_{12,\sigma}}{k_{12}^{2}}\right)~.
\end{align}
This agrees with a direct calculation of the real piece by analytically continuing (\ref{eq:YMAdS4Full}) to timelike momenta, $k_{12}^{2}<0$, and computing the discontinuity across the cut. With the exception of the polarization dependence, the analysis is the same as the scalar case considered in the previous section.

\subsection{Five-Point Tree}
The analysis for higher-point tree diagrams is similar to the four-point case. As an example, we consider the five-point tree shown in (\ref{eq:fivepointdiagram}) and use conformally coupled scalars $\phi_c$ in $d=5$. The five-point tree-diagram is
\begin{align}
W'_{\text{5-pt}}(k_1,\ldots,k_5)=&(ig)^{3}\int \frac{dz_1dz_2dz_3}{z_1^{d+1}z_2^{d+1}z_3^{d+1}}K_{\Delta_c}(k_1,z_1)K_{\Delta_c}(k_2,z_1)G_{\Delta_c}(k_{12},z_1,z_2)
\nonumber \\ &K_{\Delta_c}(k_5,z_2)G_{\Delta_c}(k_3+k_4,z_2,z_3)K_{\Delta_c}(k_3,z_3)K_{\Delta_c}(k_4,z_3)\bigg|_{\Delta_c=3,d=5}~. \label{eq:fivept_pt1}
\end{align}
To check the cutting rules, we use the five-point kinematics given in (\ref{eq:5ptKinematics}), that is we choose $k_{1}+k_2 \in V_+$, $k_3+k_4+k_5\in V_{-}$, and take all the other invariants to be spacelike. In this configuration the only non-zero cut places $G_{\Delta_c}(k_{12},z_1,z_2)$ on shell, as shown in (\ref{eq:fivepointdiagram}). Using the cutting rules, we have
\begin{align}
-2i~\Im W'_{\text{5-pt}}(k_1,\ldots&,k_5)=ig^3\int  \frac{dz_1dz_2dz_3}{z_1^{d+1}z_2^{d+1}z_3^{d+1}}K_{\Delta_c}(k_1,z_1)K_{\Delta_c}(k_2,z_1)G^+_{\Delta_c}(k_{12},z_1,z_2)
\nonumber \\ &K^*_{\Delta_c}(k_5,z_2)G^*_{\Delta_c}(k_3+k_4,z_2,z_3)K^*_{\Delta_c}(k_3,z_3)K^*_{\Delta_c}(k_4,z_3)\bigg|_{\Delta_c=3,d=5}~.
\label{eq:5ptFull}
\end{align}
As a reminder, we have a factor of $(ig)$ for the vertex to the left of the cut, a factor of $(-ig)^2$ for the two vertices to the right of the cut, and finally an overall $(-1)$ because we have an odd number of external points to the right of cut. Performing the $z$ integrals yields:
\begin{align}
-2i~\Im W'_{\text{5-pt}}(k_1,\ldots,k_5)=-8 g^3&\frac{ |k_5| |k_3+k_4| \sqrt{-k_{12}^{2} }}{\left(2 |k_5|^2 (k_{12}^{2}-k_{34}^{2})+(k_{12}^{2}+k_{34}^{2})^2+|k_5|^4\right)}
\nonumber \\
& \frac{1}{\left((|k_1|+|k_2|)^2+k_{12}^{2}\right) \left((|k_3|+|k_4|)^2-k_{34}^{2}\right) }~.\label{eq:5ptCut}
\end{align}

Next, we compute the imaginary piece of the five-point function directly from the full correlator. Evaluating the $p$ and $z$ integrals for (\ref{eq:fivept_pt1}) gives:
\begin{align}
W'_{\text{5-pt}}(k_1,\ldots,k_5)=g^3
\frac{1}{E_T} &\frac{\sqrt{k_{12}^{2}}+\sqrt{k_{34}^{2}}+|k_1|+|k_2|+|k_3|+|k_4|+2 |k_5|}{ \sqrt{k_{34}^{2}}+\sqrt{k_{12}^{2}}+|k_5|}
\nonumber \\
 & \frac{1}{\left(\sqrt{k_{34}^{2}}+|k_3|+|k_4|\right) \left(\sqrt{k_{34}^{2}}+|k_1|+|k_2|+|k_5|\right)  }
\nonumber \\ &\frac{ 1}{\left(\sqrt{k_{12}^{2}}+|k_1|+|k_2|\right)\left(\sqrt{k_{12}^{2}}+|k_3|+|k_4|+|k_5|\right) }
~
, \label{eq:fivepointtreefull}
\end{align}
where $E_T=|k_1|+\ldots+|k_5|$. 
To compute $-2i\hspace{.05cm} \Im\<T[\f(k_1)\ldots\f(k_5)]\>$, we first analytically continue $k_{12}^{2}$ to be timelike and then take the discontinuity across the branch cut. The result agrees exactly with the answer from the cutting rules in (\ref{eq:5ptCut}).

As a consistency check, our result \eqref{eq:fivepointtreefull} agrees with \cite{Albayrak:2020isk}, and we also see that on the total energy pole the five-point Witten diagram reduces to the correct five-point flat space amplitude:
\begin{align}
\lim\limits_{E_T\rightarrow 0}E_T\hspace{.1cm}W'_{\text{5-pt}}(k_1,\ldots,k_5)=&\hspace{.1cm}g^{3}\frac{1}{\left((|k_1|+|k_2|)^{2}-k_{12}^{2}\right)}\frac{1}{\left((|k_3|+|k_4|)^{2}-k_{34}^{2})\right)}
\nonumber \\=&\hspace{.1cm}g^{3}\frac{1}{s_{12}s_{34}}~.
\end{align}
One can also recover the discontinuity of the flat space five-point diagram by taking the flat space limit of \eqref{eq:5ptFull}. As with the four-point exchange diagram, it is useful to make the replacement $G^{+}_{\Delta}\rightarrow G^{+,\epsilon}_{\Delta}$ to see how the flat-space $\delta$-function emerges in this limit. The analysis is identical in form to the case of the exchange diagram. For $d=5$ the $z$ integrals can be evaluated in closed form and the $p$ integral can be extended to $(-\infty,\infty)$, and then evaluated via a contour analysis. Finally, we take the limit $E_{T}\rightarrow 0$ before taking $\epsilon\rightarrow 0$ to find the $\delta$-function. The final answer is:
\begin{align}
\lim\limits_{\epsilon\rightarrow 0}\lim\limits_{E_{T}\rightarrow 0}-2iE_{T}\Im W'_{\text{5-pt}}(k_1,\ldots,k_5)&=\lim\limits_{\epsilon\rightarrow 0}\hspace{.1cm}\frac{2 i g^3 \epsilon }{(|k_3|-|k_{34}|+|k_4|) (|k_3|+|k_{34}|+|k_4|) }
\nonumber 
\\
&\hspace{1in}\frac{\theta(k_{12}^{0})}{\left(\left(-k_{12}^2+(|k_1|+|k_2|)^2\right)^2+\epsilon ^2\right)}
\nonumber \\
&=ig^{3}\frac{2\pi}{s_{34}^{2}}\delta(s_{12}^{2})\theta(k_{12}^{0}).
\end{align}
Here we made the identifications
\begin{align}
s_{ij}=(|k_{i}|+|k_{j}|)^{2}-k_{ij}^{2},
\end{align}
where $s_{ij}$ are the flat-space Mandelstam invariants. The final result agrees with the cut flat-space amplitude.

\subsection{One-Loop Bubble}
Finally, we consider a more non-trivial example corresponding to the following one-loop bubble diagram: 
\begin{equation}
\includegraphics[scale=.25]{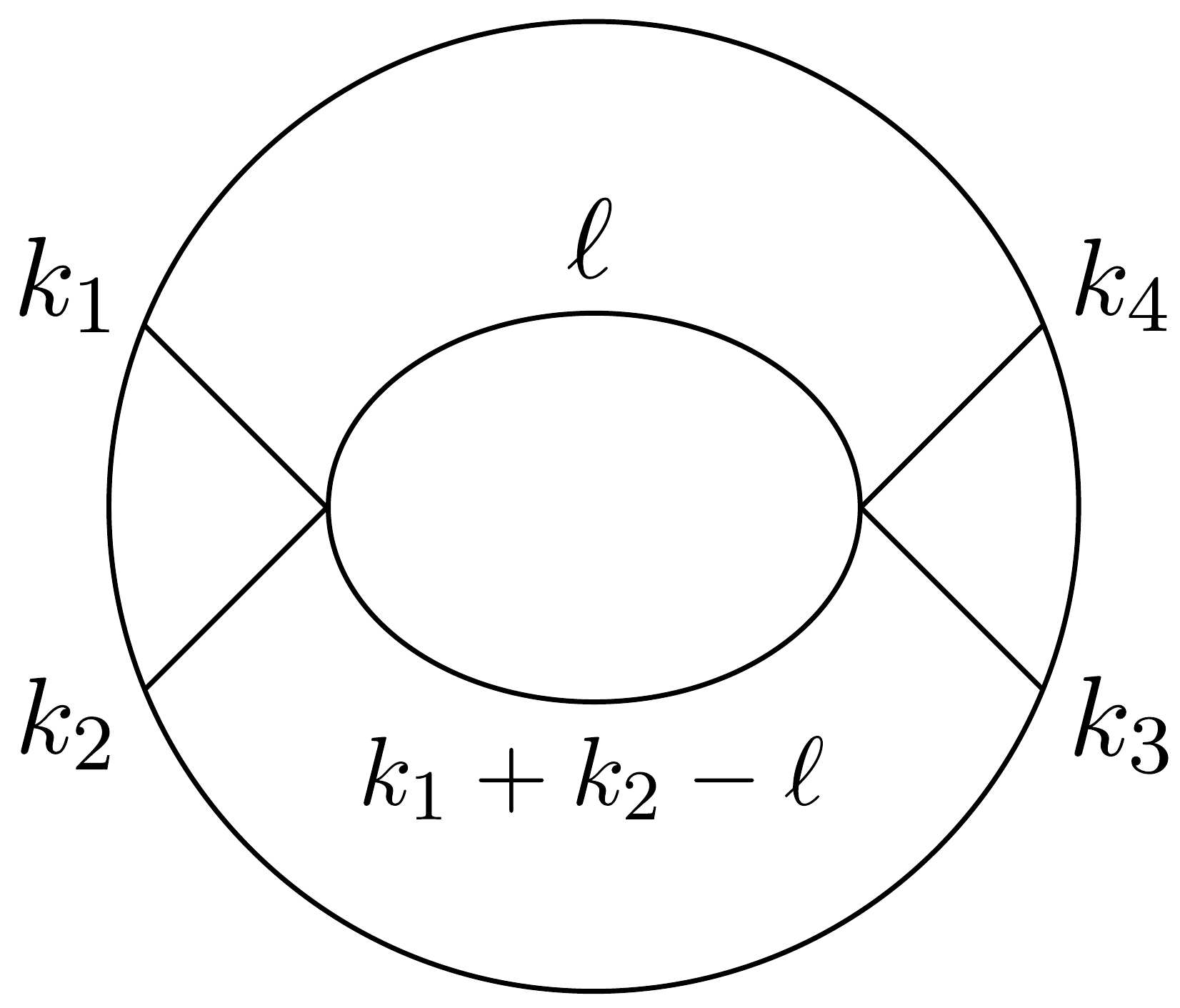}~.
\end{equation}
To verify that the cut diagram has the correct OPE limit, it is simpler to begin in position space. We will assume the external scalars are conformally coupled and that the two internal propagators are identical, but correspond to a distinct operator $\O$:
\begin{align}
W'_{\O,\text{bubble}}(x_1,\ldots,x_4)=(ig)^2&\int \limits_{AdS}\frac{d^{d}y_1d^{d}y_2dz_1dz_2}{z_1^{d+1}z_2^{d+1}}K_{\Delta_c}(x_1; y_1,z_1)K_{\Delta_c}(x_2; y_1,z_1)
\nonumber 
\\ 
& G_{\Delta_\O}(y_1,z_1;y_2,z_2)^{2}K_{\Delta_c}(x_3; y_2,z_2)K_{\Delta_c}(x_4; y_2,z_2)~.
\end{align}
The K\"all\'en-Lehmann spectral representation in AdS says \cite{Dusedau:1985ue,Fitzpatrick:2011dm}:
\begin{align}
G_{\Delta}(y_1,z_1;y_2,z_2)^{2}=\sum\limits_{n=0}^{\infty}a_{\Delta}(n)G_{2\Delta+2n}(y_1,z_1;y_2,z_2)~,
\\
a_{\Delta}(n)=\frac{(d/2)_n(2\Delta+2n)_{1-d/2}(2\Delta+n-d+1)_n}{2\pi^{d/2}n!(\Delta+n)^{2}_{1-d/2}(2\Delta+n-d/2)_n}~,
\end{align}
where $(a)_n$ is the Pochhammer symbol. In other words, the bubble diagram reduces to an infinite sum over tree-level exchange diagrams. Using this identity and then passing into momentum space, we obtain
\begin{align}
W'_{\O,\text{bubble}}(k_1,\ldots,k_4)=\sum\limits_{n}a_{\Delta}(n)W'_{[\O\O]_{n,0} \hspace{.05cm} \text{exch}}(k_1,\ldots,k_4)~.
\end{align}
We then take the real part of both sides, expand in the limit $k_{12}\rightarrow 0$, and find the expected scaling behavior for the exchange of double-trace operators:
\begin{align}
-2~\Re W'_{\O,\text{bubble}}(k_1,\ldots,k_4)\sim (-k_{12}^{2})^{2\Delta_\O-d/2}~, \label{eq:OPEBubble}
\end{align}
which we argued for in Section \ref{sec:Limits} using the integral representation of the cut diagram. This argument can be generalized to other ``bubble" type diagrams, as was done for example in \cite{Fitzpatrick:2011hu,Yuan:2018qva}.

Next, we will study the cut bubble diagram directly in momentum space for $d=3$ when all scalars are conformally coupled. We will show how the correct OPE limit emerges directly from the cutting rules and also that in the flat space limit we recover the cut flat-space bubble diagram. The cut bubble diagram is given by
\begin{align}
-2~\Re W'_{\f_c,\text{bubble}}(k_1,\ldots,k_4)=&g^{2}\int \frac{dz_1dz_2}{z_1^{d+1}z_2^{d+1}}\int \frac{d^{d}\ell}{(2\pi)^d}K_{\Delta_c}(k_1,z_1)K_{\Delta_c}(k_2,z_1)G^{+,\epsilon}_{\Delta_c}(\ell,z_1,z_2)
\nonumber \\ &G^{+,\epsilon}_{\Delta_c}(k_{12}-\ell,z_1,z_2)K^*_{\Delta_c}(k_3,z_2)K^*_{\Delta_c}(k_4,z_2)\bigg|_{\Delta_c=2,d=3}~,
\end{align}
where we used the regulated $\delta$-functions inside the cut propagators.
It will also be useful to define
\begin{align}
E_L=|k_1|+|k_2|, \qquad E_R=|k_3|+|k_4|~.
\end{align}
Performing the $z$ integrals, we find
\begin{align}
-2~\Re  W'_{\f_c,\text{bubble}}(k_1,\ldots,k_4)=64g^2&\int_0^{\infty} dp_1dp_2 \int \frac{d^3\ell}{(2\pi)^3}E_L E_R  p_1 p_2  \nonumber
\\
&
\frac{ \delta^{\epsilon}(\ell^2+p_1^2)\delta^{\epsilon}((k_{12}-\ell)^2+p_2^2) }{\left(E_L^2+(p_1-p_2)^2\right) \left(E_L^2+(p_1+p_2)^2\right)}
\nonumber 
\\ &\frac{\theta^+(-\ell^2)\theta^+(-(k_{12}-\ell)^2)}{ \left(E_R^2+(p_1-p_2)^2\right) \left(E_R^2+(p_1+p_2)^2\right)}~.\label{eq:bubbleIntegrand}
\end{align}
Here the function $\theta^+$ is defined to be a $\theta$-function for the forward lightcone $V_+$:
\begin{align}
\theta^+(-\ell^2) = \theta(-\ell^2)\theta(\ell^0)~.
\end{align}
To find the OPE limit for this diagram, we restrict to physical values for the norm, that is $|k_i|=\sqrt{k_i^2}$ with $k_i$ spacelike, and take the limit $k_{12}\rightarrow 0$. In this case we have $E_L$, $E_R>0$ and can take $\epsilon\rightarrow 0$ inside the integrand. The on-shell propagators yield $\delta$-functions that trivialize the $p$ integrals:
\begin{align}
\hspace{-.3in}-2~\Re \ W'_{\f_c,\text{bubble}}(k_1,\ldots,k_4)=&16g^2 \int \frac{d^3\ell}{(2\pi)^3} E_L E_R |\ell| |k_{12}-\ell|\theta^+(-\ell^2)\theta^+(-(k_{12}-\ell)^2)
\nonumber \\ &\frac{1}{\left(E_L^2+(|\ell|-|k_{12}-\ell|)^2\right) \left(E_L^2+(|\ell|+|k_{12}-\ell|)^2\right)}
\nonumber \\ & \frac{1}{ \left(E_R^2+(|\ell|-|k_{12}-\ell|)^2\right) \left(E_R^2+(|\ell|+|k_{12}-\ell|)^2\right)}~. \label{eq:bubbleOPEPt1}
\end{align}
To evaluate this integral in the OPE limit, we make the following change of variables,
\begin{align}
\ell=r\big(\hspace{-.05cm}\cosh(\phi),\sinh(\phi)\cos(\theta),\sinh(\phi)\sin(\theta)\big), \quad \text{ with } \quad 0\leq r, \phi \leq \infty, \quad 0\leq\theta\leq2\pi~.
\end{align}
This parameterization trivializes the $\theta^+(-\ell^2)$ function, but we still need to impose the constraint from the other $\theta^+$ function. To further simplify the analysis, we work in the center-of-mass frame,
\begin{align}
k_{12}=(k_{12}^0,0,0)~.
\end{align}
In this frame, requiring $k_{12}-\ell\in V_+$ implies
\begin{align}
0\leq r \leq e^{-\phi}k_{12}^0~.
\end{align}
Imposing these constraints, we find that the measure for the integrand becomes
\begin{align}
\int \frac{d^{3}\ell}{(2\pi)^{3}}\theta^+(-\ell^2)\theta^+(-(k_{12}-\ell)^2)= \frac{1}{(2\pi)^{3}}\int\limits_{0}^{2\pi} d\theta\int \limits_{0}^{\infty}d\phi \int\limits_{0}^{e^{-\phi}k_{12}^0}dr \hspace{.1cm} r^{2}\sinh(\phi)~,
\end{align}
where the factor of $r^{2}\sinh(\phi)$ comes from the Jacobian. To compute the OPE limit, we make the change of variables $r=k_{12}^{0}r'$ and then expand at small $k_{12}^0$ for fixed $r'$. Performing the $r'$, $\phi$, and $\theta$ integrals in this limit gives
\begin{align}
-2~\Re W'_{\f_c,\text{bubble}}(k_1,\ldots,k_4)\approx\frac{  g^2 (k_{12}^0)^5}{45 \pi^2(|k_1|+|k_2|)^3 (|k_3|+|k_4|)^3}~.
\end{align}
In the OPE limit, the exchange of an operator with dimension $\Delta$ leads to the overall scaling $(-k_{12}^2)^{\Delta-d/2}$. Here $d=3$ and $\Delta_c=2$ and we recognize that the overall $(k_{12}^0)^5$ dependence comes from the exchange of the scalar double-trace operator of dimension $4$, $[\f_c\f_c]_{0,0}$. This agrees exactly with the previous result for the bubble in the OPE limit (\ref{eq:OPEBubble}) using the K\"all\'en-Lehmann spectral representation. By expanding to higher orders in $k_{12}^0$ one can capture sub-leading terms in the OPE limit.

Finally, we will recover the cut bubble diagram in flat space from the corresponding AdS diagram. As with the four-point exchange diagram, using the regulated $\delta^{\epsilon}$-functions will make the total energy pole manifest. We will check that the flat space limit holds directly at the level of the integrand rather than working with the full integrated diagram. This approach makes manifest that in the flat space limit a $p$ integral becomes the $(d+1)^{\text{th}}$ component of the flat-space loop integral \cite{Raju:2012zr}.

To evaluate (\ref{eq:bubbleIntegrand}), we first extend the $p_{1,2}$ integrals to the entire real line. Then we can evaluate these integrals via a contour analysis. As shown in \cite{Raju:2012zr}, the total energy pole in $E_L+E_R$ comes from poles pinching the $p_i$ contours. Here we can see that closing the $p_2$ contour on the poles explicitly written in the denominator of (\ref{eq:bubbleIntegrand}) will yield the total energy pole:
\begin{align}
\hspace{-.3in}\lim\limits_{E_R\rightarrow -E_L}-2(E_L+E_R)\Re W'_{\f_c,\text{bubble}}(k_1,\ldots,k_4)=&\int\limits_{-\infty}^{\infty} dp_1 \int \frac{d^3\ell}{(2\pi)^{3}}2\pi g^2  \theta^+(\ell^0)\theta^+(k_1^0+k_2^0-\ell^0)
\nonumber \\ &\delta^{\epsilon} (\ell^{2}+p_1^2)\delta^{\epsilon} ((k_{12}-\ell)^2+(iE_L+p_1)^2)~. 
\end{align}
If we identify the $4$-dimensional external momenta as $\tilde{k}_i=(k_i,i |k_i|)$ and the internal $4$-dimensional momenta as $\tilde{\ell}=(\ell,p)$, we find
\begin{align}
\hspace{-.3in}\lim\limits_{E_R\rightarrow -E_L}-2(E_L+E_R)\Re W'_{\f_c,\text{bubble}}(k_1,\ldots,k_4)=\int \frac{d^4\tilde{\ell}}{(2\pi)^{4}}&  (2\pi)^2 g^2 \delta (\tilde{\ell}^{2})\delta((\tilde{k}_1+\tilde{k}_2-\tilde{\ell})^2)
\nonumber \\ &\theta^+(-\tilde{\ell}^2)\theta^+(-(\tilde{k}_1+\tilde{k}_2-\tilde{\ell})^2)~. 
\end{align}
This agrees with the cut flat-space bubble diagram exactly.

\section{Conclusion}
\label{sec:Discussion}

\subsection{Discussion}

In this work, we derived and applied the AdS Cutkosky rules. Together with the Lorentzian inversion formula, these cutting rules furnish a holographic unitarity method for AdS$_{d+1}$/CFT$_d$. In the process, we also provided the cutting rules for weakly-coupled CFTs. We used basic properties of Lorentzian QFTs to derive these rules, and so the results can be generalized to study QFT in other curved spaces. 

The proof of the CFT Cutkosky rules relies on the CFT optical theorem \eqref{eq:opticalV0} in combination with constraints from positivity of the spectrum and causality. Using positivity, we showed that for the restricted set of momenta \eqref{eq:kinematics},
\begin{equation}
-2\hspace{.1cm}\Re \< T[\f \f \f \f] \>  = \<[\f,\f]_{A}[\f,\f]_{R}\>.  \label{eq:summaryReT}
\end{equation}
This statement of CFT unitarity allows us to relate two seemingly different objects, the real part of a time-ordered correlator and a causal double-commutator. The right-hand side is the same double-commutator that appears in the Lorentzian inversion and CFT dispersion formulas \cite{Caron-Huot:2017vep,ssw,Kravchuk:2018htv,Carmi:2019cub}. The left-hand side is a natural generalization of $\Im(\mathcal{T})$, but now at the level of the off-shell correlation function. Like $\Im(\mathcal{T})$, the cutting rules for the real part can be derived by using the largest-time equation. Using \eqref{eq:summaryReT} and analyticity in momentum space, we can then derive the cutting rules for the double-commutator. The derivation of these rules relied on using Lorentzian momentum space, but they can also be studied using other representations of the correlator, e.g. by working in position or Mellin space.

Our method for CFT correlators is a direct generalization of the flat-space S-matrix method. In both cases, a cut replaces a time-ordered propagator with the corresponding Wightman, or on-shell, propagator and therefore factorizes the diagram into a product of on-shell sub-diagrams. Dispersion formulas can then be used to reconstruct the full diagram from its cuts. Moreover, we checked in explicit examples that the AdS unitarity cuts reduce to the usual S-matrix cuts in the flat space limit. 

The identity \eqref{eq:summaryReT} gives a notion of factorization in CFT$_d$: one can always insert a complete set of states in the right-hand side to find an infinite sum over three-point functions. The non-trivial feature of holographic CFTs is that the right-hand side can be rewritten as a phase-space integral over two AdS Witten diagrams. In other words, for holographic CFTs we have a stronger notion of factorization that comes from the locality of the bulk dual. The rules presented here make bulk locality manifest and are complementary to the previous work \cite{Meltzer:2019nbs}, where different bulk rules were derived to compute the conformal block expansion of the double-commutator. These two methods make different properties of AdS/CFT manifest -- bulk factorization and the perturbative structure of the boundary OPE -- and open new windows into $1/N$ perturbation theory via unitarity.

\subsection{Future work}

There are many open questions in the broader study of unitarity methods for CFT correlators. The appearance of the double-commutator in the real part of a time-ordered momentum-space correlator provides a hint that momentum space may be useful in the study of the CFT dispersion formula \cite{Carmi:2019cub}. We expect that the real/imaginary part of even/odd-point time-ordered correlators with spacelike external momenta will provide natural generalizations of the double-commutator. Such correlators factorize into partially time-ordered correlators and can also be computed via the cutting rules. Using momentum space may therefore clarify the structure of the higher-point inversion formula and the larger problem of bootstrapping general $n$-point functions.

It is also important to develop efficient ways of using the cutting rules in practice to determine a one-loop correlator. In this work we have given a set of rules to compute the double-commutator, but we did not introduce new tools to evaluate the dispersion formula. One possible avenue is to use the dispersion formula directly in momentum space. Another potentially useful approach is to use generalized unitarity to fix the one-loop correlator by allowing for more general cuts \cite{Bern:2011qt}. This has already been done for correlation functions in weakly-coupled $\mathcal{N}=4$ SYM \cite{Engelund:2012re}, but its application to more general weakly-coupled CFTs, such as the $O(N)$ vector models, appears to be less explored. While our work gives natural candidates for the relevant cuts, generalized unitarity has not yet been studied in AdS, and we expect its development will teach us more about a rich class of observables and theories. For example, at tree level in type IIB supergravity on AdS$_{5}\times$S$^{5}$ there exists a fascinating hidden $10d$ conformal symmetry \cite{Caron-Huot:2018kta}.\footnote{See \cite{Rastelli:2019gtj,Giusto:2020neo} for a generalization to $AdS_3\times S^3$.} This symmetry explains the simplicity of Mellin amplitudes and anomalous dimensions of the CFT dual \cite{Rastelli:2016nze,Rastelli:2017udc,Aprile:2018efk}. Recursion relations and generalized unitarity can help clarify to what extent this symmetry continues to hold at higher points and at loop level.  

Studying cutting rules for holographic CFTs in Mellin space may also provide new insight. The Mellin amplitude shares important similarities with a scattering amplitude, but it also encodes the OPE in a simple way \cite{Mack:2009gy,Penedones:2010ue,Fitzpatrick:2011ia}. This simplicity continues to hold beyond tree level in supersymmetric theories \cite{Alday:2018pdi,Alday:2018kkw,Binder:2019jwn,Chester:2019pvm,Chester:2020dja,Alday:2019nin,Alday:2020tgi,Drummond:2019hel,Bissi:2020wtv}. While we have derived the cutting rules in momentum space, it would be interesting to study their application to one-loop Mellin amplitudes. Relatedly, while most recent work on holographic correlators focuses on bootstrapping the full, integrated correlator, much of the recent progress in the study of scattering amplitudes comes from studying the integrand \cite{Elvang:2013cua,Henn:2014yza,Arkani-Hamed:2016byb}. To import this technology into AdS, it may prove useful to understand the structure of AdS integrands using Mellin space ideas. This may also help determine the class of functions that can appear in holographic correlators \cite{Aprile:2017bgs,Aprile:2017qoy,Aprile:2019rep,Drummond:2019hel}.

The cutting rules derived here contribute to the larger program of bootstrapping weakly-coupled theories in curved space via unitarity methods. Understanding unitarity constraints directly in the bulk of AdS opens up applications to other spacetimes, from deformed versions of AdS to the study of inflationary observables relevant for cosmology. We anticipate that by further generalizing S-matrix methods, we can open new avenues into this broader class of theories.

\section*{Acknowledgments}
We thank Soner Albayrak, Simon Caron-Huot, Clifford Cheung, Savan Kharel, Per Kraus, Julio Parra-Martinez, Eric Perlmutter, and David Simmons-Duffin for discussions. We also thank Julio Parra-Martinez for comments on the draft. AS thanks the Walter Burke Institute for Theoretical Physics for hospitality while this work was in progress. The research of DM is supported by Simons Foundation grant 488657, the Walter Burke Institute for Theoretical Physics and the Sherman Fairchild Foundation. AS is supported by the College of Arts and Sciences of the University of Kentucky.
\appendix
\section{Largest-Time Equation}
\label{app:largest_time}
In this appendix, we briefly review the derivation of the largest-time equation. For more details see \cite{Veltman:1963th,tHooft:1973wag,Veltman:1994wz}.
To prove the largest-time equation, we will study the integrand $\widehat{f}_{q}$ of each decorated Feynman diagram $\widehat{F}_{q}$ as a function of both internal and external points,
\begin{align}
\widehat{F}_{q}(x_1,\ldots,x_n)=\int d^{d}y_1\ldots d^dy_m \widehat{f}_{q}(x_1,\ldots,x_n;y_1,\ldots,y_m)~.
\end{align}
The largest-time equation then holds at the level of the integrand,
\begin{align}
\sum\limits_{q=1}^{2^{m+n}}\widehat{f}_{q}(x_1,\ldots,x_n;y_1,\ldots,y_m)=0~. \label{eq:unintLargesttime}
\end{align}
To prove this, we assume $x_1$ has the largest time, $x_{1}^{0}\geq x_{j}^0,y_i^{0}$. Then if we have a graph $\widehat{f}_{q}(x_1,\ldots,x_n;y_1,\ldots,y_m)$ where $x_1$ has a black vertex, it will cancel in the sum \eqref{eq:unintLargesttime} against a graph where $x_1$ has a white vertex and all the other points are the same. For example,
\begin{equation}
\includegraphics[scale=.35]{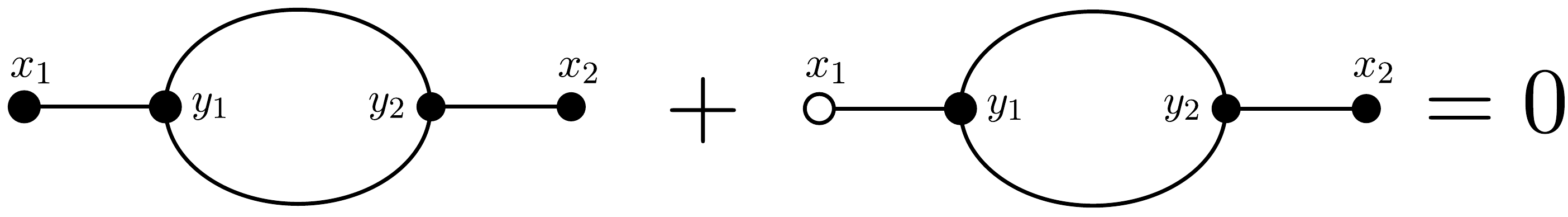}~.
\label{eq:graphscanceLargesttime}
\end{equation}
The cancellation happens because if $x_1$ has the largest time, changing the color of its vertex does not affect the propagators connected to $x_1$, but it does introduce an extra minus sign from the white vertex. To be more explicit, we can isolate the propagators connected to $x_1$:
\begin{equation}
\includegraphics[scale=.3]{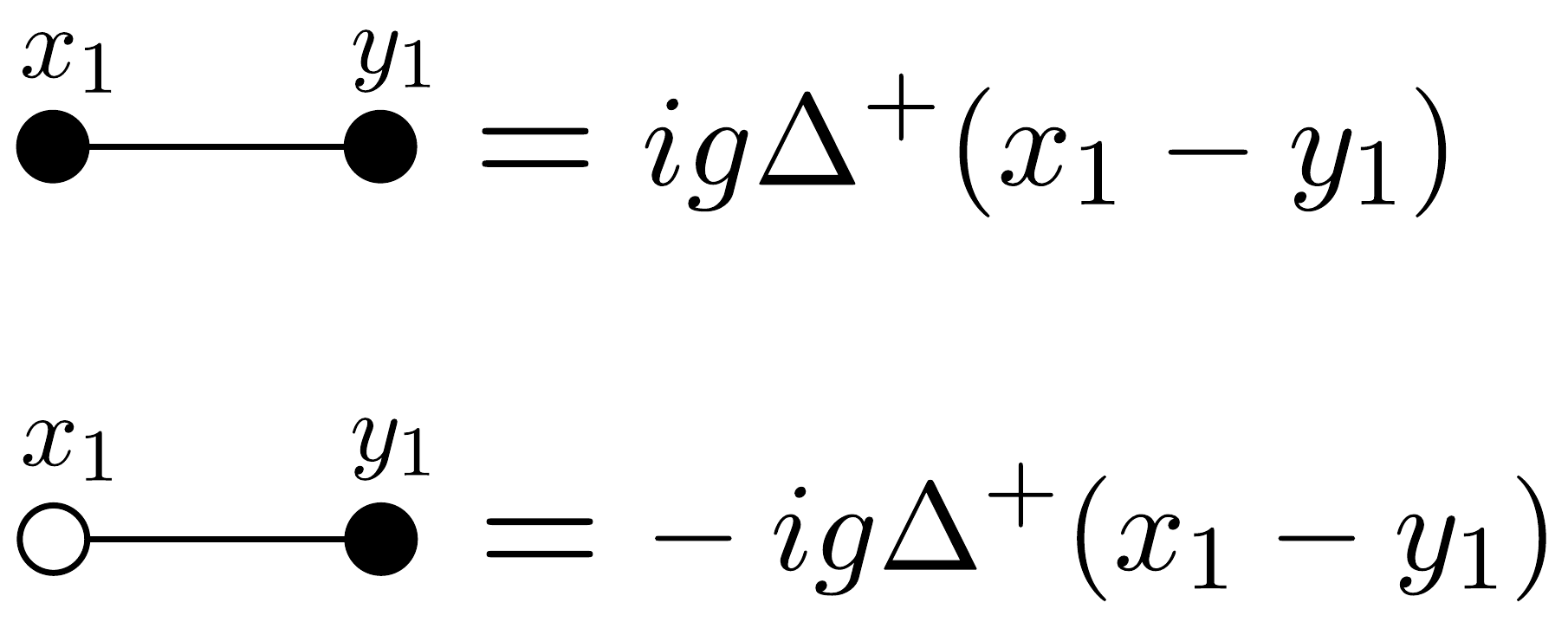}
\end{equation}
where in the first line we used $\Delta_{F}(x_{12})=\theta(x_1^{0}-x_2^0)\Delta^+(x_{12})+(1\leftrightarrow 2)$ and that $x_1$ has the largest time to drop the second term. The second line follows directly from the coloring rules. Since all other terms in the integrand are the same, it is clear the two graphs in \eqref{eq:graphscanceLargesttime} cancel. The same logic carries over if we have multiple lines connected to $x_1$ or if an internal point $y_i$ has the largest time. All graphs in \eqref{eq:graphscanceLargesttime} therefore cancel in pairs for all configurations.

To obtain the full correlator, we need to integrate over all internal points and keep the external points generic. We therefore cannot assume one point has the largest time. However, this is already taken care of by summing over all possible graphs in \eqref{eq:unintLargesttime}. Regardless of which coordinate has the largest time, the sum ensures they will cancel in pairs. We can then perform the $y_j$ integrals to find the integrated largest-time equation:
\begin{align}
\sum\limits_{q=1}^{2^{m+n}}\widehat{F}_{q}(x_1,\ldots,x_n)=0~. 
\end{align}
Once we have this equality, we can then Fourier transform to momentum space.

To see how this generalizes to AdS, it simplest to study the largest-time equation for a purely bulk correlation function $\<\Phi(x_1,z_1)\ldots\Phi(x_n,z_n)\>$. Then the structure of individual diagrams is the same as in flat space, since here we only use the bulk-to-bulk propagator. The largest-time equation then generalizes straightforwardly and one can then use the extrapolate dictionary to prove the analogous equation for the boundary correlator $\<\phi(x_1)\ldots\phi(x_n)\>$. Alternatively, one can directly prove the largest-time equation for the boundary correlator by keeping track of the two kinds of propagators, the bulk-to-bulk and bulk-to-boundary propagators, from the beginning.

\section{Analyticity in Momentum Space}
\label{app:Analyticity_k_Space}
In this appendix, we will review the analyticity properties of the double-commutator
\\
 $\<[\f(k_3),\f(k_4)]_{A}[\f(k_1),\f(k_2)]_{R}\>$ in momentum space \cite{Polyakov:1974gs}. To relate the real part of the time-ordered correlator, $\Re\<T[\f(k_1)...\f(k_4)]\>$, with this causal double-commutator, we had to assume all four momenta were spacelike, $k_{i}^{2}>0$, and only $k_1+k_2\in V_{+}$. The double-commutator is non-zero for more generic momenta, so we would like to relax some of these assumptions. Specifically, we will show that once we know the double-commutator for this set of momenta, we can analytically continue to find it for general kinematics. We first note that the double-commutator is only non-zero for $k_1+k_2\in V_{+}$ by the positive spectrum condition on the CFT Hilbert space. We therefore only need to analytically continue in $k_{1}$ and $k_4$.

To simplify the discussion and avoid the overall momentum conserving delta-function, we will write the momentum-space correlator as a Fourier transform in three out of the four positions:
\begin{align}
H(p_1,p_2,Q)=\int d^{d}r_1d^{d}r_2d^{d}R e^{i(p_1\cdot r_1+p_2\cdot r_2+R\cdot Q)}\<[\f(R),\f(R+r_2)]_{A}[\f(r_1),\f(0)]_{R}\>~.  \label{eq:partialFT}
\end{align}
Here we have adopted the notation of \cite{Polyakov:1974gs} to simplify the comparison.
The relation to the parameterization used in the body of the paper is:
\begin{eqnarray}
p_1&=&k_1~,
\\
p_2&=&k_4~,
\\
Q&=&k_1+k_2~.
\end{eqnarray}
Next, we will show $H$ is analytic in $p_{1,2}$ in the appropriate region of the complex plane. As noted originally in \cite{Polyakov:1974gs}, causal commutators in position space imply analyticity properties in $p_{1,2}$. The argument mirrors the standard proof \cite{Streater:1989vi,Haag:1992hx} that the Wightman functions $\<\f(x_1)\ldots\f(x_n)\>$ are analytic in position space. To prove analyticity of the position-space Wightman functions, one uses that the physical spectrum is in the forward lightcone, i.e. the momentum $k\in V_+$ for physical states. To prove analyticity of the causal double-commutator in momentum space, we use that the integrand of (\ref{eq:partialFT}) is only non-zero for $r_1,r_2\in V_+$. The only difference between the two cases is that here we are reversing the roles of position and momentum space.\footnote{For a similar example see section II.2.6 of \cite{Haag:1992hx} where they discuss the $r$-functions, which are given by iterated retarded commutators and are also analytic in momentum space.}

To see how this works, we make the replacement $p_i\rightarrow p_i-i\eta_{i}$ in (\ref{eq:partialFT}). Then we find:
\begin{align*}
H(p_1-i\eta_1,p_2-i\eta_2,Q)=\int d^{d}r_1d^{d}r_2d^{d}R & e^{i(p_1\cdot r_1+p_2\cdot r_2-R\cdot Q)+\eta_1\cdot r_1+\eta_2\cdot r_2}
\\
&
\<[\f(R),\f(R+r_2)]_{A}[\f(r_1),\f(0)]_{R}\>~.
\numberthis
\end{align*}
The right-hand side is only non-zero for $r_i\in V_+$. Therefore if we choose $\eta_i\in V_+$, we find the integral is exponentially damped at large $r_i$. The right-hand side is now a Laplace transform, so the left-hand side is an analytic function of $p_i-i\eta_i$ if $\eta_i\in V_+$. With this analyticity property, we can now continue the double-commutator to configurations with $p_i^{2}<0$. As we are studying the Laplace transform of a tempered distribution, the causal double-commutator for real momenta is the boundary value of this analytic function where we take $\eta_i\rightarrow 0^+$, that is we take $\eta_i$ to zero inside the forward lightcone \cite{Streater:1989vi,Haag:1992hx}. This completes the proof because once we have found $H(p_1,p_2,Q)$ for $Q\in V_{+}$ and $p_i^{2}>0$ we can find the function for general $p_i$ by analytic continuation. At this point we can also Fourier transform to recover the double-commutator in position space that enters into the inversion formula.
\section{Feynman Tree Theorem}
\label{sec:FeynmanTree}
In this appendix we will discuss the Feynman tree theorem \cite{Feynman:1963ax,CaronHuot:2010zt}, a different but related notion of cutting. This theorem follows from the fact that a closed loop of retarded propagators in a Feynman diagram must vanish by causality. Using this property, we can express the full one-loop diagram as an integral of tree-level diagrams. As opposed to the usual Cutkosky rules, the cuts do not have to split the diagram into two pieces.

In this section we can work in position space. To keep notation compact, we will use capital Latin letters, e.g. $Y=(y^{\mu},z)$, for points in the bulk of AdS$_{d+1}$ and lower-case Latin letters, $y^\mu$, for boundary points. Next, recall the retarded bulk-to-bulk propagator for a free scalar $\Phi$ is defined by:
\begin{align}
G_{\Delta}^{R}(X_1,X_2)=\<[\Phi(X_1),\Phi(X_2)]\>_{\text{free}}\theta(x^0_1-x^0_2)~,
\end{align}
where $x^0$ is the Poincar\'e time. Then we have the identity
\begin{align}
G_{\Delta}^{R}(X_1,X_2)=G_{\Delta}(X_1,X_2)-G_{\Delta}^{+}(X_2,X_1)~. \label{eq:retardedProp}
\end{align}
This can be checked by using the definition of the time-ordering symbol and retarded commutator. 

\begin{figure}
\begin{center}
\includegraphics[scale=.25]{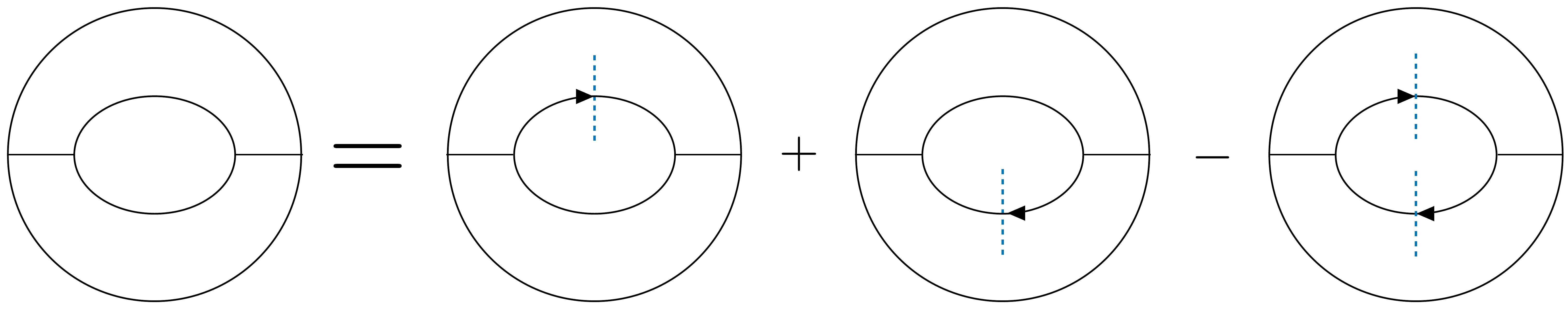}
\caption{Feynman tree theorem in AdS for a one-loop correction to the propagator in $\Phi^3$ theory. The arrow indicates the flow of positive energy across the cut.}
\label{fig:FeynmanTree}
\end{center}
\end{figure}

Next we use that any closed loop of retarded propagators has to vanish. This is manifest in position space. For example we can consider a one-loop correction to the two-point function:
\begin{align}
\int\limits_{\textrm{AdS}}dY_{1,2}K_{\Delta}(x_1;Y_1)G_{\Delta}^{R}(Y_1,Y_2)G_{\Delta}^R(Y_2,Y_1)K_{\Delta}(x_2;Y_2)=0~.
\end{align}
The retarded propagator $G_{\Delta}^{R}(Y_1,Y_2)$ is only non-zero if $Y_1$ is in the causal future of $Y_2$ and similarly $G_{\Delta}^R(Y_2,Y_1)$ is only non-zero if $Y_2$ is in the causal future of $Y_1$. The two propagators cannot be non-zero at the same time and so the integral vanishes. In momentum space, this loop vanishes because all the poles in $p^0$ are on one side of the real axis \cite{CaronHuot:2010zt}. Now we can use \eqref{eq:retardedProp} to expand this equality in terms of the positive energy on-shell propagators and the usual time-ordered propagator. Specifically we find:
\begin{align}
\int\limits_{\textrm{AdS}}dY_{1,2}K_{\Delta}(x_1;Y_1)&G_{\Delta}(Y_1,Y_2)^2K_{\Delta}(x_2;Y_2) \nonumber
\\ =\int\limits_{\textrm{AdS}}dY_{1,2} \ &K_{\Delta}(x_1;Y_1)K_{\Delta}(x_2;Y_2) \bigg(G_{\Delta}(Y_1,Y_2)(G^{+}_{\Delta}(Y_1,Y_2)+G^{+}_{\Delta}(Y_2,Y_1))
 \nonumber \\ &\hspace{1.8in}-G^{+}_{\Delta}(Y_2,Y_1)G^+_{\Delta}(Y_1,Y_2)\bigg)~.
\end{align}
The first line is the original Witten diagram while in the second and third lines there is at least one propagator put on shell, see figure \ref{fig:FeynmanTree}. In contrast to the cutting rules, we are computing the full Witten diagram, as opposed to its real or imaginary piece. Finally, we also see the momentum does not have a definite flow from the left to the right of the diagram. All these properties are exactly the same as in the original flat space Feynman tree theorem.

\section{Cutting Rules via Schwinger-Keldysh}
\label{sec:SKDerivation}
In this Appendix we give an alternative derivation of the cutting rules using Schwinger-Keldysh contours \cite{Schwinger:1960qe,Keldysh:1964ud} (see \cite{Haehl:2017qfl,Haehl:2017eob} for reviews and generalizations). One benefit of using this method is that we can avoid subtleties in the largest-time equation when we have derivative interactions \cite{Tomboulis:2017rvd}.

We start by setting $k_{i}^{2}>0$ and then from \eqref{eq:FromReToDCV2} we have
\begin{align}
\<\overline{T}[\f(k_3)\f(k_4)]T[\f(k_1)\f(k_2)]\>=\<[\f(k_3),\f(k_4)]_{A}[\f(k_1),\f(k_2)]_{R}\>~.
\end{align}
The positive spectrum condition implies these correlators are only non-zero for $k_1+k_2\in V_+$ and $k_3+k_4\in V_-$.
\begin{figure}[h]
\begin{center}
\includegraphics[scale=.3]{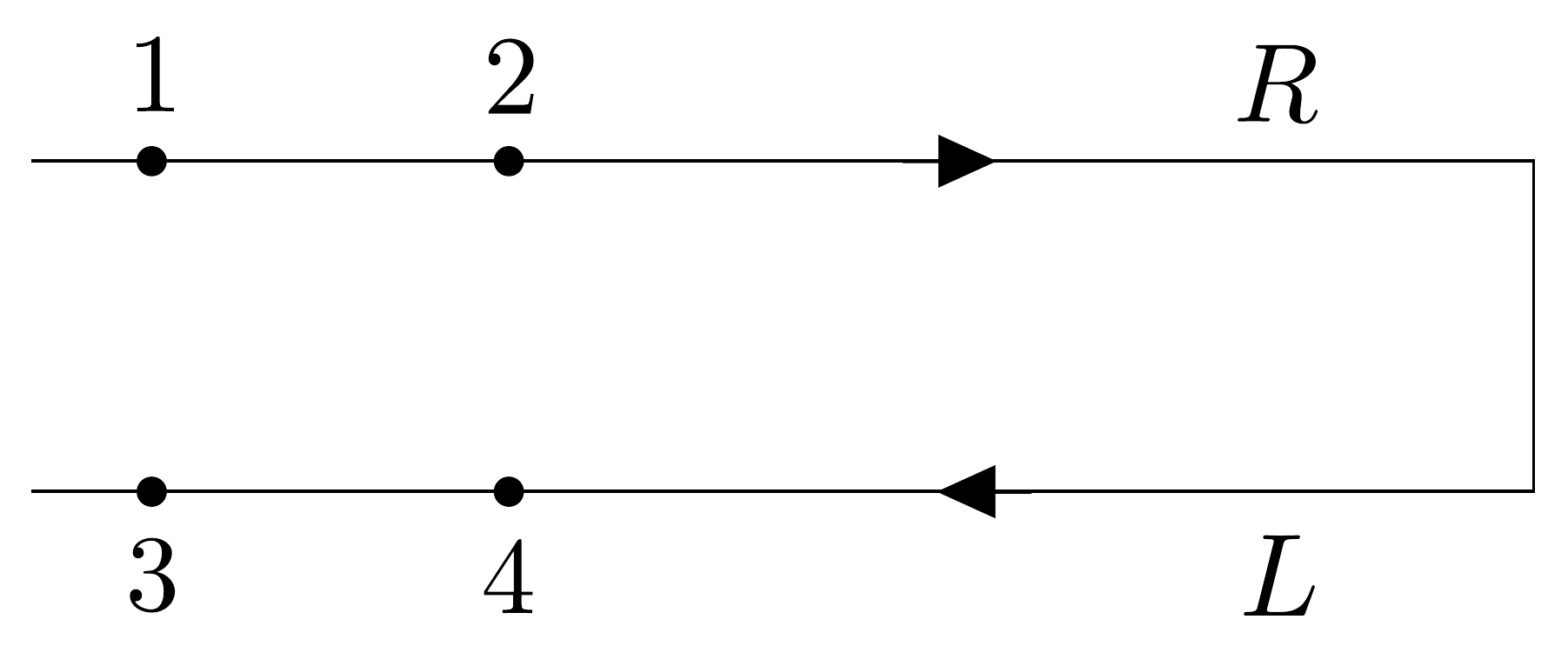}
\end{center}
\caption{Timefolded contour used for $\<\overline{T}[\f_3\f_4]T[\f_1\f_2]\>$. The arrows indicate the flow of time and the labels $R$ and $L$ are used to distinguish the two contours by the direction in which time flows.}
\label{fig:SK_Contour_1}
\end{figure}
To compute the left-hand side, we introduce a complex time contour with a single fold, as shown in figure \ref{fig:SK_Contour_1}. We now have four different propagators, depending on which segments of the contour each point lies on:
\begin{align}
\Delta_{RR}(x_1,x_2)&=\Delta_{F}(x_{12})~,
\\
\Delta_{LR}(x_1,x_2)&=\Delta^{+}(x_{12})~,
\\
\Delta_{RL}(x_1,x_2)&=\Delta^{-}(x_{12})~,
\\
\Delta_{LL}(x_1,x_2)&=\Delta_F^*(x_{12})~.
\end{align}
In free field theory these are the two-point functions, e.g. $\Delta_{LR}(x_1,x_2)=\<\f_L(x_1)\f_R(x_2)\>_{\text{free}}$ where the subscript indicates on which contour the operator sits. We can unify all four propagators by defining a contour-ordering symbol $T_{\mathcal{C}}$ such that:
\begin{align}
T_{\mathcal{C}}[ \f(x_1)\f(x_2)]&=T[\f(x_1)\f(x_2) ]\quad  \text{if} \quad x_{1,2}\in R~,
\\
T_{\mathcal{C}} [  \f(x_1)\f(x_2)]&=\f(x_1)\f(x_2) \quad \hspace{.51cm} \text{if} \quad x_{1}\in L, \ x_2\in R~,
\\
T_{\mathcal{C}} [ \f(x_1)\f(x_2)]&=\f(x_2)\f(x_1) \quad \hspace{.51cm}  \text{if} \quad x_{1}\in R, \ x_2\in L~,
\\
T_{\mathcal{C}} [ \f(x_1)\f(x_2)]&=\overline{T}[\f(x_1)\f(x_2) ] \quad\text{if} \quad x_{1,2}\in L~.
\end{align}
The four propagators are encoded in the free-field two-point function defined using the contour-ordering symbol, $\<T_{\mathcal{C}}[\f(x_1)\f(x_2)]\>$. For the interaction vertices, we use $ig$ for a vertex inserted on the $R$ contour and $-ig$ for a vertex inserted on the $L$ contour. To compute $\<\overline{T}[\f(k_3)\f(k_4)]T[\f(k_1)\f(k_2)]\>$, we use the time-folded contour shown in figure \ref{fig:SK_Contour_1}, place the operators $\f(k_1)$ and $\f(k_2)$ on the $R$-contour, place $\f(k_3)$ and $\f(k_4)$ on the $L$-contour, and use the Feynman rules given above. 

It is now apparent that these Feynman rules are exactly the same as the coloring rules introduced by Veltman to compute the real part of a four-point function. To make this map clear, we associate black dots with vertices on the $R$-contour and use white dots for vertices on the $L$-contour. To make the connection with the double-commutator, we need to impose that the external momenta are spacelike, $k_i^{2}>0$. This implies that any propagator connected to an external point has to end on the same contour as that point. If we connect it to a different contour we need to use the Wightman functions, $\Delta^{\pm}(k)$, which vanish for spacelike momenta. This is equivalent to the statement that when external momenta are spacelike, we cannot cut external lines.

To illustrate the equivalence, we can consider some specific examples. For example, for the cut tree diagram the mapping is
\begin{center}
\begin{eqnarray}
\includegraphics[scale=.4]{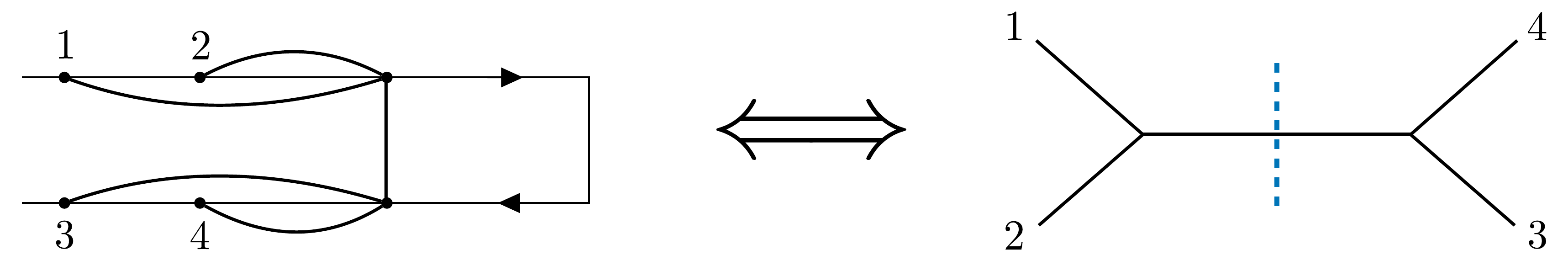}\nonumber,
\\
\end{eqnarray}
\end{center} 
and for the triangle, we have
\begin{center}
\begin{eqnarray}
\includegraphics[scale=.4]{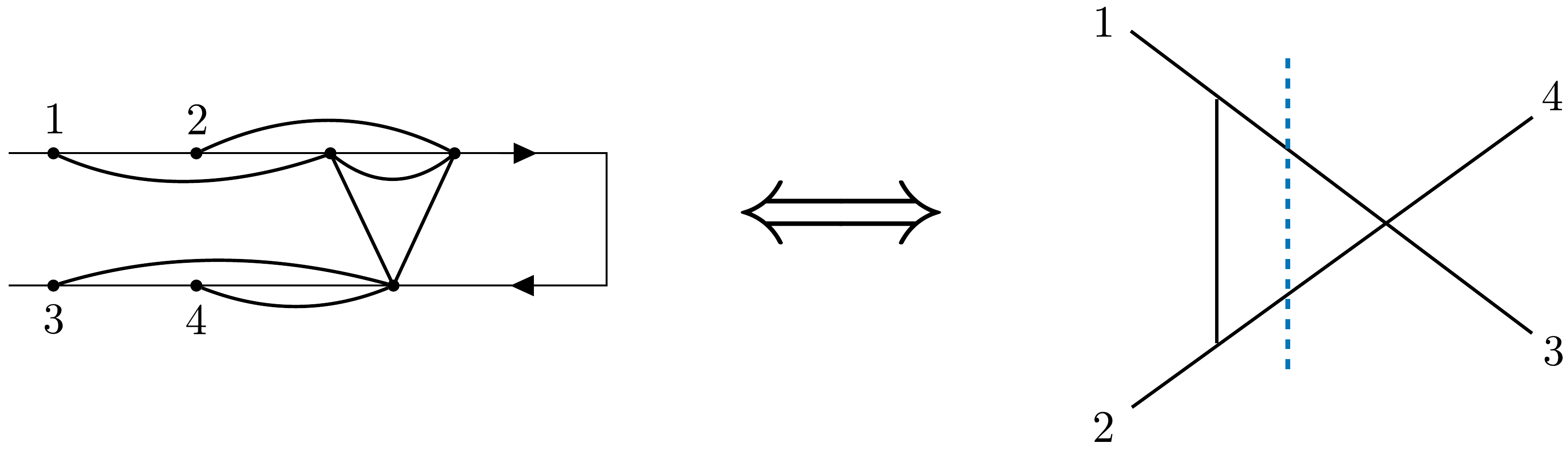}~\nonumber . \\
\end{eqnarray}
\end{center} 
In these pictures, we integrate the interaction vertices only on either the $L$ or $R$-contours, but not both. While we have only given the map for flat-space Feynman diagrams, the generalization to AdS is straightforward by using the corresponding Schwinger-Keldysh rules \cite{Herzog:2002pc}.

\bibliographystyle{JHEP}
\bibliography{biblio}

\providecommand{\href}[2]{#2}\begingroup\raggedright\begin{thebibliography}{100}

\bibitem{Bern:1994zx}
Z.~Bern, L.~J. Dixon, D.~C. Dunbar, and D.~A. Kosower, {\it {One loop n point
  gauge theory amplitudes, unitarity and collinear limits}},  {\em Nucl. Phys.
  B} {\bf 425} (1994) 217--260,
  [\href{http://arxiv.org/abs/hep-ph/9403226}{{\tt hep-ph/9403226}}].

\bibitem{Bern:1994cg}
Z.~Bern, L.~J. Dixon, D.~C. Dunbar, and D.~A. Kosower, {\it {Fusing gauge
  theory tree amplitudes into loop amplitudes}},  {\em Nucl. Phys. B} {\bf 435}
  (1995) 59--101, [\href{http://arxiv.org/abs/hep-ph/9409265}{{\tt
  hep-ph/9409265}}].

\bibitem{Bern:2011qt}
Z.~Bern and Y.-t. Huang, {\it {Basics of Generalized Unitarity}},  {\em J.
  Phys. A} {\bf 44} (2011) 454003, [\href{http://arxiv.org/abs/1103.1869}{{\tt
  arXiv:1103.1869}}].

\bibitem{Elvang:2013cua}
H.~Elvang and Y.-t. Huang, {\it {Scattering Amplitudes}},
  \href{http://arxiv.org/abs/1308.1697}{{\tt arXiv:1308.1697}}.

\bibitem{Dixon:2013uaa}
L.~J. Dixon, {\it {A brief introduction to modern amplitude methods}},  in {\em
  {Theoretical Advanced Study Institute in Elementary Particle Physics}:
  {Particle Physics: The Higgs Boson and Beyond}}, pp.~31--67, 2014.
\newblock \href{http://arxiv.org/abs/1310.5353}{{\tt arXiv:1310.5353}}.

\bibitem{Henn:2014yza}
J.~M. Henn and J.~C. Plefka, {\em {Scattering Amplitudes in Gauge Theories}},
  vol.~883.
\newblock Springer, Berlin, 2014.

\bibitem{Arkani-Hamed:2016byb}
N.~Arkani-Hamed, J.~L. Bourjaily, F.~Cachazo, A.~B. Goncharov, A.~Postnikov,
  and J.~Trnka, {\em {Grassmannian Geometry of Scattering Amplitudes}}.
\newblock Cambridge University Press, 4, 2016.

\bibitem{Cheung:2017pzi}
C.~Cheung, {\em {TASI Lectures on Scattering Amplitudes}}, pp.~571--623.
\newblock 2018.
\newblock \href{http://arxiv.org/abs/1708.03872}{{\tt arXiv:1708.03872}}.

\bibitem{ArkaniHamed:2010kv}
N.~Arkani-Hamed, J.~L. Bourjaily, F.~Cachazo, S.~Caron-Huot, and J.~Trnka, {\it
  {The All-Loop Integrand For Scattering Amplitudes in Planar N=4 SYM}},  {\em
  JHEP} {\bf 01} (2011) 041, [\href{http://arxiv.org/abs/1008.2958}{{\tt
  arXiv:1008.2958}}].

\bibitem{Bern:2008qj}
Z.~Bern, J.~Carrasco, and H.~Johansson, {\it {New Relations for Gauge-Theory
  Amplitudes}},  {\em Phys. Rev. D} {\bf 78} (2008) 085011,
  [\href{http://arxiv.org/abs/0805.3993}{{\tt arXiv:0805.3993}}].

\bibitem{Bern:2010ue}
Z.~Bern, J.~J.~M. Carrasco, and H.~Johansson, {\it {Perturbative Quantum
  Gravity as a Double Copy of Gauge Theory}},  {\em Phys. Rev. Lett.} {\bf 105}
  (2010) 061602, [\href{http://arxiv.org/abs/1004.0476}{{\tt
  arXiv:1004.0476}}].

\bibitem{Bern:2018jmv}
Z.~Bern, J.~J. Carrasco, W.-M. Chen, A.~Edison, H.~Johansson,
  J.~Parra-Martinez, R.~Roiban, and M.~Zeng, {\it {Ultraviolet Properties of
  $\mathcal N = 8$ Supergravity at Five Loops}},  {\em Phys. Rev. D} {\bf 98}
  (2018), no.~8 086021, [\href{http://arxiv.org/abs/1804.09311}{{\tt
  arXiv:1804.09311}}].

\bibitem{Susskind:1998vk}
L.~Susskind, {\it {Holography in the flat space limit}},  {\em AIP Conf. Proc.}
  {\bf 493} (1999), no.~1 98--112,
  [\href{http://arxiv.org/abs/hep-th/9901079}{{\tt hep-th/9901079}}].

\bibitem{Polchinski:1999ry}
J.~Polchinski, {\it {S matrices from AdS space-time}},
  \href{http://arxiv.org/abs/hep-th/9901076}{{\tt hep-th/9901076}}.

\bibitem{Heemskerk:2009pn}
I.~Heemskerk, J.~Penedones, J.~Polchinski, and J.~Sully, {\it {Holography from
  Conformal Field Theory}},  {\em JHEP} {\bf 0910} (2009) 079,
  [\href{http://arxiv.org/abs/0907.0151}{{\tt arXiv:0907.0151}}].

\bibitem{Gary:2009ae}
M.~Gary, S.~B. Giddings, and J.~Penedones, {\it {Local bulk S-matrix elements
  and CFT singularities}},  {\em Phys. Rev. D} {\bf 80} (2009) 085005,
  [\href{http://arxiv.org/abs/0903.4437}{{\tt arXiv:0903.4437}}].

\bibitem{Penedones:2010ue}
J.~Penedones, {\it {Writing CFT correlation functions as AdS scattering
  amplitudes}},  {\em JHEP} {\bf 1103} (2011) 025,
  [\href{http://arxiv.org/abs/1011.1485}{{\tt arXiv:1011.1485}}].

\bibitem{Raju:2012zr}
S.~Raju, {\it {New Recursion Relations and a Flat Space Limit for AdS/CFT
  Correlators}},  {\em Phys. Rev.} {\bf D85} (2012) 126009,
  [\href{http://arxiv.org/abs/1201.6449}{{\tt arXiv:1201.6449}}].

\bibitem{Balasubramanian:1999ri}
V.~Balasubramanian, S.~B. Giddings, and A.~E. Lawrence, {\it {What do CFTs tell
  us about Anti-de Sitter space-times?}},  {\em JHEP} {\bf 03} (1999) 001,
  [\href{http://arxiv.org/abs/hep-th/9902052}{{\tt hep-th/9902052}}].

\bibitem{Mack:2009mi}
G.~Mack, {\it {D-independent representation of Conformal Field Theories in D
  dimensions via transformation to auxiliary Dual Resonance Models. Scalar
  amplitudes}},  \href{http://arxiv.org/abs/0907.2407}{{\tt arXiv:0907.2407}}.

\bibitem{Rastelli:2016nze}
L.~Rastelli and X.~Zhou, {\it {Mellin amplitudes for $AdS_5\times S^5$}},  {\em
  Phys. Rev. Lett.} {\bf 118} (2017), no.~9 091602,
  [\href{http://arxiv.org/abs/1608.06624}{{\tt arXiv:1608.06624}}].

\bibitem{Cutkosky:1960sp}
R.~E. Cutkosky, {\it {Singularities and discontinuities of Feynman
  amplitudes}},  {\em J. Math. Phys.} {\bf 1} (1960) 429--433.

\bibitem{Eden:1966dnq}
R.~J. Eden, P.~V. Landshoff, D.~I. Olive, and J.~C. Polkinghorne, {\em {The
  analytic S-matrix}}.
\newblock Cambridge Univ. Press, Cambridge, 1966.

\bibitem{Abreu_2014}
S.~Abreu, R.~Britto, C.~Duhr, and E.~Gardi, {\it From multiple unitarity cuts
  to the coproduct of feynman integrals},  {\em Journal of High Energy Physics}
  {\bf 2014} (Oct, 2014).

\bibitem{Abreu_2017}
S.~Abreu, R.~Britto, C.~Duhr, and E.~Gardi, {\it Cuts from residues: the
  one-loop case},  {\em Journal of High Energy Physics} {\bf 2017} (Jun, 2017).

\bibitem{Bourjaily:2020wvq}
J.~L. Bourjaily, H.~Hannesdottir, A.~J. McLeod, M.~D. Schwartz, and C.~Vergu,
  {\it {Sequential Discontinuities of Feynman Integrals and the Monodromy
  Group}},  \href{http://arxiv.org/abs/2007.13747}{{\tt arXiv:2007.13747}}.

\bibitem{Veltman:1963th}
M.~J.~G. Veltman, {\it {Unitarity and causality in a renormalizable field
  theory with unstable particles}},  {\em Physica} {\bf 29} (1963) 186--207.

\bibitem{Caron-Huot:2017vep}
S.~Caron-Huot, {\it {Analyticity in Spin in Conformal Theories}},
  \href{http://arxiv.org/abs/1703.00278}{{\tt arXiv:1703.00278}}.

\bibitem{Gribov:1961fr}
V.~Gribov, {\it {Partial waves with complex orbital angular momenta and the
  asymptotic behavior of the scattering amplitude}},  {\em Sov. Phys. JETP}
  {\bf 14} (1962) 1395.

\bibitem{Froissart:1961ux}
M.~Froissart, {\it {Asymptotic behavior and subtractions in the Mandelstam
  representation}},  {\em Phys. Rev.} {\bf 123} (1961) 1053--1057.

\bibitem{Carmi:2019cub}
D.~Carmi and S.~Caron-Huot, {\it {A Conformal Dispersion Relation: Correlations
  from Absorption}},  \href{http://arxiv.org/abs/1910.12123}{{\tt
  arXiv:1910.12123}}.

\bibitem{Meltzer:2019nbs}
D.~Meltzer, E.~Perlmutter, and A.~Sivaramakrishnan, {\it {Unitarity Methods in
  AdS/CFT}},  \href{http://arxiv.org/abs/1912.09521}{{\tt arXiv:1912.09521}}.

\bibitem{Schwinger:1960qe}
J.~S. Schwinger, {\it {Brownian motion of a quantum oscillator}},  {\em J.
  Math. Phys.} {\bf 2} (1961) 407--432.

\bibitem{Keldysh:1964ud}
L.~Keldysh, {\it {Diagram technique for nonequilibrium processes}},  {\em Zh.
  Eksp. Teor. Fiz.} {\bf 47} (1964) 1515--1527.

\bibitem{Chou:1984es}
K.-c. Chou, Z.-b. Su, B.-l. Hao, and L.~Yu, {\it {Equilibrium and
  Nonequilibrium Formalisms Made Unified}},  {\em Phys. Rept.} {\bf 118} (1985)
  1--131.

\bibitem{Stanford:2015owe}
D.~Stanford, {\it {Many-body chaos at weak coupling}},  {\em JHEP} {\bf 10}
  (2016) 009, [\href{http://arxiv.org/abs/1512.07687}{{\tt arXiv:1512.07687}}].

\bibitem{Haehl:2016pec}
F.~M. Haehl, R.~Loganayagam, and M.~Rangamani, {\it {Schwinger-Keldysh
  formalism. Part I: BRST symmetries and superspace}},  {\em JHEP} {\bf 06}
  (2017) 069, [\href{http://arxiv.org/abs/1610.01940}{{\tt arXiv:1610.01940}}].

\bibitem{Haehl:2017qfl}
F.~M. Haehl, R.~Loganayagam, P.~Narayan, and M.~Rangamani, {\it {Classification
  of out-of-time-order correlators}},  {\em SciPost Phys.} {\bf 6} (2019),
  no.~1 001, [\href{http://arxiv.org/abs/1701.02820}{{\tt arXiv:1701.02820}}].

\bibitem{Murugan:2017eto}
J.~Murugan, D.~Stanford, and E.~Witten, {\it {More on Supersymmetric and 2d
  Analogs of the SYK Model}},  {\em JHEP} {\bf 08} (2017) 146,
  [\href{http://arxiv.org/abs/1706.05362}{{\tt arXiv:1706.05362}}].

\bibitem{Maldacena:2002vr}
J.~M. Maldacena, {\it {Non-Gaussian features of primordial fluctuations in
  single field inflationary models}},  {\em JHEP} {\bf 05} (2003) 013,
  [\href{http://arxiv.org/abs/astro-ph/0210603}{{\tt astro-ph/0210603}}].

\bibitem{Maldacena:2011nz}
J.~M. Maldacena and G.~L. Pimentel, {\it {On graviton non-Gaussianities during
  inflation}},  {\em JHEP} {\bf 09} (2011) 045,
  [\href{http://arxiv.org/abs/1104.2846}{{\tt arXiv:1104.2846}}].

\bibitem{Mata:2012bx}
I.~Mata, S.~Raju, and S.~Trivedi, {\it {CMB from CFT}},  {\em JHEP} {\bf 07}
  (2013) 015, [\href{http://arxiv.org/abs/1211.5482}{{\tt arXiv:1211.5482}}].

\bibitem{Kundu:2014gxa}
N.~Kundu, A.~Shukla, and S.~P. Trivedi, {\it {Constraints from Conformal
  Symmetry on the Three Point Scalar Correlator in Inflation}},  {\em JHEP}
  {\bf 04} (2015) 061, [\href{http://arxiv.org/abs/1410.2606}{{\tt
  arXiv:1410.2606}}].

\bibitem{Ghosh:2014kba}
A.~Ghosh, N.~Kundu, S.~Raju, and S.~P. Trivedi, {\it {Conformal Invariance and
  the Four Point Scalar Correlator in Slow-Roll Inflation}},  {\em JHEP} {\bf
  07} (2014) 011, [\href{http://arxiv.org/abs/1401.1426}{{\tt
  arXiv:1401.1426}}].

\bibitem{Arkani-Hamed:2015bza}
N.~Arkani-Hamed and J.~Maldacena, {\it {Cosmological Collider Physics}},
  \href{http://arxiv.org/abs/1503.08043}{{\tt arXiv:1503.08043}}.

\bibitem{Kundu:2015xta}
N.~Kundu, A.~Shukla, and S.~P. Trivedi, {\it {Ward Identities for Scale and
  Special Conformal Transformations in Inflation}},  {\em JHEP} {\bf 01} (2016)
  046, [\href{http://arxiv.org/abs/1507.06017}{{\tt arXiv:1507.06017}}].

\bibitem{Sleight:2019mgd}
C.~Sleight, {\it {A Mellin Space Approach to Cosmological Correlators}},  {\em
  JHEP} {\bf 01} (2020) 090, [\href{http://arxiv.org/abs/1906.12302}{{\tt
  arXiv:1906.12302}}].

\bibitem{Sleight:2019hfp}
C.~Sleight and M.~Taronna, {\it {Bootstrapping Inflationary Correlators in
  Mellin Space}},  {\em JHEP} {\bf 02} (2020) 098,
  [\href{http://arxiv.org/abs/1907.01143}{{\tt arXiv:1907.01143}}].

\bibitem{Sleight:2020obc}
C.~Sleight and M.~Taronna, {\it {From AdS to dS Exchanges: Spectral
  Representation, Mellin Amplitudes and Crossing}},
  \href{http://arxiv.org/abs/2007.09993}{{\tt arXiv:2007.09993}}.

\bibitem{Arkani-Hamed:2017fdk}
N.~Arkani-Hamed, P.~Benincasa, and A.~Postnikov, {\it {Cosmological Polytopes
  and the Wavefunction of the Universe}},
  \href{http://arxiv.org/abs/1709.02813}{{\tt arXiv:1709.02813}}.

\bibitem{Arkani-Hamed:2018bjr}
N.~Arkani-Hamed and P.~Benincasa, {\it {On the Emergence of Lorentz Invariance
  and Unitarity from the Scattering Facet of Cosmological Polytopes}},
  \href{http://arxiv.org/abs/1811.01125}{{\tt arXiv:1811.01125}}.

\bibitem{Benincasa:2018ssx}
P.~Benincasa, {\it {From the flat-space S-matrix to the Wavefunction of the
  Universe}},  \href{http://arxiv.org/abs/1811.02515}{{\tt arXiv:1811.02515}}.

\bibitem{Benincasa:2019vqr}
P.~Benincasa, {\it {Cosmological Polytopes and the Wavefuncton of the Universe
  for Light States}},  \href{http://arxiv.org/abs/1909.02517}{{\tt
  arXiv:1909.02517}}.

\bibitem{Arkani-Hamed:2018kmz}
N.~Arkani-Hamed, D.~Baumann, H.~Lee, and G.~L. Pimentel, {\it {The Cosmological
  Bootstrap: Inflationary Correlators from Symmetries and Singularities}},
  {\em JHEP} {\bf 04} (2020) 105, [\href{http://arxiv.org/abs/1811.00024}{{\tt
  arXiv:1811.00024}}].

\bibitem{Baumann:2019oyu}
D.~Baumann, C.~Duaso~Pueyo, A.~Joyce, H.~Lee, and G.~L. Pimentel, {\it {The
  Cosmological Bootstrap: Weight-Shifting Operators and Scalar Seeds}},
  \href{http://arxiv.org/abs/1910.14051}{{\tt arXiv:1910.14051}}.

\bibitem{Baumann:2020dch}
D.~Baumann, C.~Duaso~Pueyo, A.~Joyce, H.~Lee, and G.~L. Pimentel, {\it {The
  Cosmological Bootstrap: Spinning Correlators from Symmetries and
  Factorization}},  \href{http://arxiv.org/abs/2005.04234}{{\tt
  arXiv:2005.04234}}.

\bibitem{Raju:2010by}
S.~Raju, {\it {BCFW for Witten Diagrams}},  {\em Phys. Rev. Lett.} {\bf 106}
  (2011) 091601, [\href{http://arxiv.org/abs/1011.0780}{{\tt
  arXiv:1011.0780}}].

\bibitem{Raju:2011mp}
S.~Raju, {\it {Recursion Relations for AdS/CFT Correlators}},  {\em Phys. Rev.}
  {\bf D83} (2011) 126002, [\href{http://arxiv.org/abs/1102.4724}{{\tt
  arXiv:1102.4724}}].

\bibitem{Raju:2012zs}
S.~Raju, {\it {Four Point Functions of the Stress Tensor and Conserved Currents
  in AdS$_4$/CFT$_3$}},  {\em Phys. Rev. D} {\bf 85} (2012) 126008,
  [\href{http://arxiv.org/abs/1201.6452}{{\tt arXiv:1201.6452}}].

\bibitem{Isono:2018rrb}
H.~Isono, T.~Noumi, and G.~Shiu, {\it {Momentum space approach to crossing
  symmetric CFT correlators}},  {\em JHEP} {\bf 07} (2018) 136,
  [\href{http://arxiv.org/abs/1805.11107}{{\tt arXiv:1805.11107}}].

\bibitem{Isono:2019wex}
H.~Isono, T.~Noumi, and G.~Shiu, {\it {Momentum space approach to crossing
  symmetric CFT correlators. Part II. General spacetime dimension}},  {\em
  JHEP} {\bf 10} (2019) 183, [\href{http://arxiv.org/abs/1908.04572}{{\tt
  arXiv:1908.04572}}].

\bibitem{Farrow:2018yni}
J.~A. Farrow, A.~E. Lipstein, and P.~McFadden, {\it {Double copy structure of
  CFT correlators}},  {\em JHEP} {\bf 02} (2019) 130,
  [\href{http://arxiv.org/abs/1812.11129}{{\tt arXiv:1812.11129}}].

\bibitem{Lipstein:2019mpu}
A.~E. Lipstein and P.~McFadden, {\it {Double copy structure and the flat space
  limit of conformal correlators in even dimensions}},
  \href{http://arxiv.org/abs/1912.10046}{{\tt arXiv:1912.10046}}.

\bibitem{Albayrak:2018tam}
S.~Albayrak and S.~Kharel, {\it {Towards the higher point holographic momentum
  space amplitudes}},  {\em JHEP} {\bf 02} (2019) 040,
  [\href{http://arxiv.org/abs/1810.12459}{{\tt arXiv:1810.12459}}].

\bibitem{Albayrak:2019asr}
S.~Albayrak, C.~Chowdhury, and S.~Kharel, {\it {New relation for Witten
  diagrams}},  {\em JHEP} {\bf 10} (2019) 274,
  [\href{http://arxiv.org/abs/1904.10043}{{\tt arXiv:1904.10043}}].

\bibitem{Albayrak:2019yve}
S.~Albayrak and S.~Kharel, {\it {Towards the higher point holographic momentum
  space amplitudes II: Gravitons}},  {\em JHEP} {\bf 12} (2019) 135,
  [\href{http://arxiv.org/abs/1908.01835}{{\tt arXiv:1908.01835}}].

\bibitem{Albayrak:2020isk}
S.~Albayrak, C.~Chowdhury, and S.~Kharel, {\it {An étude of momentum space
  scalar amplitudes in AdS}},  {\em Phys. Rev. D} {\bf 101} (2020) 124043,
  [\href{http://arxiv.org/abs/2001.06777}{{\tt arXiv:2001.06777}}].

\bibitem{Albayrak:2020bso}
S.~Albayrak and S.~Kharel, {\it {On spinning loop amplitudes in Anti-de Sitter
  space}},  \href{http://arxiv.org/abs/2006.12540}{{\tt arXiv:2006.12540}}.

\bibitem{Aharony:2016dwx}
O.~Aharony, L.~F. Alday, A.~Bissi, and E.~Perlmutter, {\it {Loops in AdS from
  Conformal Field Theory}},  \href{http://arxiv.org/abs/1612.03891}{{\tt
  arXiv:1612.03891}}.

\bibitem{Alday:2017vkk}
L.~F. Alday and S.~Caron-Huot, {\it {Gravitational S-matrix from CFT dispersion
  relations}},  {\em JHEP} {\bf 12} (2018) 017,
  [\href{http://arxiv.org/abs/1711.02031}{{\tt arXiv:1711.02031}}].

\bibitem{Ponomarev:2019ofr}
D.~Ponomarev, {\it {From bulk loops to boundary large-N expansion}},
  \href{http://arxiv.org/abs/1908.03974}{{\tt arXiv:1908.03974}}.

\bibitem{Leonhardt:2003qu}
T.~Leonhardt, R.~Manvelyan, and W.~Ruhl, {\it {The Group approach to AdS space
  propagators}},  {\em Nucl. Phys.} {\bf B667} (2003) 413--434,
  [\href{http://arxiv.org/abs/hep-th/0305235}{{\tt hep-th/0305235}}].

\bibitem{Costa:2014kfa}
M.~S. Costa, V.~Goncalves, and J.~Penedones, {\it {Spinning AdS Propagators}},
  {\em JHEP} {\bf 09} (2014) 064, [\href{http://arxiv.org/abs/1404.5625}{{\tt
  arXiv:1404.5625}}].

\bibitem{Paulos:2016fap}
M.~F. Paulos, J.~Penedones, J.~Toledo, B.~C. van Rees, and P.~Vieira, {\it {The
  S-matrix bootstrap. Part I: QFT in AdS}},  {\em JHEP} {\bf 11} (2017) 133,
  [\href{http://arxiv.org/abs/1607.06109}{{\tt arXiv:1607.06109}}].

\bibitem{Gillioz:2016jnn}
M.~Gillioz, X.~Lu, and M.~A. Luty, {\it {Scale Anomalies, States, and Rates in
  Conformal Field Theory}},  {\em JHEP} {\bf 04} (2017) 171,
  [\href{http://arxiv.org/abs/1612.07800}{{\tt arXiv:1612.07800}}].

\bibitem{Gillioz:2018kwh}
M.~Gillioz, X.~Lu, and M.~A. Luty, {\it {Graviton Scattering and a Sum Rule for
  the c Anomaly in 4D CFT}},  {\em JHEP} {\bf 09} (2018) 025,
  [\href{http://arxiv.org/abs/1801.05807}{{\tt arXiv:1801.05807}}].

\bibitem{Gillioz:2018mto}
M.~Gillioz, {\it {Momentum-space conformal blocks on the light cone}},  {\em
  JHEP} {\bf 10} (2018) 125, [\href{http://arxiv.org/abs/1807.07003}{{\tt
  arXiv:1807.07003}}].

\bibitem{Polyakov:1974gs}
A.~M. Polyakov, {\it {Nonhamiltonian approach to conformal quantum field
  theory}},  {\em Zh. Eksp. Teor. Fiz.} {\bf 66} (1974) 23--42.

\bibitem{ssw}
D.~Simmons-Duffin, D.~Stanford, and E.~Witten, {\it {A spacetime derivation of
  the Lorentzian OPE inversion formula}},
  \href{http://arxiv.org/abs/1711.03816}{{\tt arXiv:1711.03816}}.

\bibitem{Kravchuk:2018htv}
P.~Kravchuk and D.~Simmons-Duffin, {\it {Light-ray operators in conformal field
  theory}},  \href{http://arxiv.org/abs/1805.00098}{{\tt arXiv:1805.00098}}.

\bibitem{Streater:1989vi}
R.~F. Streater and A.~S. Wightman, {\em {PCT, spin and statistics, and all
  that}}.
\newblock 1989.

\bibitem{Haag:1992hx}
R.~Haag, {\em {Local quantum physics: Fields, particles, algebras}}.
\newblock 1992.

\bibitem{Balasubramanian:1998sn}
V.~Balasubramanian, P.~Kraus, and A.~E. Lawrence, {\it {Bulk versus boundary
  dynamics in anti-de Sitter space-time}},  {\em Phys. Rev.} {\bf D59} (1999)
  046003, [\href{http://arxiv.org/abs/hep-th/9805171}{{\tt hep-th/9805171}}].

\bibitem{Balasubramanian:1998de}
V.~Balasubramanian, P.~Kraus, A.~E. Lawrence, and S.~P. Trivedi, {\it
  {Holographic probes of anti-de Sitter space-times}},  {\em Phys. Rev.} {\bf
  D59} (1999) 104021, [\href{http://arxiv.org/abs/hep-th/9808017}{{\tt
  hep-th/9808017}}].

\bibitem{Bogolyubov:1990kw}
N.~Bogolyubov, A.~Logunov, A.~Oksak, and I.~Todorov, {\em {General principles
  of quantum field theory}}.
\newblock 12, 1990.

\bibitem{Meltzer:2021bmb}
D.~Meltzer, {\it {Dispersion Formulas in QFTs, CFTs, and Holography}},
  \href{http://arxiv.org/abs/2103.15839}{{\tt arXiv:2103.15839}}.

\bibitem{Schweber:1961zz}
S.~S. Schweber, {\it {An Introduction to Relativistic Quantum Field Theory}}, .

\bibitem{tHooft:1973wag}
G.~'t~Hooft and M.~J.~G. Veltman, {\it {DIAGRAMMAR}},  {\em NATO Sci. Ser. B}
  {\bf 4} (1974) 177--322.

\bibitem{Veltman:1994wz}
M.~J.~G. Veltman, {\it {Diagrammatica: The Path to Feynman rules}},  {\em
  Cambridge Lect. Notes Phys.} {\bf 4} (1994) 1--284.

\bibitem{Liu:1998ty}
H.~Liu and A.~A. Tseytlin, {\it {On four point functions in the CFT / AdS
  correspondence}},  {\em Phys. Rev.} {\bf D59} (1999) 086002,
  [\href{http://arxiv.org/abs/hep-th/9807097}{{\tt hep-th/9807097}}].

\bibitem{Freedman:1998tz}
D.~Z. Freedman, S.~D. Mathur, A.~Matusis, and L.~Rastelli, {\it {Correlation
  functions in the CFT(d) / AdS(d+1) correspondence}},  {\em Nucl. Phys.} {\bf
  B546} (1999) 96--118, [\href{http://arxiv.org/abs/hep-th/9804058}{{\tt
  hep-th/9804058}}].

\bibitem{Fitzpatrick:2011dm}
A.~Fitzpatrick and J.~Kaplan, {\it {Unitarity and the Holographic S-Matrix}},
  {\em JHEP} {\bf 10} (2012) 032, [\href{http://arxiv.org/abs/1112.4845}{{\tt
  arXiv:1112.4845}}].

\bibitem{Avis:1977yn}
S.~Avis, C.~Isham, and D.~Storey, {\it {Quantum Field Theory in anti-De Sitter
  Space-Time}},  {\em Phys. Rev. D} {\bf 18} (1978) 3565.

\bibitem{Breitenlohner:1982jf}
P.~Breitenlohner and D.~Z. Freedman, {\it {Stability in Gauged Extended
  Supergravity}},  {\em Annals Phys.} {\bf 144} (1982) 249.

\bibitem{Breitenlohner:1982bm}
P.~Breitenlohner and D.~Z. Freedman, {\it {Positive Energy in anti-De Sitter
  Backgrounds and Gauged Extended Supergravity}},  {\em Phys. Lett. B} {\bf
  115} (1982) 197--201.

\bibitem{Gillioz:2019lgs}
M.~Gillioz, {\it {Conformal 3-point functions and the Lorentzian OPE in
  momentum space}},  \href{http://arxiv.org/abs/1909.00878}{{\tt
  arXiv:1909.00878}}.

\bibitem{Gillioz:2019iye}
M.~Gillioz, X.~Lu, M.~A. Luty, and G.~Mikaberidze, {\it {Convergent
  Momentum-Space OPE and Bootstrap Equations in Conformal Field Theory}},
  \href{http://arxiv.org/abs/1912.05550}{{\tt arXiv:1912.05550}}.

\bibitem{Yuan:2017vgp}
E.~Y. Yuan, {\it {Loops in the Bulk}},
  \href{http://arxiv.org/abs/1710.01361}{{\tt arXiv:1710.01361}}.

\bibitem{Yuan:2018qva}
E.~Y. Yuan, {\it {Simplicity in AdS Perturbative Dynamics}},
  \href{http://arxiv.org/abs/1801.07283}{{\tt arXiv:1801.07283}}.

\bibitem{Okuda:2010ym}
T.~Okuda and J.~Penedones, {\it {String scattering in flat space and a scaling
  limit of Yang-Mills correlators}},  {\em Phys. Rev. D} {\bf 83} (2011)
  086001, [\href{http://arxiv.org/abs/1002.2641}{{\tt arXiv:1002.2641}}].

\bibitem{Alday:2018kkw}
L.~F. Alday, {\it {On Genus-one String Amplitudes on $AdS_5 \times S^5$}},
  \href{http://arxiv.org/abs/1812.11783}{{\tt arXiv:1812.11783}}.

\bibitem{Bissi:2020wtv}
A.~Bissi, G.~Fardelli, and A.~Georgoudis, {\it {Towards All Loop Supergravity
  Amplitudes on $AdS_5 \times S^5$}},
  \href{http://arxiv.org/abs/2002.04604}{{\tt arXiv:2002.04604}}.

\bibitem{Dusedau:1985ue}
D.~W. Dusedau and D.~Z. Freedman, {\it {Lehmann Spectral Representation for
  Anti-de Sitter Quantum Field Theory}},  {\em Phys. Rev.} {\bf D33} (1986)
  389.

\bibitem{Fitzpatrick:2011hu}
A.~L. Fitzpatrick and J.~Kaplan, {\it {Analyticity and the Holographic
  S-Matrix}},  {\em JHEP} {\bf 1210} (2012) 127,
  [\href{http://arxiv.org/abs/1111.6972}{{\tt arXiv:1111.6972}}].

\bibitem{Engelund:2012re}
O.~T. Engelund and R.~Roiban, {\it {Correlation functions of local composite
  operators from generalized unitarity}},  {\em JHEP} {\bf 03} (2013) 172,
  [\href{http://arxiv.org/abs/1209.0227}{{\tt arXiv:1209.0227}}].

\bibitem{Caron-Huot:2018kta}
S.~Caron-Huot and A.-K. Trinh, {\it {All tree-level correlators in $AdS_5\times
  S_5$ supergravity: hidden ten-dimensional conformal symmetry}},  {\em JHEP}
  {\bf 01} (2019) 196, [\href{http://arxiv.org/abs/1809.09173}{{\tt
  arXiv:1809.09173}}].

\bibitem{Rastelli:2019gtj}
L.~Rastelli, K.~Roumpedakis, and X.~Zhou, {\it {$\mathbf{AdS_3\times S^3}$
  Tree-Level Correlators: Hidden Six-Dimensional Conformal Symmetry}},
  \href{http://arxiv.org/abs/1905.11983}{{\tt arXiv:1905.11983}}.

\bibitem{Giusto:2020neo}
S.~Giusto, R.~Russo, A.~Tyukov, and C.~Wen, {\it {The CFT$_6$ origin of all
  tree-level 4-point correlators in AdS$_3 \times S^3$}},
  \href{http://arxiv.org/abs/2005.08560}{{\tt arXiv:2005.08560}}.

\bibitem{Rastelli:2017udc}
L.~Rastelli and X.~Zhou, {\it {How to Succeed at Holographic Correlators
  Without Really Trying}},  {\em JHEP} {\bf 04} (2018) 014,
  [\href{http://arxiv.org/abs/1710.05923}{{\tt arXiv:1710.05923}}].

\bibitem{Aprile:2018efk}
F.~Aprile, J.~Drummond, P.~Heslop, and H.~Paul, {\it {Double-trace spectrum of
  $N=4$ supersymmetric Yang-Mills theory at strong coupling}},  {\em Phys.
  Rev.} {\bf D98} (2018), no.~12 126008,
  [\href{http://arxiv.org/abs/1802.06889}{{\tt arXiv:1802.06889}}].

\bibitem{Mack:2009gy}
G.~Mack, {\it {D-dimensional Conformal Field Theories with anomalous dimensions
  as Dual Resonance Models}},  {\em Bulg.J.Phys.} {\bf 36} (2009) 214--226,
  [\href{http://arxiv.org/abs/0909.1024}{{\tt arXiv:0909.1024}}].

\bibitem{Fitzpatrick:2011ia}
A.~L. Fitzpatrick, J.~Kaplan, J.~Penedones, S.~Raju, and B.~C. van Rees, {\it
  {A Natural Language for AdS/CFT Correlators}},  {\em JHEP} {\bf 1111} (2011)
  095, [\href{http://arxiv.org/abs/1107.1499}{{\tt arXiv:1107.1499}}].

\bibitem{Alday:2018pdi}
L.~F. Alday, A.~Bissi, and E.~Perlmutter, {\it {Genus-One String Amplitudes
  from Conformal Field Theory}},  {\em JHEP} {\bf 06} (2019) 010,
  [\href{http://arxiv.org/abs/1809.10670}{{\tt arXiv:1809.10670}}].

\bibitem{Binder:2019jwn}
D.~J. Binder, S.~M. Chester, S.~S. Pufu, and Y.~Wang, {\it {$\mathcal{N}=4$
  Super-Yang-Mills Correlators at Strong Coupling from String Theory and
  Localization}},  \href{http://arxiv.org/abs/1902.06263}{{\tt
  arXiv:1902.06263}}.

\bibitem{Chester:2019pvm}
S.~M. Chester, {\it {Genus-2 Holographic Correlator on AdS5 x S5 from
  Localization}},  \href{http://arxiv.org/abs/1908.05247}{{\tt
  arXiv:1908.05247}}.

\bibitem{Chester:2020dja}
S.~M. Chester and S.~S. Pufu, {\it {Far Beyond the Planar Limit in
  Strongly-Coupled $\mathcal{N}=4$ SYM}},
  \href{http://arxiv.org/abs/2003.08412}{{\tt arXiv:2003.08412}}.

\bibitem{Alday:2019nin}
L.~F. Alday and X.~Zhou, {\it {Simplicity of AdS Supergravity at One Loop}},
  \href{http://arxiv.org/abs/1912.02663}{{\tt arXiv:1912.02663}}.

\bibitem{Alday:2020tgi}
L.~F. Alday, S.~M. Chester, and H.~Raj, {\it {6d (2,0) and M-theory at
  1-loop}},  \href{http://arxiv.org/abs/2005.07175}{{\tt arXiv:2005.07175}}.

\bibitem{Drummond:2019hel}
J.~Drummond and H.~Paul, {\it {One-loop string corrections to AdS amplitudes
  from CFT}},  \href{http://arxiv.org/abs/1912.07632}{{\tt arXiv:1912.07632}}.

\bibitem{Aprile:2017bgs}
F.~Aprile, J.~Drummond, P.~Heslop, and H.~Paul, {\it {Quantum Gravity from
  Conformal Field Theory}},  {\em JHEP} {\bf 01} (2018) 035,
  [\href{http://arxiv.org/abs/1706.02822}{{\tt arXiv:1706.02822}}].

\bibitem{Aprile:2017qoy}
F.~Aprile, J.~Drummond, P.~Heslop, and H.~Paul, {\it {Loop corrections for
  Kaluza-Klein AdS amplitudes}},  {\em JHEP} {\bf 05} (2018) 056,
  [\href{http://arxiv.org/abs/1711.03903}{{\tt arXiv:1711.03903}}].

\bibitem{Aprile:2019rep}
F.~Aprile, J.~Drummond, P.~Heslop, and H.~Paul, {\it {One-loop amplitudes in
  AdS5xS5 supergravity from N = 4 SYM at strong coupling}},  {\em JHEP} {\bf
  03} (2020) 190, [\href{http://arxiv.org/abs/1912.01047}{{\tt
  arXiv:1912.01047}}].

\bibitem{Feynman:1963ax}
R.~P. Feynman, {\it {Quantum theory of gravitation}},  {\em Acta Phys. Polon.}
  {\bf 24} (1963) 697--722. [,272(1963)].

\bibitem{CaronHuot:2010zt}
S.~Caron-Huot, {\it {Loops and trees}},  {\em JHEP} {\bf 05} (2011) 080,
  [\href{http://arxiv.org/abs/1007.3224}{{\tt arXiv:1007.3224}}].

\bibitem{Haehl:2017eob}
F.~M. Haehl, R.~Loganayagam, P.~Narayan, A.~A. Nizami, and M.~Rangamani, {\it
  {Thermal out-of-time-order correlators, KMS relations, and spectral
  functions}},  {\em JHEP} {\bf 12} (2017) 154,
  [\href{http://arxiv.org/abs/1706.08956}{{\tt arXiv:1706.08956}}].

\bibitem{Tomboulis:2017rvd}
E.~Tomboulis, {\it {Causality and Unitarity via the Tree-Loop Duality
  Relation}},  {\em JHEP} {\bf 05} (2017) 148,
  [\href{http://arxiv.org/abs/1701.07052}{{\tt arXiv:1701.07052}}].

\bibitem{Herzog:2002pc}
C.~Herzog and D.~Son, {\it {Schwinger-Keldysh propagators from AdS/CFT
  correspondence}},  {\em JHEP} {\bf 03} (2003) 046,
  [\href{http://arxiv.org/abs/hep-th/0212072}{{\tt hep-th/0212072}}].

\end{thebibliography}\endgroup
\end{document}